\documentclass[twocolumn]{aastex631}

\usepackage{amsmath}
\usepackage{multirow}
\usepackage{xspace}  
\usepackage[caption=false]{subfig}
\usepackage{graphicx}
\usepackage[]{threeparttable}

\newcommand{\prospector}{{\sc Prospector}\xspace}
\newcommand{\tlookback}{{$t_{\rm{lookback}}$}\xspace}

\begin{document}

\title{JADES $+$ JEMS: A Detailed Look at the Buildup of Central Stellar Cores and Suppression of Star Formation in Galaxies at Redshifts $3<z<4.5$}

\correspondingauthor{Zhiyuan Ji}
\email{zhiyuanji@arizona.edu}

\author[0000-0001-7673-2257]{Zhiyuan Ji}
\affiliation{Steward Observatory, University of Arizona, 933 N. Cherry Avenue, Tucson, AZ 85721, USA}

\author[0000-0003-2919-7495]{Christina C.\ Williams}
\affiliation{NSF’s National Optical-Infrared Astronomy Research Laboratory, 950 North Cherry Avenue, Tucson, AZ 85719, USA}
\affiliation{Steward Observatory, University of Arizona, 933 N. Cherry Avenue, Tucson, AZ 85721, USA}

\author[0000-0002-8224-4505]{Sandro Tacchella}
\affiliation{Kavli Institute for Cosmology, University of Cambridge, Madingley Road, Cambridge, CB3 0HA, UK}
\affiliation{Cavendish Laboratory, University of Cambridge, 19 JJ Thomson Avenue, Cambridge, CB3 0HE, UK}

\author[0000-0002-1714-1905]{Katherine A.\ Suess}
\affiliation{Department of Astronomy and Astrophysics, University of California, Santa Cruz, 1156 High Street, Santa Cruz, CA 95064 USA}
\affiliation{Kavli Institute for Particle Astrophysics and Cosmology and Department of Physics, Stanford University, Stanford, CA 94305, USA}

\author[0000-0003-0215-1104]{William M.\ Baker}
\affiliation{Kavli Institute for Cosmology, University of Cambridge, Madingley Road, Cambridge, CB3 0HA, UK}
\affiliation{Cavendish Laboratory, University of Cambridge, 19 JJ Thomson Avenue, Cambridge, CB3 0HE, UK}

\author[0000-0002-8909-8782]{Stacey Alberts}
\affiliation{Steward Observatory, University of Arizona, 933 N. Cherry Avenue, Tucson, AZ 85721, USA}

\author[0000-0002-8651-9879]{Andrew J. Bunker}
\affiliation{Department of Physics, University of Oxford, Denys Wilkinson Building, Keble Road, Oxford OX1 3RH, UK}

\author[0000-0002-9280-7594]{Benjamin D.\ Johnson}
\affiliation{Center for Astrophysics $|$ Harvard \& Smithsonian, 60 Garden St., Cambridge MA 02138 USA}

\author[0000-0002-4271-0364]{Brant Robertson}
\affiliation{Department of Astronomy and Astrophysics, University of California, Santa Cruz, 1156 High Street, Santa Cruz, CA 95064, USA}

\author[0000-0002-4622-6617]{Fengwu Sun}
\affiliation{Steward Observatory, University of Arizona, 933 N. Cherry Avenue, Tucson, AZ 85721, USA}

\author[0000-0002-2929-3121]{Daniel J.\ Eisenstein}
\affiliation{Center for Astrophysics $|$ Harvard \& Smithsonian, 60 Garden St., Cambridge MA 02138 USA}

\author[0000-0002-7893-6170]{Marcia Rieke}
\affiliation{Steward Observatory, University of Arizona, 933 N. Cherry Avenue, Tucson, AZ 85721, USA}

\author[0000-0003-0695-4414]{Michael V.\ Maseda}
\affiliation{Department of Astronomy, University of Wisconsin-Madison, 475 N. Charter St., Madison, WI 53706 USA}

\author[0000-0003-4565-8239]{Kevin Hainline}
\affiliation{Steward Observatory, University of Arizona, 933 N. Cherry Avenue, Tucson, AZ 85721, USA}

\author[0000-0002-8543-761X]{Ryan Hausen}
\affiliation{Department of Physics and Astronomy, The Johns Hopkins University, 3400 N. Charles St.,
Baltimore, MD 21218, USA}

\author[0000-0003-2303-6519]{George Rieke}
\affiliation{Steward Observatory, University of Arizona, 933 N. Cherry Avenue, Tucson, AZ 85721, USA}

\author[0000-0001-9262-9997]{Christopher N. A. Willmer}
\affiliation{Steward Observatory, University of Arizona, 933 N. Cherry Avenue, Tucson, AZ 85721, USA}

\author[0000-0003-1344-9475]{Eiichi Egami}
\affiliation{Steward Observatory, University of Arizona, 933 N. Cherry Avenue, Tucson, AZ 85721, USA}

\author[0000-0003-4702-7561]{Irene Shivaei}
\affiliation{Steward Observatory, University of Arizona, 933 N. Cherry Avenue, Tucson, AZ 85721, USA}

\author[0000-0002-6719-380X]{Stefano Carniani}
\affiliation{Scuola Normale Superiore, Piazza dei Cavalieri 7, I-56126 Pisa, Italy}

\author[0000-0003-3458-2275]{Stephane Charlot}
\affiliation{Sorbonne Universit\'e, CNRS, UMR 7095, Institut d'Astrophysique de Paris, 98 bis bd Arago, 75014 Paris, France}

\author[0000-0002-7636-0534]{Jacopo Chevallard}
\affiliation{Department of Physics, University of Oxford, Denys Wilkinson Building, Keble Road, Oxford OX1 3RH, UK}

\author[0000-0002-9551-0534]{Emma Curtis-Lake}
\affiliation{Centre for Astrophysics Research, Department of Physics, Astronomy and Mathematics, University of Hertfordshire, Hatfield AL10 9AB, UK}

\author[0000-0002-3642-2446]{Tobias J.\ Looser}
\affiliation{Kavli Institute for Cosmology, University of Cambridge, Madingley Road, Cambridge, CB3 0HA, UK}
\affiliation{Cavendish Laboratory, University of Cambridge, 19 JJ Thomson Avenue, Cambridge, CB3 0HE, UK}

\author[0000-0002-4985-3819]{Roberto Maiolino}
\affiliation{Kavli Institute for Cosmology, University of Cambridge, Madingley Road, Cambridge, CB3 0HA, UK}
\affiliation{Cavendish Laboratory, University of Cambridge, 19 JJ Thomson Avenue, Cambridge, CB3 0HE, UK}
\affiliation{Department of Physics and Astronomy, University College London, Gower Street, London WC1E 6BT, UK}

\author[0000-0002-4201-7367]{Chris Willott}
\affiliation{NRC Herzberg, 5071 West Saanich Rd, Victoria, BC V9E 2E7, Canada}

\author[0000-0002-2178-5471]{Zuyi Chen}
\affiliation{Steward Observatory, University of Arizona, 933 N. Cherry Avenue, Tucson, AZ 85721, USA}

\author[0000-0003-4337-6211]{Jakob M. Helton}
\affiliation{Steward Observatory, University of Arizona, 933 N. Cherry Avenue, Tucson, AZ 85721, USA}

\author[0000-0002-6221-1829]{Jianwei Lyu}
\affiliation{Steward Observatory, University of Arizona, 933 N. Cherry Avenue, Tucson, AZ 85721, USA}

\author[0000-0002-7524-374X]{Erica Nelson}
\affiliation{Department for Astrophysical and Planetary Science, University of Colorado, Boulder, CO 80309, USA}

\author[0000-0003-0883-2226]{Rachana Bhatawdekar}
\affiliation{European Space Agency (ESA), European Space Astronomy Centre (ESAC), Camino Bajo del Castillo s/n, 28692 Villanueva de la Cañada, Madrid, Spain; European Space Agency, ESA/ESTEC, Keplerlaan 1, 2201 AZ Noordwijk, NL}

\author[0000-0003-4109-304X]{Kristan Boyett}
\affiliation{School of Physics, University of Melbourne, Parkville 3010, VIC, Australia}
\affiliation{ARC Centre of Excellence for All Sky Astrophysics in 3 Dimensions (ASTRO 3D), Australia}

\author[0000-0001-9276-7062]{Lester Sandles}
\affiliation{Kavli Institute for Cosmology, University of Cambridge, Madingley Road, Cambridge, CB3 0HA, UK}
\affiliation{Cavendish Laboratory, University of Cambridge, 19 JJ Thomson Avenue, Cambridge, CB3 0HE, UK}

\begin{abstract}

We present a spatially resolved study of stellar populations in 6 galaxies with stellar masses $M_*\sim10^{10}M_\sun$ at $z\sim3.7$ using 14-filter JWST/NIRCam imaging from the JADES and JEMS surveys. The 6 galaxies are visually selected to have clumpy substructures with distinct colors over rest-frame $3600-4100$ \AA, including a red, dominant stellar core that is close to their stellar-light centroids. With 23-filter photometry from HST to JWST, we measure the stellar-population properties of individual structural components via SED fitting using \prospector. We find that the central stellar cores are $\gtrsim2$ times more massive than the Toomre mass, indicating they may not form via single in-situ fragmentation. The stellar cores have stellar ages of $0.4-0.7$ Gyr that are similar to the timescale of clump inward migration due to dynamical friction, suggesting that they likely instead formed through the coalescence of giant stellar clumps. While they have not yet quenched, the 6 galaxies are below the star-forming main sequence by $0.2-0.7$ dex. Within each galaxy, we find that the specific star formation rate is lower in the central stellar core, and the stellar-mass surface density of the core is already similar to quenched galaxies of the same masses and redshifts. Meanwhile, the stellar ages of the cores are either comparable to or younger than the extended, smooth parts of the galaxies. Our findings are consistent with model predictions of the gas-rich compaction scenario for the buildup of galaxies' central regions at high redshifts. We are likely witnessing the coeval formation of  dense central cores, along with the onset of galaxy-wide quenching at $z>3$.  

\end{abstract}


\section{Introduction} \label{sec:intro}

Over the last two decades, deep imaging surveys with the Hubble Space Telescope (HST), such as GOODS \citep[][]{Giavalisco2004}, CANDELS \citep{Grogin2011,Koekemoer2011} and Hubble Frontier Fields \citep{Lotz2017}, have obtained more than $100,000$ rest-frame optical images for mass complete samples of galaxies up to redshift $z=3$. These observations have not only greatly advanced our understanding of galaxies' structural evolution across cosmic time, but also raised a number of new questions. 

It immediately becomes apparent after analyzing those HST images of $1<z<3$ galaxies that they have distinct structural and morphological properties. Relative to galaxies in the local universe, massive galaxies at $z\sim2$ are on average smaller and denser \citep[e.g.,][]{Ferguson2004,vanderwel2014,Shibuya2015,Barro2017}, and their morphologies are irregular and often show large-scale asymmetries \citep[e.g.,][]{Cowie1995,vandenbergh1996}. This latter morphological feature is also present in $z\sim0$ galaxies, but it has been predominately observed in galaxy mergers \citep[e.g.,][]{Conselice2000, Lotz2004}. Despite that galaxy major merger rate increases with redshift \citep[][]{Conselice2014}, spatially resolved spectroscopy revealed that at least half of the $z\sim2$ star-forming galaxies have a rotationally supported disk, arguing against they being interacting systems \citep[e.g.,][]{ForsterSchreiber2006,Genzel2006,Weiner2006,Shapiro2008,ForsterSchreiber2009,Law2009,Kassin2012,Simons2017,ForsterSchreiber2018,Wisnioski2019,ForsterSchreiber2022}. Moreover, most of the $z\sim2$ star-forming disks show several giant, kpc-scale stellar clumps \citep[e.g.,][]{Elmegreen2005,Elmegreen2007,Genzel2008,Elmegreen2009,ForsterSchreiber2011,Genzel2011,Guo2012,Genzel2014,Guo2015}, a morphological characteristic rarely seen in the local Universe. Understanding this substantial change in galaxies' structures across cosmic time is the key to constraining models not only of galaxy evolution but also of the interplay between baryonic and dark matter \citep[e.g.,][]{Genzel2020}. 

Another important process tied to  structural transformation is galaxy quenching --  the physical processes that control the transition from widespread star formation to quiescence in galaxies. Since $z=3$, star-forming galaxies showed a gradual transformation in their structures from disturbed to normal, smooth disks \citep{HuertasCompany2015}, while the fraction of clumpy star-forming galaxies decreased from $\approx50\%$ to $\approx5\%$ \citep[e.g.,][]{Murata2014,Guo2015}. In the meantime, there was a rapid buildup of quiescent galaxies \citep[e.g.,][]{Ilbert2010,Ilbert2013,Muzzin2013,Tomczak2014} which notably have much larger stellar-mass surface densities than star-forming galaxies of the same masses and redshifts \citep[e.g.,][]{Kauffmann2003,Brinchmann2004,Franx2008,Cheung2012,Mosleh2017}. Is there a causal link between structural transformations and quenching? While a physical link between the two phenomena has been seen in cosmological simulations \citep{Wellons2015,Zolotov2015,Tacchella2016}, this question still eludes us observationally, because the structure of galaxies is only known at the time of observation, which makes empirical constraints on the relative timing sequence of the two events difficult \citep[e.g.,][]{Bundy2010,vanderwel2011,Bruce2012,Newman2015,Toft2017,Newman2018,Ji2022a,Ji2023}. 
 
There is growing evidence that, as they head toward quiescence, galaxies develop a dense stellar core, possibly a bulge \citep[e.g.,][]{Kauffmann2003,Brinchmann2004,Cheung2012,Bruce2014,Lang2014,Nelson2014,Williams2014,Tacchella2015,Barro2017,Whitaker2017,Suess2021,Dimauro2022,Ji2023}, but the exact physical mechanism responsible for this buildup of central regions remains unknown.While major mergers and weak, non-axisymmetric instabilities (e.g., bars, spiral arms) are believed to drive the development of central structures in $z\sim0$ disks \citep{Kormendy2004}, the physical conditions, hydrodynamical instabilities in particular, are very different in high-redshift galaxies. One mechanism to effectively and efficiently grow the central regions of a galaxy at high redshift is actually a class of processes, generically referred to as ``wet/gas-rich compaction'' whose main feature is highly dissipative gas accretion toward the galactic center which, in turn, promotes enhanced activity of star formation at a higher rate relative to the average SFR \citep{Dekel2009}. Consequently, right after a compaction event, a galaxy will have a younger center relative to the rest smooth parts of the galaxy. In the absence of further gas inflow, this leads to the central gas depletion and finally the quenching of central star formation \citep[e.g.][]{Tacchella2016}. Hiydrodynamical simulations with adequate resolution to model relevant gaseous physics show that gas-rich major mergers \citep{Zolotov2015,Inoue2016}, collisions of counter-rotating gas streams \citep{Danovich2015} and violent disc instability (VDI) are all associated with gas-rich compaction. 

Regarding VDI in particular, it can be understood in a framework of Toomre instability \citep{Toomre1964,Binney2008}, where a rotating gaseous disk becomes unstable when
\begin{equation}
    Q = \frac{\sigma\kappa}{\pi G \Sigma} < Q_c \sim 1\footnote{For thick disks, $Q_c$ is about 0.7 rather than 1 \citep{Behrendt2015}.}
\end{equation}
where $\sigma$ is local gas velocity dispersion, $\kappa$ is epicyclic frequency and $\Sigma$ is local gas surface density. Observations found that star-forming galaxies at cosmic noon are characterized by large $\sigma$, with a typical level of 50 km s$^{-1}$ \citep[e.g.,][]{Genzel2008,ForsterSchreiber2009,Genzel2011} compared to only a few km s$^{-1}$ observed for nearby spiral galaxies. In addition, high-redshift galaxies have higher gas fractions, with a typical value of $\approx50\%$ compared to a few to 10\% at $z\sim0$ \citep[][]{Tacconi2020}. With these properties, high-redshift disks have $Q\sim1$ (i.e., the entire disk is unstable/marginally stable), while low-redshift disks are largely stable with $Q>1$. These lead to two direct consequences in high-redshift galaxies. First,  high-redshift disks are subject to VDI that lead to the formation of giant stellar clumps. Second, the timescale for the inward migration of clumps toward the gravitational center\footnote{The timescale is set by dynamical friction, hence it is proportional to the square of the ratio of disk rotation velocity divided by velocity dispersion, i.e. $(v_{\rm{rot}}/\sigma)^2$.} is much shorter at high redshifts than in the nearby universe \citep{Dekel2009,Genzel2011}.

A popular hypothesis is thus that massive stellar cores in high-redshift galaxies are the result of  unstable, gas-rich disks, which first fragment into giant stellar clumps that then rapidly migrate inward and finally coalesce to form dense stellar cores \citep[e.g.,][]{Noguchi1999,Immeli2004,Bournaud2007,Elmegreen2008,Dekel2009,Ceverino2010,Bournaud2014}. On the one hand, the observed radial color gradients of star-forming clumps seem to support this hypothesis \citep[][]{ForsterSchreiber2011,Guo2012,Guo2015,Shibuya2016,Soto2017,Mandelker2017}. On the other, studies of gas kinematics find evidence that strong outflows can launch from the clump sites \citep{Genzel2011}, arguing that clumps might be self-disrupted by strong stellar feedback which can significantly reduce the growth rate of central stellar cores \citep[][]{Murray2010,Genel2012,Hopkins2012,Buck2017}. Thus, it remains unknown to what extent violent disk instabilities drive the growth of central regions in high-redshift galaxies. 

The James Webb Space Telescope \citep[JWST;][]{Gardner2023} is revolutionary for studying the detailed physics of structural transformations and quenching at cosmic noon. For the first time, we can get images at rest-frame optical and NIR wavelengths, with a sensitivity that is not attainable with HST, for galaxies beyond $z>3$, closer to the time of early core/bulge formation and quenching. This guarantees more direct constraints on the underlying physics.With sub-kpc spatial resolution, JWST's Near Infrared Camera (NIRCam, \citealt{Rieke2023})  can resolve a physical scale comparable to the typical Toomre length of gaseous disks at cosmic noon \citep[e.g.,][]{Elmegreen2009,Escala2008,Dekel2009,Genzel2011}, promising to identify individual stellar substructures. Together with the ancillary HST data, imaging that covers truly panchromatic swathes from rest-frame UV through NIR enable us to put unprecedented constraints on the detailed, spatially resolved mass assembly history of galaxies at high redshifts. 

In this paper, we present the spatially resolved stellar populations in 6 galaxies at $3<z<4.5$. This redshift range represents one of the key epochs of the universe when quenching becomes a significant process \citep{Muzzin2013,Schreiber2018,Merlin2019,Shahidi2020,Carnall2023b,Carnall2023}, which can provide unique insight into the role that substructures among massive galaxies may play in the transition from star-forming disks to bulge-dominated quiescent objects. The galaxies are visually selected to have bright and red central regions. We show that these galaxies are very likely in the act of developing their stellar cores/bulges. We discuss their implications on structural transformations and quenching at cosmic noon. Throughout this paper, we adopt the $\Lambda$CDM cosmology with \citealt{Planck2020} parameters, i.e., $\Omega_m = 0.315$ and $\rm{h = H_0/(100 km\,s^{-1}\,Mpc^{-1}) = 0.673}$.

\section{Observations \& Analysis} 

\subsection{Sample Selection} \label{sec:sample}

\begin{table*}
    \centering
    \caption{Physical Properties of the Final Sample Presented in this Work.}
    \begin{tabular}{|| c c c | c c c c c||}
    \toprule
      ID  & RA & DEC  & Redshift & Redshift Ref. $^{(a)}$ & log M$_*$ $^{(b)}$ & SFR $^{(b)}$ & Stellar Age $^{(b),(c)}$\\
       & (Deg) & (Deg) &  &  & (M$_\sun$) & (M$_\sun$ yr$^{-1}$) & (Gyr) \\
    \hline
    JADES-JEMS-13396 & 53.14921 & -27.79158 & 3.559 & MUSE & 10.00 $^{+0.02}_{-0.04}$ & 13.5 $^{+0.8}_{-0.6}$ & 0.8 $^{+0.1}_{-0.1}$ \\
    JADES-JEMS-15157 & 53.13943 & -27.78009  & 3.591 & MUSE \& FRESCO $^{(d)}$ & 9.36 $^{+0.09}_{-0.11}$ & 1.3 $^{+1.0}_{-0.4}$ & 0.4 $^{+0.2}_{-0.1}$ \\
    JADES-JEMS-6885 & 53.12832 & -27.84573  & 3.698 & FRESCO & 9.87 $^{+0.03}_{-0.07}$ & 7.4 $^{+1.1}_{-0.9}$ & 0.4 $^{+0.2}_{-0.2}$  \\
    JADES-JEMS-16296 & 53.15651 & -27.77227  & 3.53 & Photometric & 9.65 $^{+0.05}_{-0.06}$ & 5.4 $^{+1.2}_{-0.9}$ & 0.5 $^{+0.3}_{-0.2}$  \\
    JADES-JEMS-14436 & 53.13306 & -27.78456  & 3.67 & Photometric & 10.34 $^{+0.07}_{-0.09}$ & 11.8 $^{+2.5}_{-1.9}$  & 0.8 $^{+0.1}_{-0.1}$  \\
    JADES-JEMS-11059 & 53.13484 & -27.80887  & 4.07 & Photometric & 9.59 $^{+0.08}_{-0.07}$ & 9.6 $^{+0.7}_{-0.6}$ & 0.4 $^{+0.1}_{-0.2}$ \\
    \hline
    \end{tabular}
    \begin{tablenotes}
        \item[](a) Redshifts are from either the MUSE/HUDF Data Release 2 \citep{Bacon2022}, or [SIII]$\lambda$9531 emission line detected in the NIRCam WFSS spectra (Sun, F. private communication) obtained by the FRESCO survey \citep{Oesch2023}, or photometric redshifts when spectroscopic ones are not available. 
        (b) Reported values are from the \prospector SED fitting with our fiducial model (Section \ref{sec:sed}).
        (c) Mass-weighted stellar age. (d) The redshift of this galaxy derived using the [SIII]$\lambda$9531 in its FRESCO spectrum is sightly different from using the Ly$\alpha$ in its MUSE spectrum, with the former being 3.591 and latter being 3.605.  
    \end{tablenotes}
    \label{tab:basic_info}
\end{table*}

This work is a pilot study focused only on 6 galaxies at $3 < z < 4.5$ having visually identified prominent, red stellar cores at rest-frame 4000\AA. Basic information of the 6 galaxies is present in Table \ref{tab:basic_info}. Utilizing the power of multi-band JWST$+$HST images, we provide a spatially resolved view of the stellar-population properties of these 6 galaxies, aiming at shedding light on the physical mechanisms responsible for the formation of central spheroidal structure in massive galaxies at cosmic noon. 

The 6 galaxies are drawn from an initial sample of HST/F160W-selected galaxies with $3<z<4.5$ from the latest CANDELS/GOODS-S catalog obtained by the ASTRODEEP project \citep{Merlin2021}. By selection, we first retain 361 galaxies in the footprint of the JWST Extragalactic Medium-band Survey (JEMS) \citep[$\approx 10$ arcmin$^2$,][JWST Cycle 1, PID: 1963]{Williams2023}, which also appear in the footprint of the JWST Advanced Deep Extragalactic Survey (JADES, \citealt{jadesprop}), a collaborative program between the NIRCam and NIRSpec GTO teams. These yield a total of 14-filter NIRCam imaging at $\lambda\approx0.8-5\micron$ (Section \ref{sec:obs}). Together with the ancillary data from HST, this high angular resolution imaging data across a total of 23 filters enables us to perform detailed stellar-population analysis in these galaxies. With the new photometry from NIRCam imaging (Section \ref{sec:obs}), we then run SED fitting with the code \prospector \citep{Johnson2021} using the fiducial model described in Section \ref{sec:sed}, during which we set redshift as a free parameter. Finally, we only retain 301 galaxies as the parent sample whose best-fit redshifts from our new SED fitting are still in the range of $3 < z_{\rm{phot}} < 4.5$. 

\begin{figure}
    \centering
    \includegraphics[width=0.47\textwidth]{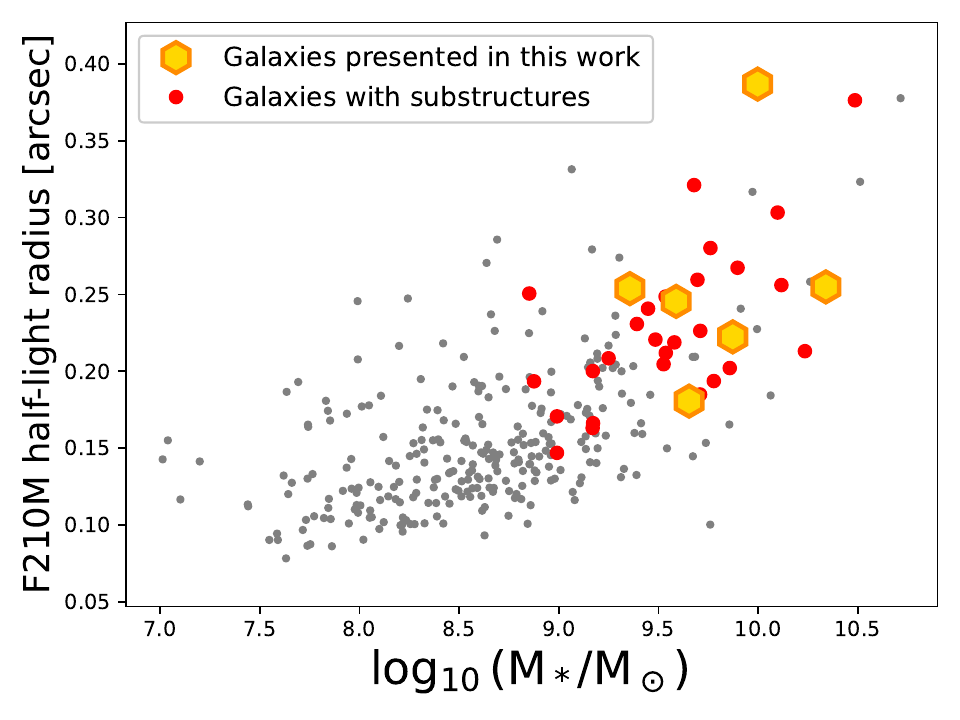}
    \caption{The parent sample of $3<z<4.5$ galaxies from JEMS. The $x$-axis shows the stellar mass from our \prospector SED fitting (Section \ref{sec:sed}), the $y$-axis shows the half-light radius derived directly from the F210M surface brightness profile (no PSF-convolved analysis included). Red circles show the 37 galaxies with visually identified substructures having distinct rest-frame colors at 4000 \AA. The golden hexagons show the final sample of 6 galaxies with red, massive cores presented in this work (see Section \ref{sec:sample} for the detailed sample selection). }
    \label{fig:size_mass}
\end{figure}

For each one of the galaxies in the parent sample, we visually check the color maps produced by its NIRCam F150W, F182M and F210M images, which probe rest-frame light at $\sim$ 3000, 4000 and 4500 \AA, respectively. This wavelength range, i.e. around the 4000 \AA\ break, is chosen because it contains key information about stellar-population properties, especially stellar ages \citep[e.g.,][]{Bruzual1983,Kauffmann2003}. Our visual identification finds 37 (out of 301) galaxies having substructures with distinct F150W/F182M/F210M colors. As Figure \ref{fig:size_mass} shows, these 37 galaxies are at the massive end of our JEMS parent sample, i.e. the majority of them have stellar masses $>10^9 M_\sun$, which is expected because the visual selection is biased toward bright substructures in relatively bright galaxies. Among 36 galaxies in the parent sample with masses $>10^{9.5}M_\sun$, 22 of them are visually identified to have substructures, corresponding to a fraction of $61\pm17\%$. Interestingly, at face value this fraction is in quantitative agreement with earlier HST studies of the fraction of galaxies with UV clumps at slightly lower redshifts $2.5<z<3$ \citep[e.g.,][]{Shibuya2016,Sattari2023}. While we caution that the methods used to identify clumps can significantly affect the results, we defer more detailed investigation of this impact to a future work with larger samples and more uniform selection methods. 

\begin{figure*}
    \centering
    \includegraphics[width=0.97\textwidth]{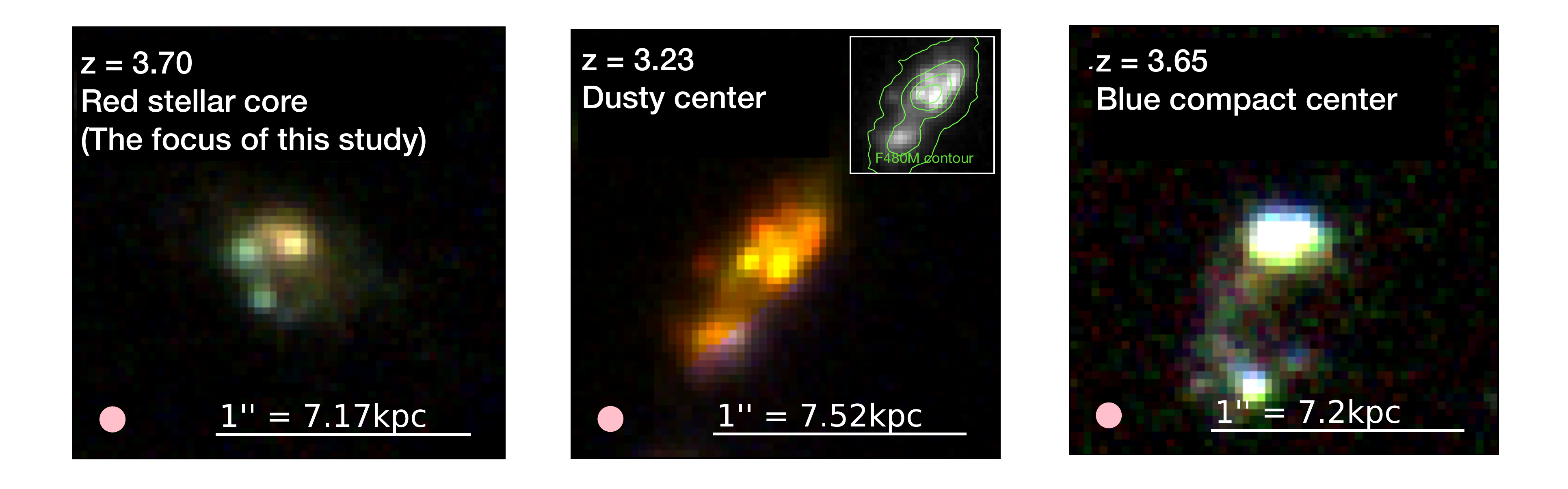}
    \caption{Examples of galaxies with different types of substructures identified by our visual selection (see Section \ref{sec:sample}). The RGB cutouts shown here are made with JWST/NIRCam F150W (B), F182M (G) and F210M (R) images. The pink circle marks the angular resolution (FWHM) of NIRCam/F210M imaging. Our visual identifications find a sample of 37 galaxies having substructures with distinct colors at rest-frame 4000\AA. These galaxies show diverse substructures, and can be generally categorized into three groups. The left panel shows an example galaxy with a red, prominent stellar core. The middle panel shows an example galaxy with a dusty core, where in the RGB (made with images at rest-4000\AA) cutout a flux deficit is seen near the center of the NIRCam/F480M image (rest-1$\mu m$, starlight).  The right panel shows an example galaxy with a compact blue center. In this paper, we will only focus on 6 galaxies having visually identified prominent, red stellar cores at rest-frame 4000\AA\ (similar to the example galaxy shown in the left panel), while the latter (middle and right) two types of galaxies will be studied in forthcoming papers. }
    \label{fig:vs_example}
\end{figure*}

As Figure \ref{fig:vs_example} shows, our analysis finds that the 37 galaxies with visually identified substructures have diverse spatially resolved stellar populations, representing different stages of galaxy structural transformations. This motivates this and upcoming papers in this series, where we address relervant physical mechanisms associated with those different stages of structural transformations. 

\begin{figure*}
    \centering
    \subfloat[{\bf JADES--JEMS--13396}]{
        \includegraphics[width=\textwidth]{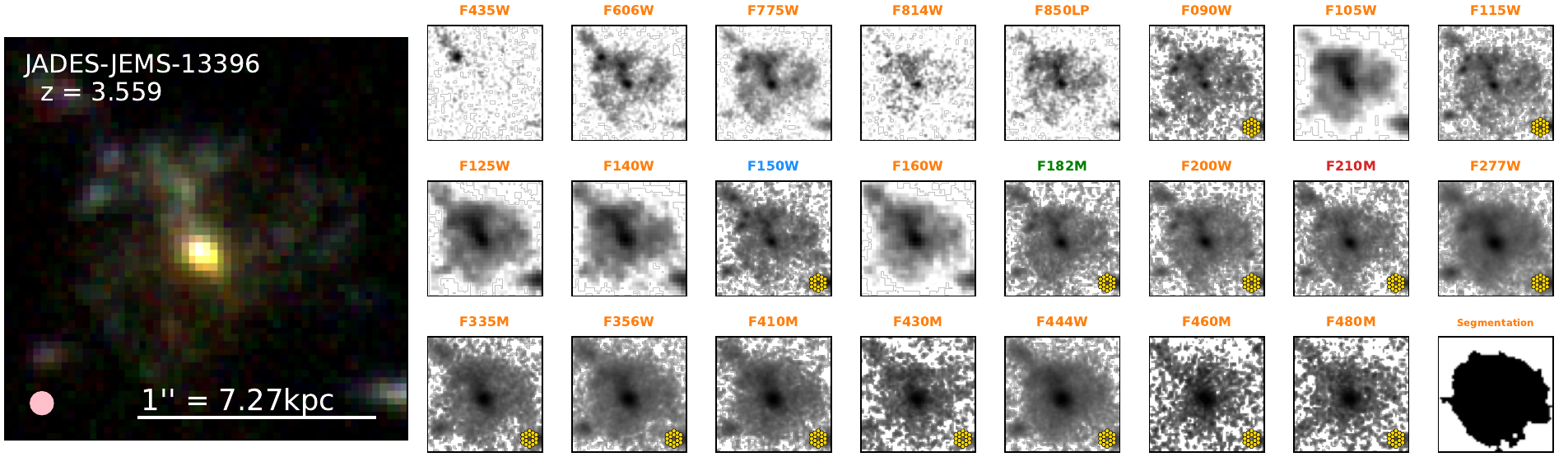}
        }\\
    \subfloat[{\bf JADES--JEMS--15157}]{
        \includegraphics[width=\textwidth]{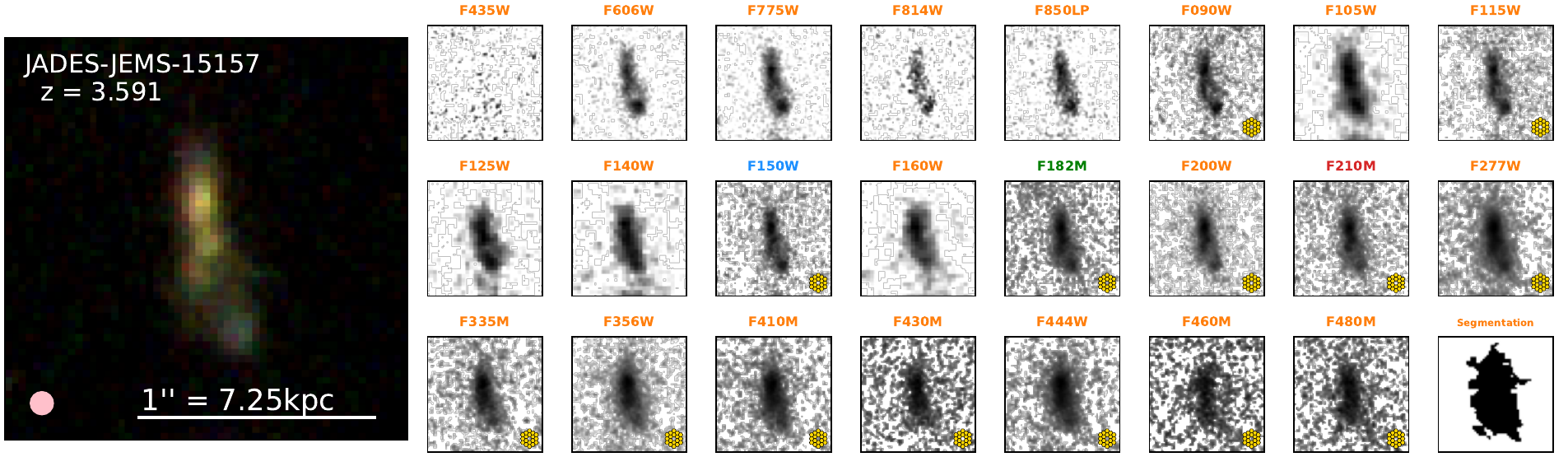}
        }\\
    \subfloat[{\bf JADES--JEMS--6885}]{
        \includegraphics[width=\textwidth]{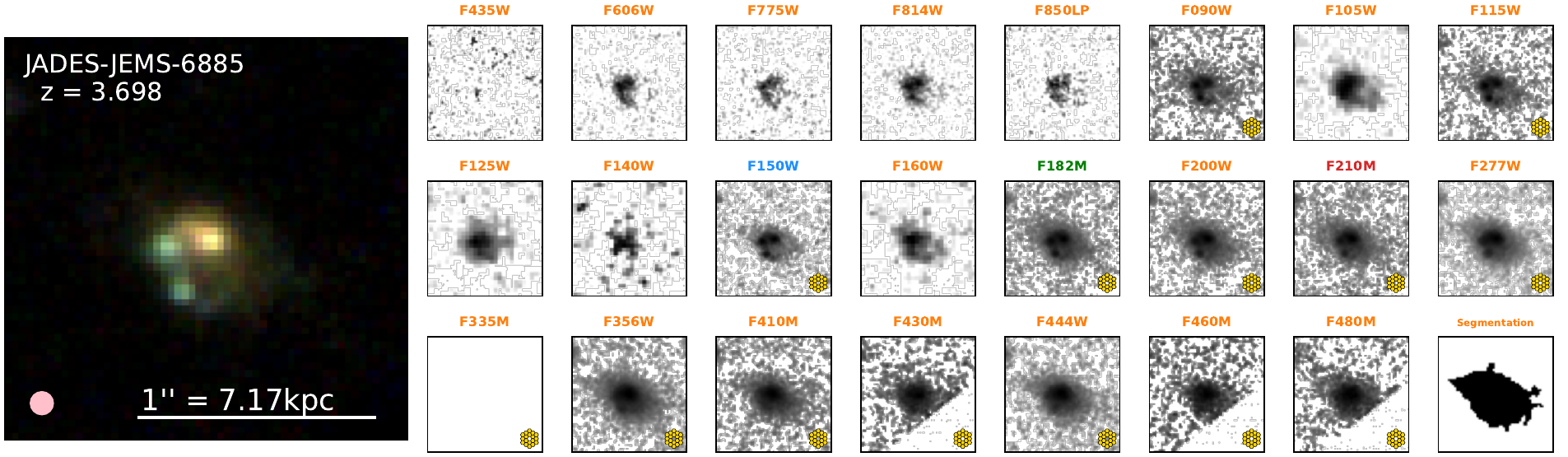}
        }
    \caption{Panchromatic HST and JWST/NIRCam images of the three galaxies with spectroscopically confirmed redshifts $z_{\rm{spec}}$. The RGB image shown on the left is made with JWST/NIRCam F150W (B), F182M (G) and F210M (R) images. We use a linear color scale for the RGB cutout to better show the substructures. The pink circle at the bottom-left marks the FWHM of the NIRCam/F210M PSF (Appendix \ref{app:psf}). Small panels show cutouts of individual 23 filters covering observed $\lambda \approx 0.35-5 \micron$ included in the analysis of this work, where the newly obtained 14-filter NIRCam images are labelled with a hexagon symbol at the bottom-right. The last small panel shows the segmentation map.  The cutouts of individual filters are shown in a log color scale, highlighting the smooth component of the galaxies.}
    \label{fig:poststamps}
\end{figure*}

\begin{figure*}
    \centering
    \subfloat[{\bf JADES--JEMS--16296}]{
        \includegraphics[width=\textwidth]{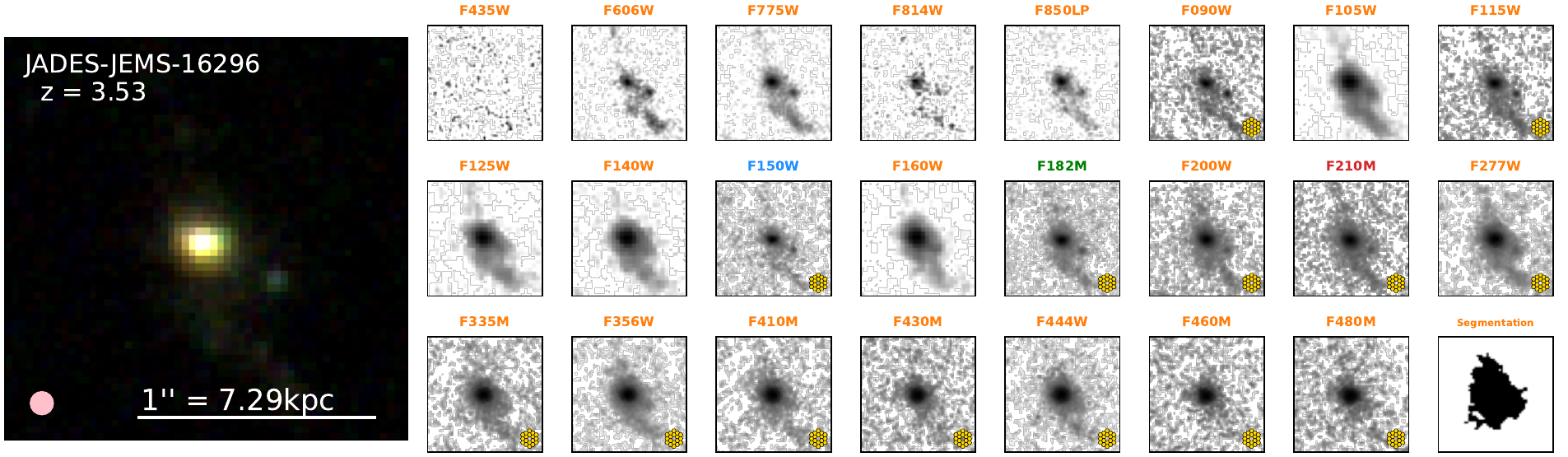}
        } \\
        \subfloat[{\bf JADES--JEMS--14436}]{
        \includegraphics[width=\textwidth]{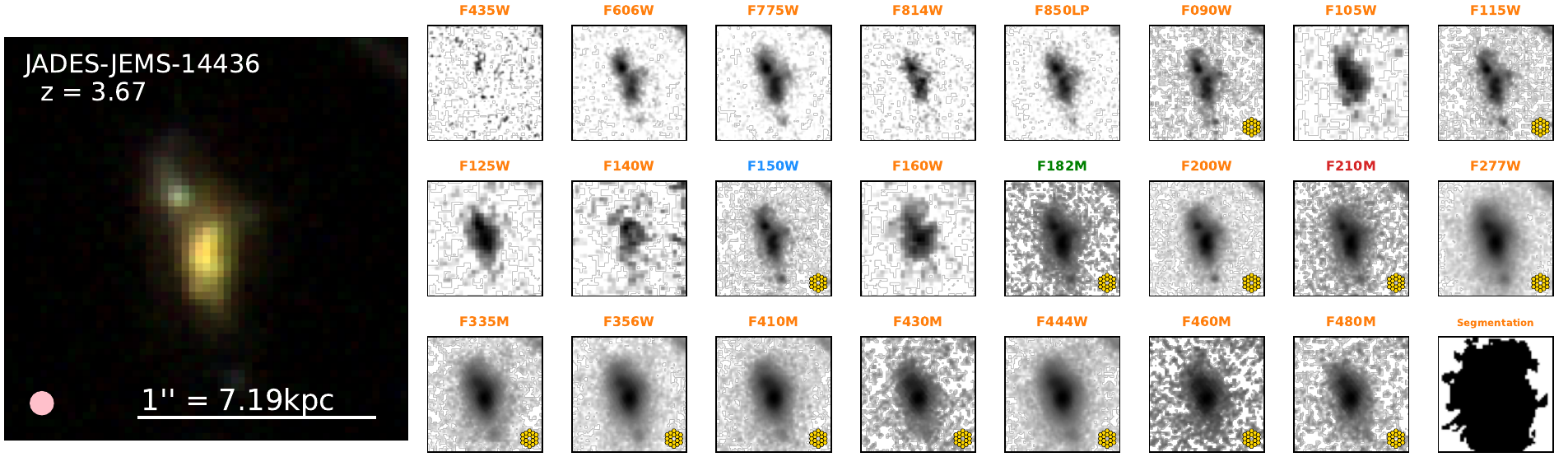}
        } \\
        \subfloat[{\bf JADES--JEMS--11059}]{
        \includegraphics[width=\textwidth]{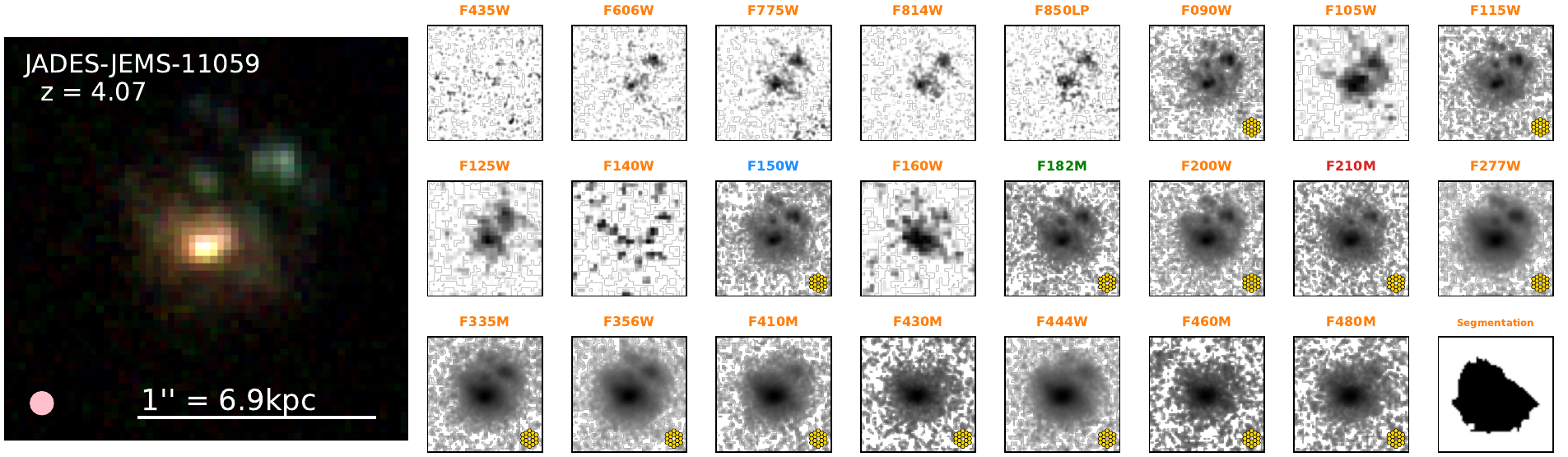}
        }
     \caption{Similar to Figure \ref{fig:poststamps}, but for the remaining three galaxies with photometric redshifts $z_{\rm{phot}}$.}
    \label{fig:poststamps_cont}
\end{figure*}

In this first paper, we only focus on a group of 6 galaxies whose main characteristic is to have a bright and red (at rest-frame $\sim$ 4000 \AA) stellar core which we refer to as C1 in the following. C1 is close to the center of stellar-light distribution, and it is often accompanied by  additional off-center, minor clumpy substructures (named as C2 and C3, if present).  Figure \ref{fig:poststamps} and \ref{fig:poststamps_cont} show their 23-filter cutouts. Our analysis below shows that C1 is more massive than what one would expect from a single collapse due to the Toomre instability, and has a star formation history (SFH) that shows a recent major star-formation episode started $\sim0.5-1$ Gyr ago that peaked at $\sim0.1-0.3$ Gyr prior to the time of observation. The formation of the stellar core C1 and its impact, if any, on star formation will be the main focus of this paper. 

Finally, we note that in the future, instead of using visual selection, we will use a more uniform selection method to study resolved properties of the more general population of galaxies at $3 < z < 4.5$ with deep JADES NIRCam imaging.

\subsection{Observations \& Data Reduction} \label{sec:obs}

The 6 galaxies have been observed with deep\footnote{The sensitivity for a 5$\sigma$ detection using a $r=0.15''$ circular aperture is $\sim 30$ magnitude in the F200W filter \citep{Robertson2022}.} imaging in 14 JWST/NIRCam filters. These include 9-filter images (F090W, F115W, F150W, F200W, F277W, F335M, F356W, F410M, F444W) acquired as part in the GOODS-S portion of JADES, and 5 additional medium-filter images (F182M, F210M, F430M, F460M and F480M) acquired by JEMS. 

The NIRCam observations of all 14 filters are reduced using a consistent method, which has been briefly summarized in \citet{Robertson2022} and will be presented in detail in Rieke et al. (2023 in preparation) and Tacchella et al. (2023 in preparation). In short, we process the raw images with the JWST Calibration Pipeline v1.8.1 with the CRDS pipeline mapping (pmap) context 1009. To perform the detector-level corrections and obtain count-rate images, we run the Stage 1 of the JWST pipeline with default parameters, during which we also mask and correct for the ``snowballs'' effect caused by charge deposition arising from cosmic ray hits. We then run the Stage 2 of the JWST pipeline, with default parameters, to perform the flat-fielding and flux calibration, after which we perform several custom corrections including the removal of $1/f$ noise and the subtraction of 2D background from the images, and the correction for the ``wisp'' features in the short-wavelength (SW, i.e. filters bluer than F277W) images. Afterwards, for individual exposures of a given visit, we match sources to the World Coordinate System of a reference catalog constructed from HST F160W mosaics in the GOODS-S field with astrometry tied to Gaia-EDR3 \citep{GaiaEDR3}, calculate astrometric corrections and then run the Stage 3 of the JWST pipeline to combine together all those individual exposures. Finally, we combine individual visit-level mosaics together to create the final mosaic, where we choose a pixel scale of 0.03 arcsec pixel$^{-1}$ for all filters. 

The ancillary HST data used in this work are taken from the latest Hubble Legacy Fields Data Release\footnote{https://archive.stsci.edu/prepds/hlf/} in the GOODS-S field \citep{Illingworth2016,Whitaker2019}, including the imaging data of the ACS/WFC F435W, F606W, F775W, F850LP and F814W filters, and of the WFC3 F105W, F125W, F140W and F160W filters. For all those HST images, we have (1) tied them to the same astrometry and (2) resampled them to the same pixel scale as those of our NIRCam images (Robertson et al 2023 in preparation).

\subsection{Photometry} \label{sec:phot}
For each galaxy, we perform aperture-matched photometry for (1) integrated flux, (2) individual substructures and (3) a smooth component. The integrated flux is defined as total flux in the pixel aperture generated from a galaxy's segmentation map generated by JADES \citep{Robertson2022}. For fluxes of individual substructures, we use customized elliptical apertures which are illustrated in the RGB images in Figure \ref{fig:seds_specz} and \ref{fig:seds_photz}. Finally, the smooth component is defined as integrated flux minus co-added fluxes of individual substructures.

Point Spread Functions (PSFs) of all images are first homogenised to that of NIRCam/F444W. To do so, in the footprint of JEMS, we visually identify 13 isolated, bright stars which are unsaturated across all filters from HST through JWST. We then build effective PSFs (ePSFs) following the method from \citet{Anderson2000}. For NIRCam filters, we have also generated model PSFs (mPSFs) using the package {\sc Webbpsf}\citep[v1.1.1,][]{Perrin2012,Perrin2014}, during which we also took into account the observational (e.g., source location on detectors, optical path difference) and data reduction (e.g., mosaicking) effects. As shown in Appendix \ref{app:psf}, we find excellent agreement between ePSFs and mPSFs, with a typical difference of $\lesssim 1\%$. We then homogenise ePSFs using the package {\sc Pypher} which calculates the convolution kernel based on Wiener filtering with a regularisation factor \citep{Boucaud2016}. Finally, we perform aperture-matched photometry in the PSF-matched images.

To estimate photometric uncertainties, we follow \citet{Labbe2005} to first use randomly placed apertures in empty (i.e., no sources) regions in the images to estimate the growth of noise as a function of the linear size of aperture. We then fit the function using the parametrization of \citet{Quadri2007} and finally estimate photometric uncertainties using equation 5 in \citet{Whitaker2011}. This procedure will be presented in detail in Robertson et al. (2023 in preparation).

\subsection{Measuring Stellar-population Properties} \label{sec:sed}

We model panchromatic Spectral Energy Distributions (SEDs) to derive stellar-population properties of individual structural components, with the emphasis on robustly measuring their stellar masses, ages and SFHs. For each one of the galaxies, we perform two rounds of SED fitting. The first round aims at checking the physical association of all structural components, as opposed to the case when sources at different distances are projected onto the regions of sky in close proximity of each other. During the SED fitting of this round, we set redshift as a free parameter with a flat prior of $z\in(0,10)$. We then check posterior distributions of redshift, and consider structural components as being physically linked when the difference in the median of their redshift posteriors is $\le0.1$. Afterwards, we fix redshifts of all components to either the spectroscopic redshift if available or the best-fit value that we derive from the first-round SED fitting using the galaxy's integrated photometry\footnote{We notice that, for galaxies with spectroscopic redshifts, our photometric redshifts are highly consistent with the spectroscopic ones, with a typical difference of $|z_{\rm{spec}}-z_{\rm{phot}}|<0.1$, thanks to the high S/N photometric coverage around both the Lyman Break and Balmer Break.}, and run the second-round SED fitting. The fitting results are shown in Figure \ref{fig:seds_specz} and \ref{fig:seds_photz}, and will be discussed in Section \ref{sec:sfh_ind}. Below, we detail the model assumptions of our SED fitting.

\subsubsection{Basic setup of the Prospector fitting}
The SED fitting code \prospector \citep{Johnson2021} is elected to use in this work. It is built upon a fully Bayesian framework that makes it possible to fit galaxies' SEDs with complex models of stellar population synthesis.  

Regarding the basic setup of \prospector fitting, we adopt the Flexible Stellar Population Synthesis (FSPS) code \citep{Conroy2009,Conroy2010} where we use the stellar isochrone libraries MIST \citep{Choi2016,Dotter2016} and the stellar spectral libraries MILES \citep{Falcon-Barroso2011}. During the modeling, we use the MCMC sampling code {\sc dynesty} \citep{Speagle2020} which adopts the nested sampling procedure \citep{Skilling2004}. We assume the \citealt{Kroupa2001} initial mass function (IMF) for  consistency with the IMF used in the \citet{Byler2017} nebular continuum and line emission model which we adopt in our SED modeling. We adopt the \citet{Madau1995} IGM absorption model.

We set the stellar metallicity as a free parameter and assume a flat prior in logarithmic space $\log(Z_*/Z_\sun) \in (-2, 0.19)$, where $Z_\sun = 0.0142$ is solar metallicity. The upper limit of the prior is chosen because it is the highest metallicity that the MILES library has. We also leave the gas-phase metallicity $Z_{\rm{gas}}$ and ionization parameter $U$ as free parameters during the fitting, where we use flat priors of $\log(Z_{\rm{gas}}/Z_\sun)\in(-2,0.5)$ and $\log U\in(-4,1)$.  

Following \citet{Tacchella2022}, we assume a two-component dust attenuation model where the dust attenuation of nebular emission and young stellar populations, and of old stellar populations, are treated differently \citep{Charlot2000}. For stellar populations older than 10 Myr, we assume the dust attenuation using the parametrization from \citet{Noll2009}, i.e.,
\begin{equation}
\tau_{\rm{dust,2}}(\lambda) = \frac{\hat{\tau}_{\rm{dust,2}}}{4.05}(k^{'}(\lambda)+D(\lambda))\left(\frac{\lambda}{5500 \rm{\AA}}\right )^{\delta}
\end{equation}
where $\rm{\hat{\tau}_{dust,2}}$ and $\delta$ correspond to \texttt{dust2} and \texttt{dust\_index} in the FSPS, $k^{'}(\lambda)$ is the \citet{Calzetti2000} dust attenuation law and $D(\lambda)$ is the Lorentzian-like profile used to describe the UV dust bump at 2175 \AA. Instead of setting $D(\lambda)$ as a free parameter, we tie it to $\delta$ following the results of \citet{Kriek2013}. Therefore, the dust attenuation for the older stellar populations only has two free parameters, $\delta$ and $\rm{\hat{\tau}_{dust,2}}$. We assume flat priors for both parameters, i.e., $\delta\in(-1,0.4)$ and $\rm{\hat{\tau}_{dust,2}}\in(0,10)$. For stellar populations younger than 10 Myr, we assume the same dust attenuation law as for the nebular emission, which has a functional form of
\begin{equation}
\tau_{\rm{dust,1}} (\lambda) = \hat{\tau}_{\rm{dust,1}} \left(\frac{\lambda}{5500 \rm{\AA}}\right )^{-1},
\end{equation}
where $\rm{\hat{\tau}_{dust,1}}$ corresponds to the parameter \texttt{dust1} in the FSPS. Instead of modeling $\rm{\hat{\tau}_{dust,1}}$ as an independent free parameter, we tie it to $\rm{\hat{\tau}_{dust,2}}$ and model their ratio using a clipped normal prior centered at 1, with a width of 0.3 and in the range of $\rm{\hat{\tau}_{dust,1}}/\rm{\hat{\tau}_{dust,2}} \in (0,2)$. 

We also include AGN dust torus templates from \citet{Nenkova2008} and \citet{Nenkova2008b}, but we stress that the purpose of including this AGN component is not to quantitatively constrain the AGN strength, since the data coverage we currently have does not have power to do so. Instead, the purpose simply is to marginalize over it in order to check how the AGN component can affect the fitting results. This adds two free parameters: the ratio of bolometric luminosity from the galaxy divided by that from the AGN ($f_{\rm{AGN}}$), and the optical depth of clumps in AGN dust torus at 5500 \AA\ ($\tau_{\rm{AGN}}$). We assume flat priors in logarithmic space for both parameters, i.e., $f_{\rm{AGN}}\in(10^{-5},3)$ and $\tau_{\rm{AGN}}\in(5,150)$. We find that $f_{\rm{AGN}}$ is consistent with zero for all sample galaxies, which is in line with the fact that these galaxies are not classified as AGN using other different selection methods, including X-ray \citep{Luo2017}, IRAC colors \citep{Ji2022} and multi-wavelength selections \citep{Lyu2022}.

\subsubsection{Reconstructing star formation histories}

One key feature of \prospector is that it allows flexible parameterizations of galaxies’ SFHs, both in parametric and nonparametric forms. This has been demonstrated as crucial to reducing systematic biases in the measurements of stellar mass, star formation rate and stellar age \citep{Leja2019}. Several recent studies have further shown that statistically, \prospector is capable of reconstructing high-fidelity nonparametric SFHs for synthetic galaxies generated from cosmological simulations 
\citep{Leja2019, Johnson2021, Tacchella2022,Ji2022a}.

For the fiducial SFH, we assume a piece-wise, nonparametric form composed of $N_{\rm{SFH}}=9$ lookback time bins (\tlookback, i.e., the time prior to the time of observation), where star formation rate is constant in each bin. Among the 9 lookback time bins, the first two bins are fixed to be $0-30$ and $30-100$ Myr in order to capture the recent star formation activity; the last bin is assumed to be $\rm{0.85t_H - t_H}$ where $\rm{t_{H}}$ is the Hubble Time at the time of observation; and the remaining 6 bins are evenly spaced in logarithmic space between $\rm{100\, Myr - 0.85t_H}$. As have been extensively tested by \citet{Leja2019} using mock observations of simulated galaxies (see their Figure 15), the recovered physical properties are largely insensitive to $N_{\rm{SFH}}$ when it is greater than 5. This is also found later through our analysis presented in Section \ref{sec:sfh_ind} and Figure \ref{fig:coadd_total}. 

The procedure above to reconstruct nonparametric SFHs potentially suffers from overfitting problems, because it includes ``more bins than the data warrant'' \citep{Leja2019}. An effective solution to mitigate the issue is to choose a prior weight for physically plausible forms, as opposed to letting the SFHs have a fully arbitrary shape \citep{Carnall2019,Leja2019}. Our fiducial measures use the continuity prior that has been demonstrated to work well across various galaxy types \citep{Leja2019}. Recent studies have noted the systematics introduced by assumed priors to reconstructed SFHs \citep{Tacchella2022b,Ji2022a,Suess2022}. To check possible systematics in our measurements, we therefore also experiment using another well-tested Dirichlet prior \citep{leja2017}. In addition, we have also tested our results using the parametric delayed-tau model. As we show in Appendix \ref{app:prior}, the results of this work remain qualitatively unchanged using these different methods of SFH reconstructions. In the remainder of main text, our discussion thus will only focus on measurements from the continuity prior.

\subsection{Measuring Morphological Properties of the Substructures} \label{sec:morph_mea}

To estimate intrinsic stellar masses (Section \ref{sec:masses}) and stellar-mass surface densities (Section \ref{dis:quench}) of the clumpy substructures, we perform PSF-convolved morphological fitting. For each one of the galaxies, we simultaneously model the 2D light distributions of all clumpy substructures with the fitting code {\sc forcepho} (Johnson et al. 2023 in preparation), which adopts a fully MCMC framework and performs the fitting on the basis of individual NIRCam exposures, as opposed to performing the fitting in a drizzled image (i.e., a mosaic). In this way, parameter uncertainties are better estimated because correlations of adjacent image pixels, and, more importantly here given the complex morphologies of the galaxies, correlations of fluxes from neighbouring substructures are naturally taken into account (see \citealt{Robertson2022} and \citealt{Tacchella2023} for a brief discussion). 

A redder NIRCam filter generally better probes the stellar-mass distribution. Meanwhile, a better angular resolution also helps to better separate fluxes from individual substructures. The combination of the poorer imaging angular resolution of longer-wavelength filters and the younger stellar populations of off-center clumps (i.e. their SEDs are bluer) makes  it  increasingly difficult to morphologically decompose individual substructures using images of longer-wavelength filters. We thus perform the {\sc forcepho} analysis using the reddest NIRCam's SW filter we have, i.e., F210M which probes rest-frame $\sim$ 5000\AA\ of the six galaxies. Finally, we model each clumpy substructure using a single S\'{e}rsic profile. 

Figure \ref{fig:fpho} shows results of the {\sc forcepho} fitting. As the residual map shows, each one of the substructures is well fit by a single S\'{e}rsic profile. The $\chi$ map also confirms the presence of the Smooth component surrounding the clumpy substructures in each one of the galaxies. Ideally, in the {\sc forcepho} fitting we could add an additional morphological component to model this Smooth component together with the clumpy substructures. However, we find that a single S\'{e}rsic profile is unable to describe the Smooth component, because its morphology is rather complex with large-scale asymmetries. In principle, we could assume a more complex model for it, e.g., a summation of several S\'{e}rsic profiles. But such a model is purely arbitrary and it would be hard to understand and control the systematic errors behind it. We therefore decide not to model the Smooth component in the {\sc forcepho} fitting. One concern is then how this can affect the morphological fits to the clumpy substructures. While it can be significant for faint clumps C2 and C3, we argue that the effect should be minor for C1 (which is the focus of this work) because the ratio of the median residual flux (primarily from the Smooth component) to the peak flux of C1 is only $\sim 0.1-4\%$.

\section{Results} \label{sec:res}

In this Section, we present the spatially resolved stellar populations of the 6 galaxies. In particular, we focus on the properties of the red, massive core C1 in each one of the galaxies, and discuss the constraints on its formation mechanisms.

\begin{table*}[!ht]
    \centering
    \caption{Properties of Spatially Resolved Stellar Populations}
    \begin{tabular}{||c|c|c|ccccc||}
    \toprule
      ID & Redshift & Component & F$_\nu^{\rm{F150W}}$ $^{(a)}$ & log M$_*$ $^{(b)}$ & SFR $^{(b),(c)}$ & Stellar Age $^{(b)}$ & sSFR $^{(b)}$ \\
       &  & &  (nJy) & (M$_\sun$) & (M$_\sun$ yr$^{-1}$) & (Gyr) & (Gyr$^{-1}$) \\
    \hline
    
    \multirow{3}{*}{JADES-JEMS-13396} & \multirow{3}{*}{3.559} & C1 & 24.6 $\pm$0.8 & 9.40 $^{+0.03}_{-0.04}$ & 2.7 $^{+0.2}_{-0.3}$ & 0.7 $^{+0.1}_{-0.2}$ & 1.07 $^{+0.10}_{-0.15}$ \\
    & & Smooth & 212.8 $\pm$11.4  & 9.88 $^{+0.06}_{-0.13}$ & 13.4 $^{+2.3}_{-1.6}$ & 0.6 $^{+0.1}_{-0.2}$ & 1.76 $^{+0.38}_{-0.56}$ \\
    & & Integrated & 237.4 $\pm$11.7 & 10.00 $^{+0.02}_{-0.04}$ & 13.5 $^{+0.8}_{-0.6}$ & 0.8 $^{+0.1}_{-0.1}$ & 1.35 $^{+0.10}_{-0.13}$\\
    \hline
    
    \multirow{5}{*}{JADES-JEMS-15157} & \multirow{5}{*}{3.591} & C1 & 11.4 $\pm$0.7 & 8.71 $^{+0.10}_{-0.08}$ & 0.2 $^{+0.1}_{-0.1}$ & 0.6 $^{+0.1}_{-0.2}$ & 0.38 $^{+0.21}_{-0.20}$ \\
    & & C2 & 9.2 $\pm$0.9 & 8.02 $^{+0.15}_{-0.43}$ & 0.3 $^{+0.1}_{-0.1}$ & 0.5 $^{+0.1}_{-0.4}$ & 2.86 $^{+1.37}_{-2.99}$  \\
    & & C3 & 9.8 $\pm$0.8 & 8.55 $^{+0.09}_{-0.15}$ & 0.2 $^{+0.1}_{-0.1}$ & 0.5 $^{+0.1}_{-0.3}$ & 0.56 $^{+0.30}_{-0.34}$ \\
    & & Smooth  & 58.7 $\pm$7.3 & 9.35 $^{+0.07}_{-0.05}$ & 0.4 $^{+0.4}_{-0.2}$ &  0.7 $^{+0.1}_{-0.2}$ & 0.17 $^{+0.18}_{-0.09}$  \\
    & & Integrated & 89.2 $\pm$8.3 & 9.36 $^{+0.09}_{-0.11}$ & 1.3 $^{+1.0}_{-0.4}$ & 0.4 $^{+0.2}_{-0.1}$ & 0.56 $^{+0.45}_{-0.22}$ \\ 
    \hline
    
    \multirow{5}{*}{JADES-JEMS-6885} & \multirow{5}{*}{3.698} & C1 & 39.9 $\pm$1.6 & 9.19 $^{+0.07}_{-0.03}$ & 1.5 $^{+0.1}_{-0.2}$ & 0.6 $^{+0.1}_{-0.1}$ & 0.96 $^{+0.16}_{-0.14}$ \\
    & & C2 & 16.8 $\pm$0.8 & 8.30 $^{+0.10}_{-0.15}$ & 1.2 $^{+0.1}_{-0.1}$ & 0.3 $^{+0.1}_{-0.1}$ & 6.01 $^{+1.47}_{-2.13}$ \\
    & & C3 & 12.2 $\pm$0.8 & 8.13 $^{+0.08}_{-0.09}$ & 1.2 $^{+0.1}_{-0.2}$ & 0.1 $^{+0.1}_{-0.1}$ & 8.89 $^{+1.79}_{-2.36}$ \\
    & & Smooth & 169.0 $\pm$13.0 & 9.62 $^{+0.07}_{-0.21}$ & 3.6 $^{+1.1}_{-0.6}$ & 0.4 $^{+0.2}_{-0.2}$ & 0.86 $^{+0.29}_{-0.14}$\\
    & & Integrated & 238.1 $\pm$14.0 & 9.87 $^{+0.03}_{-0.07}$ & 7.4 $^{+1.1}_{-0.9}$ & 0.4 $^{+0.2}_{-0.2}$ & 0.99 $^{+0.16}_{-0.20}$ \\ 
    \hline
    
    \multirow{4}{*}{JADES-JEMS-16296} & \multirow{5}{*}{3.53} & C1 & 42.5 $\pm$1.3 &  8.82 $^{+0.02}_{-0.03}$ & 2.7 $^{+0.1}_{-0.1}$ & 0.5 $^{+0.1}_{-0.1}$ & 4.08 $^{+0.24}_{-0.32}$ \\
    & & C2 & 7.1 $\pm$0.9 & 7.88 $^{+0.05}_{-0.08}$ & 0.4 $^{+0.1}_{-0.1}$ & 0.1 $^{+0.1}_{-0.1}$ & 5.27 $^{+1.45}_{-1.63}$ \\
    & & Smooth & 80.8 $\pm$7.3  & 9.37 $^{+0.02}_{-0.01}$ & 0.9 $^{+0.1}_{-0.2}$ & 0.9 $^{+0.1}_{-0.1}$ & 0.38 $^{+0.04}_{-0.08}$ \\
    & & Integrated & 130.6 $\pm$8.4 & 9.65 $^{+0.05}_{-0.06}$ & 5.4 $^{+1.2}_{-0.9}$ & 0.5 $^{+0.3}_{-0.2}$ & 1.20 $^{+0.30}_{-0.26}$ \\ 
    \hline
    
    \multirow{4}{*}{JADES-JEMS-14436} & \multirow{5}{*}{3.67} & C1 & 87.8 $\pm$1.8 & 9.58 $^{+0.05}_{-0.02}$ & 3.9 $^{+0.2}_{-0.2}$ & 0.4 $^{+0.1}_{-0.1}$ & 1.02 $^{+0.12}_{-0.07}$ \\
    & & C2 & 38.5 $\pm$0.9 & 9.32 $^{+0.11}_{-0.06}$ & 5.8 $^{+0.6}_{-0.7}$ & 0.3 $^{+0.2}_{-0.1}$ & 2.77 $^{+0.75}_{-0.50}$ \\
    & & Smooth & 169.9 $\pm$17.0  & 10.05 $^{+0.04}_{-0.06}$ & 2.8 $^{+0.6}_{-0.6}$ & 0.9 $^{+0.1}_{-0.2}$ & 0.24 $^{+0.05}_{-0.06}$ \\
    & & Integrated & 296.3 $\pm$18.0 & 10.34 $^{+0.07}_{-0.09}$ & 11.8 $^{+2.5}_{-1.9}$  & 0.8 $^{+0.1}_{-0.1}$ & 0.53 $^{+0.14}_{-0.14}$ \\ 
    \hline
     
    \multirow{5}{*}{JADES-JEMS-11059} & \multirow{5}{*}{4.07} & C1 & 47.8 $\pm$1.2 & 9.32 $^{+0.02}_{-0.03}$ & 2.0 $^{+0.1}_{-0.1}$ & 0.7 $^{+0.1}_{-0.1}$ & 0.95 $^{+0.06}_{-0.08}$ \\
    & & C2 & 24.0 $\pm$1.1 & 7.97 $^{+0.05}_{-0.07}$ & 1.2 $^{+0.1}_{-0.1}$ & 0.4 $^{+0.1}_{-0.1}$ & 12.85 $^{+1.82}_{-2.33}$ \\
    & & C3 & 8.2 $\pm$0.5 & 8.22 $^{+0.22}_{-0.39}$ & 0.4 $^{+0.1}_{-0.1}$ & 0.5 $^{+0.1}_{-0.3}$ & 2.41 $^{+1.36}_{-2.24}$ \\
    & & Smooth & 164.0 $\pm$11.6  & 9.81 $^{+0.07}_{-0.12}$ & 7.3 $^{+1.7}_{-2.5}$ & 0.3 $^{+0.4}_{-0.2}$ & 1.13 $^{+0.32}_{-0.49}$  \\
    & & Integrated & 244.2 $\pm$12.9 & 9.59 $^{+0.08}_{-0.07}$ & 9.6 $^{+0.7}_{-0.6}$ & 0.4 $^{+0.1}_{-0.2}$ & 2.46 $^{+0.48}_{-0.42}$\\ 
    \hline
    
    \end{tabular}
    \begin{tablenotes}
        \item[](a) Fluxes from aperture photometry in PSF-matched NIRCam/F150W images. (b) Reported values are from the \prospector SED fitting with our fiducial model (Section \ref{sec:sed}). (c) Instantaneous SFR from SED fitting, i.e., SFR in the first lookback time bin (30 Myr) of a nonparametric SFH.
    \end{tablenotes}
    \label{tab:sed_info}
\end{table*}

\subsection{Spatially Resolved Stellar Populations} \label{sec:sfh_ind}

Table \ref{tab:sed_info} presents the physical properties derived from the fiducial \prospector model described in Section \ref{sec:sed}. To check the SED fitting results, we first compare the stellar mass and star formation rate derived using the integrated flux with the sum of that measured for individual components, i.e., substructures plus the Smooth component. We note that these two values are not necessarily consistent with each other, because summation of the measurements of $N$ individual components is equivalent to fitting a galaxy's SED with $(N-1)$ $\times$ more free parameters than using its integrated flux alone. Yet, excellent agreement between the two measurements is seen in Figure \ref{fig:coadd_total}, showing the consistency and robustness of our SED fitting. In Appendix \ref{app:sfh_comparison}, we further compare the coadded SFH of individual structural components, with the SFH obtained from SED fitting of integrated flux. The two SFHs are consistent with each other within uncertainty.

\begin{figure}
    \centering
    \includegraphics[width=0.47\textwidth]{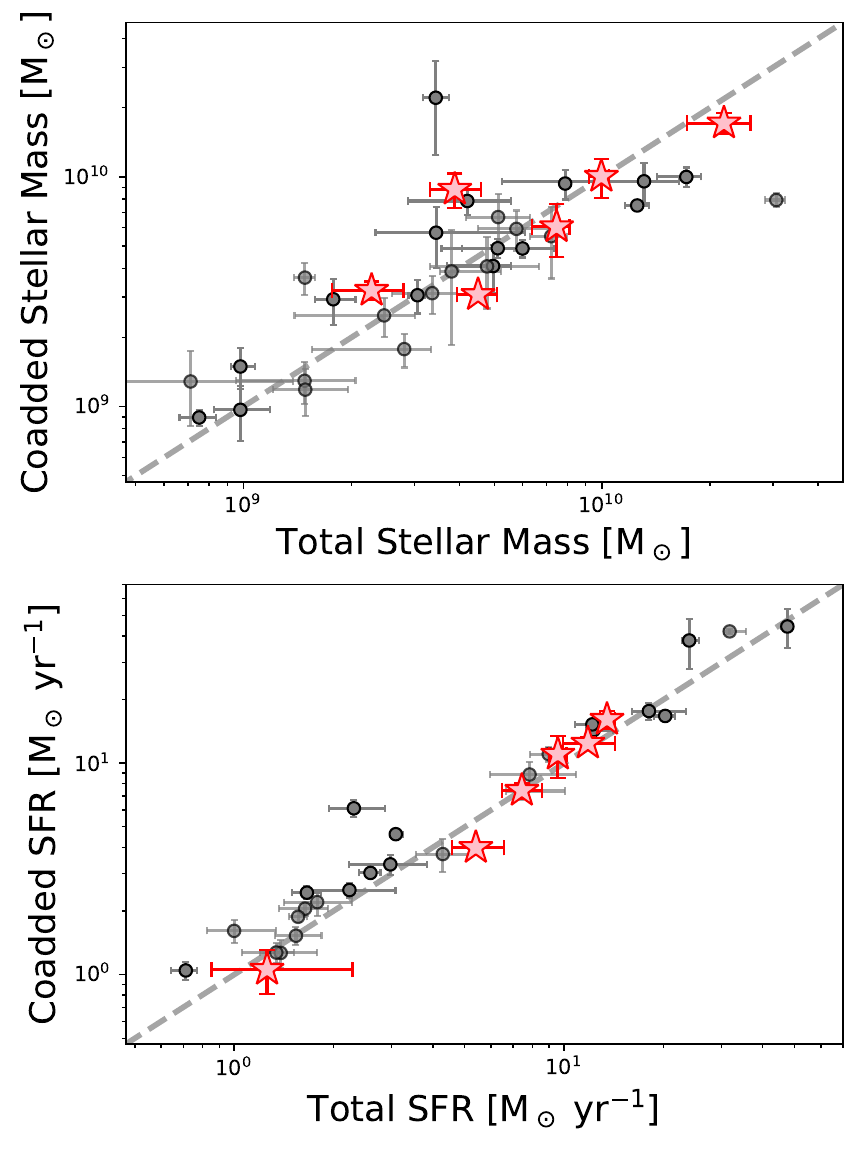}
    \caption{Comparisons of the stellar-mass (top panel) and star formation rate (bottom panel) measures. The $x$-axis shows the measure using integrated flux, and the $y$-axis shows the results by coadding the measures of individual structural components (i.e. substructures+smooth) together. In the figure we plot all 37 galaxies showing distinct substructures around the rest-frame 4000 \AA\ (Section \ref{sec:sample}), and the final sample of 6 galaxies presented in this paper are marked with the red star symbols. We see excellent agreement between the two measurements. }
    \label{fig:coadd_total}
\end{figure}

\begin{figure*}
    \centering
    \includegraphics[width=\textwidth]{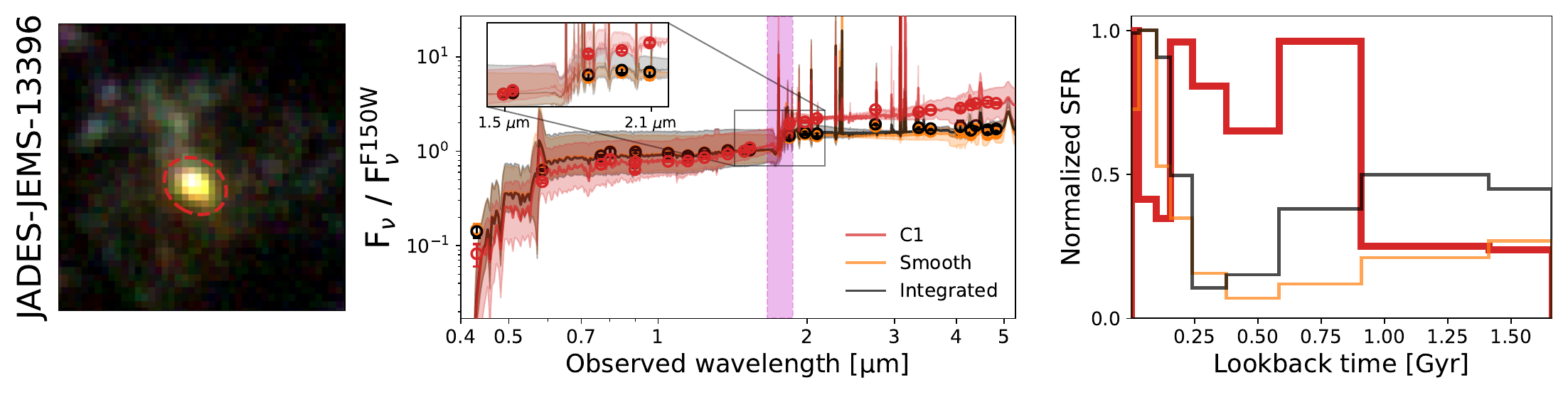}
    \includegraphics[width=\textwidth]{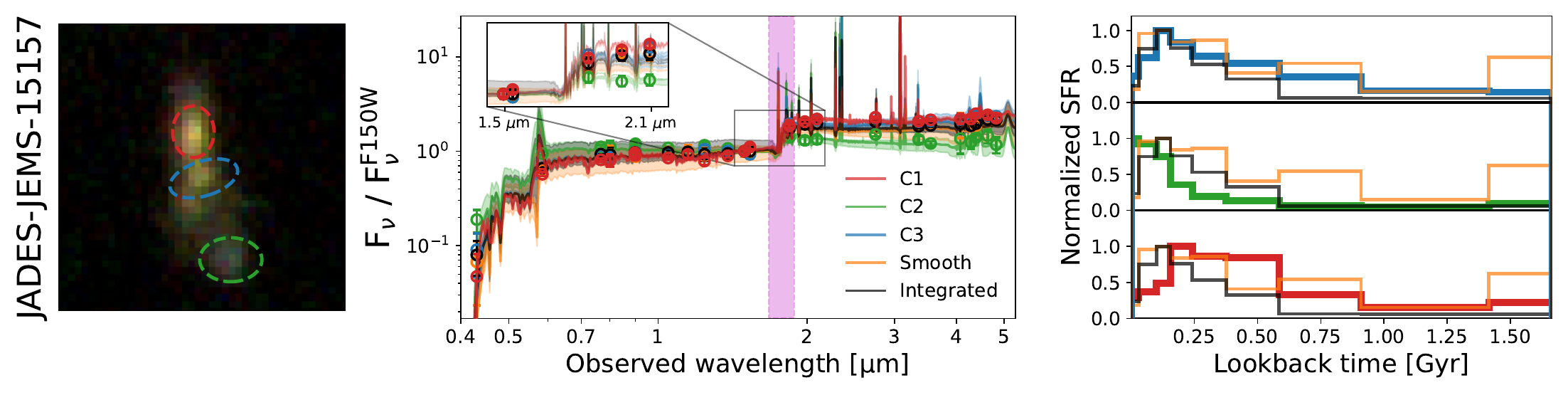}
    \includegraphics[width=\textwidth]{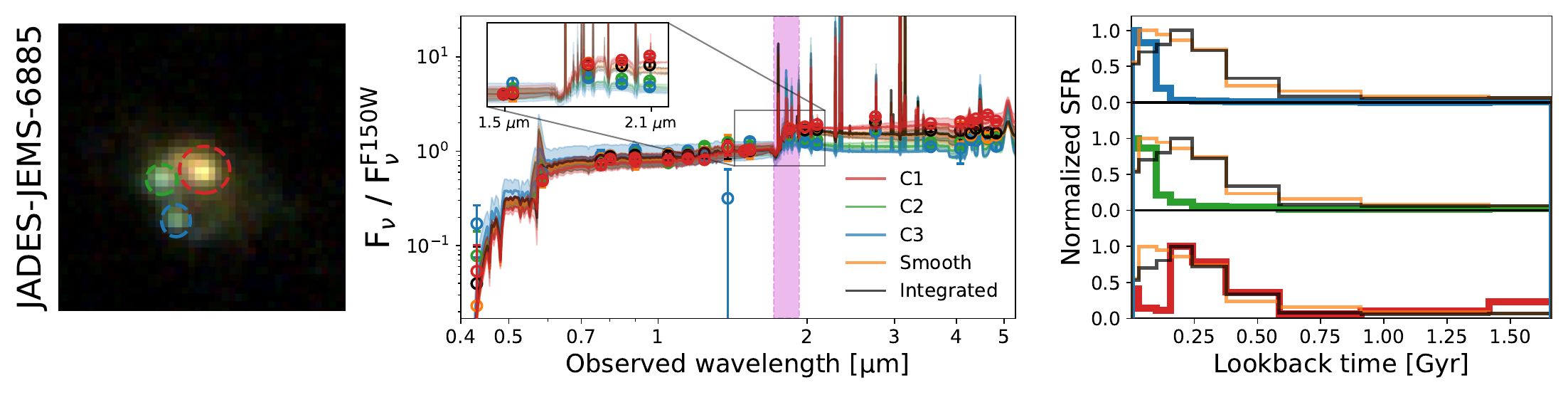}
    \caption{Spatially resolved stellar populations of the three galaxies with $z_{\rm{spec}}$. {\bf Left:} The RGB images are the same as those shown in Figure \ref{fig:poststamps}, except that here we zoom in to the central 1.3"$\times$1.3" region. The middle and right panels show the SED fitting results from our fiducial \prospector model (Section \ref{sec:sed}). {\bf Middle:} The normalized (i.e., F$_{\nu}^{\rm{F150W}}=1$) photometry and best-fit SEDs of individual structural components defined in Section \ref{sec:phot}. These include best-fit SEDs for (1) visually identified clumpy substructures which are color-coded accordingly based on the colors used in the RGB images for individual apertures, (2) the Smooth component (orange) and (3) integrated flux (black). The inset highlights the wavelength range around 2$\micron$ covered by NIRCam F150W, F182M, F200W and F210M, and HST/WFC3 F160W filters. The magenta shaded region marks the rest-frame wavelength range between 3645 \AA\ and 4100 \AA\ which contains key spectral features sensitive to stellar age such as the Balmer jump and D4000 break \citep{Bruzual1983,Balogh1999}. {\bf Right:} Reconstructed nonparametric SFHs of individual substructures. Each SFH is normalized with its peak SFR, i.e., Normalized SFR(t) = SFR(t)/max(SFR(t)), to better show the difference in the shape of SFHs. In each subpanel, we overplot the best-fit nonparametric SFHs for the Smooth component (orange) and integrated flux (black). For better illustration, here we only plot the best-fit SFHs with solid lines, while in Appendix \ref{app:prior} we also plot the 1$\sigma$ uncertainties.}
    \label{fig:seds_specz}
\end{figure*}

\begin{figure*}
    \centering
    \includegraphics[width=\textwidth]{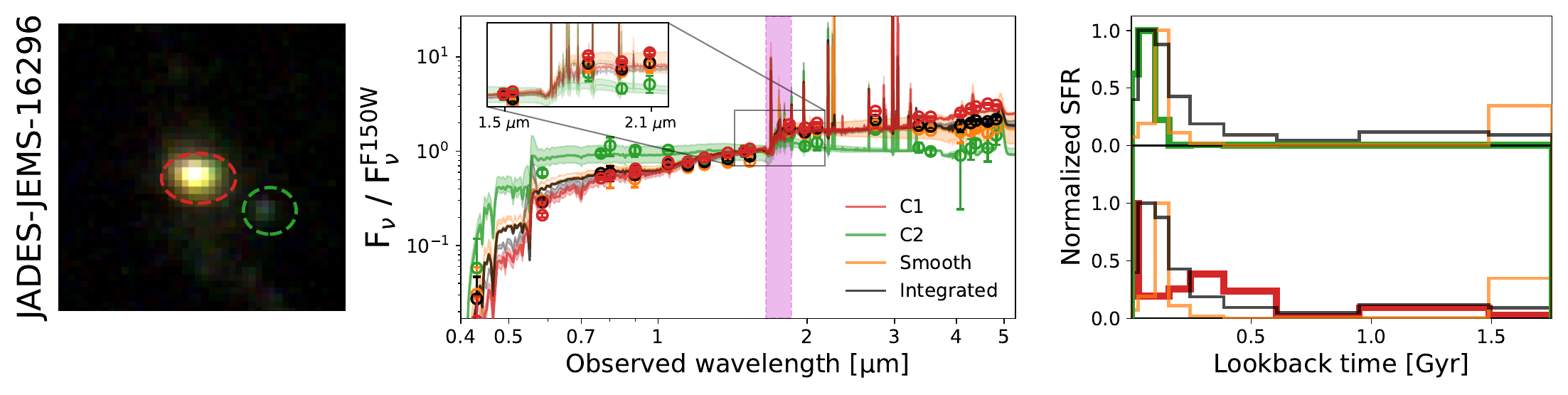}
    \includegraphics[width=\textwidth]{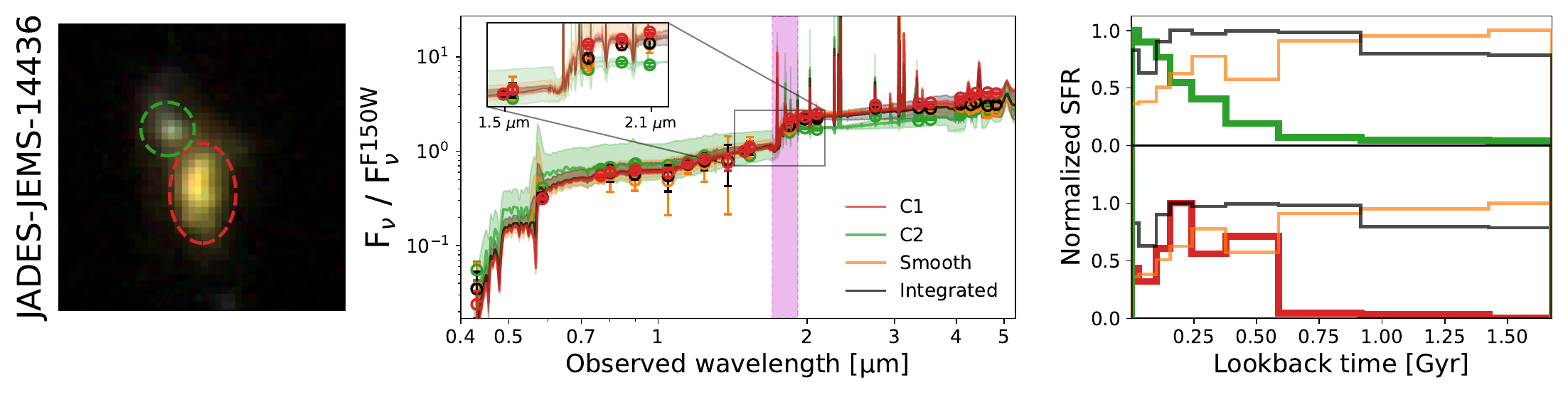}
    \includegraphics[width=\textwidth]{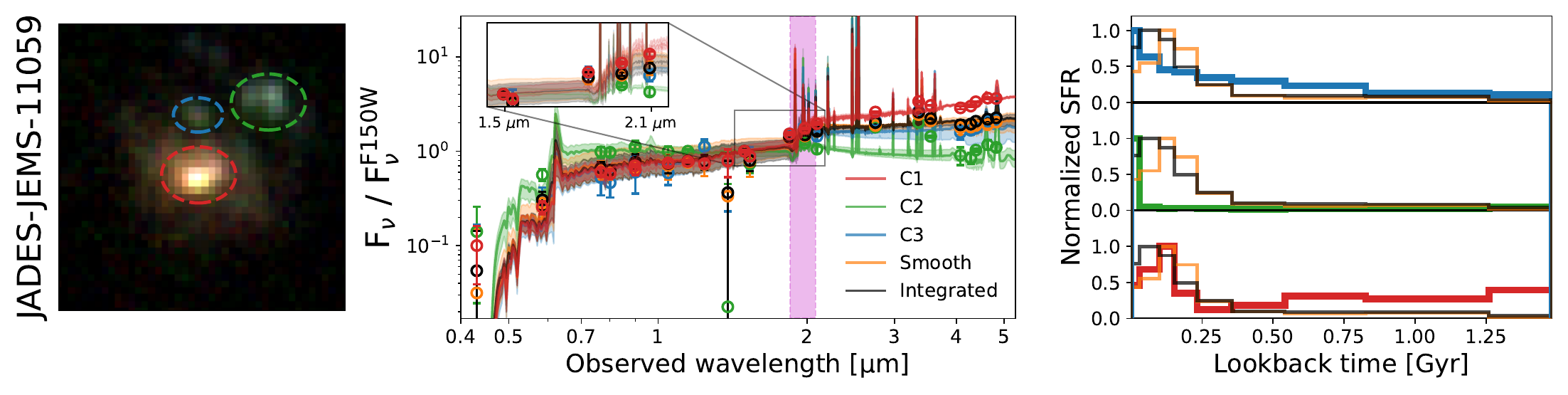}
    \caption{Similar to Figure \ref{fig:seds_specz}, but for the three galaxies with $z_{\rm{phot}}$ shown in Figure \ref{fig:poststamps_cont}. Note that the C2 of JADES-JEMS-11059 is not covered by its segmentation map (the bottom row of Figure \ref{fig:poststamps_cont}), but we still include it to the analysis because its photometric redshift is consistent with that of the central galaxy.}
    \label{fig:seds_photz}
\end{figure*}

Figure \ref{fig:seds_specz} and \ref{fig:seds_photz} show the spatially resolved SED fitting results in detail. The SED shapes of individual structural components differ across the entire rest-frame UV to NIR wavelengths, suggesting that stellar populations vary significantly within each one of the galaxies. This highlights the importance of having spatially resolved measures of stellar populations to get a comprehensive view on the mass assembly processes in high-redshift galaxies.

We refer the readers to Appendix \ref{app:ind} for detailed descriptions of SED fitting results of individual galaxies. In the list below, we only summarize the key findings. In each one of the galaxies,
\begin{itemize}
    \item the red stellar core C1 is the dominant, most massive one among all identified clumpy substructures. Relative to the off-center, minor clumps C2 and C3, C1 contains older stellar populations, spanning a range in mass-weighted stellar age of $0.4-0.7$ Gyr, and has a lower specific star formation rate. In all galaxies, C1 has a common feature in its reconstructed SFH -- it experienced a recent major star-formation episode that started $\sim 0.5-1$ Gyr and peaked at $\sim 0.1-0.3$ Gyr prior to the time of observation. 
    \item compared to C1, the minor, off-center clumps C2 and C3 are less massive by $\gtrsim0.5$ dex. They have bluer colors and larger specific star formation rates, and feature a generally rising SFH. 
    \item the stellar population of the Smooth component has a comparable or older stellar age than C1. The Smooth component contributes the most stellar mass and star formation rate to the entire galaxy, but it has smaller surface densities (i.e. mass/SFR per area) compared to the clumpy substructures.

\end{itemize}

\subsection{Stellar Masses of the Substructures} \label{sec:masses}

The stellar mass of clumpy substructures contains key information on their physical origins \citep[e.g.,][]{Dekel2009,Genzel2011,Guo2012,Bournaud2014,Guo2015,Dessauges-Zavadsky2018}. Directly comparing their stellar masses with the mass threshold of gravitational collapse due to local hydrodynamical instabilities \citep[$M_{\rm{Toomre}}$,][see Section \ref{sec:ori} for details]{Toomre1964} will immediately tell us that if their formation is consistent with in-situ fragmentation in gaseous disks. 

Accurately measuring the stellar mass of clumps, however, is very challenging, because they are embedded in a smooth component, e.g., a galaxy disk. We thus need to properly subtract the contribution from the smooth component at the clump locations, otherwise the stellar-mass measures will be biased \citep{Huertas-Company2020}. In our case, this can be formularized as 
\begin{equation}
    \rm{M_C^{int}= \eta\cdot(M_C-M_{S}^{C}})
    \label{eqn:m_int}
\end{equation}
where $\rm{M_C^{int}}$ is the intrinsic (i.e., true) stellar mass of a clump; $\rm{M_C}$ is the stellar mass of that clump derived from SED fitting with fluxes directly from aperture photometry (Section \ref{sec:phot} and Section \ref{sec:sed}); $\rm{M_{S}^{C}}$ is the contribution from the smooth component to the stellar-mass measure at the clump location and finally $\eta$ accounts for aperture correction. 

Reliably estimating $\rm{M_{S}^{C}}$ and $\eta$ requires information on intrinsic shapes of clumps and the smooth component, which unfortunately are not known a priori, and it becomes increasingly challenging when clumps are fainter. However, we reiterate that the primary goal of this work is to understand the formation of massive stellar cores (i.e., C1), rather than to, e.g., measure the stellar-mass function of clumps. If the stellar mass of a clump is less than the threshold mass even before subtracting $\rm{M_{S}^{C}}$, i.e., $\eta M_{\rm{C}} < M_{\rm{Toomre}}$, its formation will then be fully consistent with in-situ fragmentation. Robustly determining the stellar masses for these low-mass, faint clumps (mostly C2 and C3) is beyond the scope of this paper, and in such cases we simply use $\eta M_{\rm{C}}$ as a strict upper limit (Figure \ref{fig:mt} and Section \ref{sec:ori}). 

In the remainder of this Section, the main focus will be on estimating intrinsic stellar masses of clumps (C1 in particular) more massive than $M_{\rm{Toomre}}$. We estimate them using two different empirical methods (Section \ref{sec:mcint_submass} and \ref{sec:mcint_subflux}), and with different aperture correction factors (Section \ref{sec:mcint_mu}). In short, 
our conclusion, that intrinsic stellar masses of C1 are larger than $M_{\rm{Toomre}}$, does not depend on the choice of method.

\subsubsection{Estimating $\rm{M_C^{int}}$ by Directly Subtracting the Mass Contribution from the Smooth Component} \label{sec:mcint_submass}

As we showed earlier in Figure \ref{fig:coadd_total}, because the coadded measurements of stellar mass and star formation rate are consistent with those derived using integrated flux, this supports our decision to perform the direct subtraction in mass. If we assume that the mass-to-light (M/L) ratio is constant and stellar mass is uniformly distributed within the Smooth component, then $\rm{M_{S}^{C}}$ simply is the stellar mass of the Smooth component derived via SED fitting using fluxes directly from aperture photometry ($\rm{M_S}$) and scaled by the area ratio of the clump (A$_{\rm{C}}$) to the Smooth component (A$_{\rm{S}}$). The intrinsic stellar mass of the clump is then
\begin{equation}
    \rm{M_C^{int}= \eta\cdot \left( M_C - \frac{A_C}{A_S} M_S\right)}.
\end{equation}
Because stellar masses of individual components were derived by SED fitting (Section \ref{sec:sed}), the biggest advantage of this approach is that the spatial variation in M/L has already been largely taken into account, despite not being on a pixel-by-pixel basis. In Figure \ref{fig:mt}, we plot the resulting $\rm{M_C^{int}}$ as filled symbols with error bars. For all C1, we find $\rm{M_{C}^{int}/M_{Toomre}\gtrsim2}$, arguing against formation via in-situ fragmentation. We use the intrinsic stellar masses of C1 derived from this approach as the fiducial ones for the discussion in Section \ref{sec:diss}. 

\subsubsection{Estimating $\rm{M_C^{int}}$ by Subtracting the Flux Contribution from the Smooth Component} \label{sec:mcint_subflux}

Differing from directly subtracting mass, alternatively we can first subtract the flux contribution from the Smooth component to the clump location, and then estimate $\rm{M_C^{int}}$. The Equation (4) thus becomes 
\begin{equation}
    \rm{M_C^{int}= \eta\cdot \left(M_C-\frac{A_C\cdot f_S^{C}}{F_C}M_C\right)}
    \label{eqn:sublight}
\end{equation} 
where $\rm{f_S^{C}}$ is the flux contribution per pixel from the Smooth component to the clump location, and $\rm{F_C}$ is the flux of the clump from aperture photometry.

For each one of the galaxies, we estimate $\rm{f_S^{C}}$  by subtracting from its F210M image the best-fit {\sc forcepho} models (Section \ref{sec:morph_mea}) of all clumpy substructures. We then do a 3$\sigma$ clipping on the residual image and use the median residual pixel value as $\rm{f_S^{C}}$. Because we ignore the Smooth component during the {\sc forcepho} fitting (Section \ref{sec:morph_mea}), instead of using the residual image, we also estimate $\rm{f_S^{C}}$ in a more aggressive way. We first mask out pixels within $r=2\times$R$_e$ circular regions of individual clumpy substructures from the galaxy's segmentation map, and then estimate $\rm{f_S^{C}}$ as the median pixel value of the masked image (i.e., without subtracting the best-fit {\sc forcepho} models of clumpy substructures). This latter estimate certainly overestimates $\rm{f_S^{C}}$ because the flux contribution from clumps to the Smooth component is not fully removed. Even so, we find that $\rm{f_S^{C}}$ only increases by $\sim3-7\%$, which is small and will not cause any substantial changes in our results.

In Figure \ref{fig:mt}, we use hollow symbols with dashed edges to show the $\rm{M_C^{int}}$ of C1 from this approach. The $\rm{M_C^{int}}$ from this approach are larger than the fiducial ones, which is expected because Equation \ref{eqn:sublight} assumes the same M/L for C1 and the Smooth component. As Figure \ref{fig:seds_specz}
and \ref{fig:seds_photz} show, in each one of the galaxies, C1 has a redder rest-frame (B $-$ V) color, as probed by F182M $-$ F277W, than the Smooth component. Because M/L increases as (B $-$ V) becomes redder \citep[e.g.,][]{McGaugh2014}, using Equation \ref{eqn:sublight} can lead to an under-subtraction of $\rm{M_{S}^{C}}$. This issue can be mitigated by doing the {\sc forcepho} analysis across all filters, as opposed to only using the F210M filter, and then taking the M/L variation into account with the {\sc forcepho}-derived model photometry. Following this, we find that the intrinsic masses of C1 can change by $\lesssim 20\%$, a sufficiently small amount that does not substantially change any of our results\footnote{We could use the {\sc forcepho} model photometry from the outset to do SED fitting, instead of using aperture photometry. But we argue that using the model photometry would then make the accuracy of the measured SED sensitive to the assumption of the intrinsic shapes of the clumps, which could deviate from a single S\'{e}rsic profile and may also   depend on wavelength. These systematics are difficult to control, and can be particularly significant for the inferences of stellar age and SFH. Therefore, in this work we decide to stay empirical and use the properties of stellar populations measured with fluxes from aperture photometry.}.

\subsubsection{Estimating the Aperture Correction Factor $\mu$} \label{sec:mcint_mu}

\begin{figure*}
    \centering
    \includegraphics[width=0.87\textwidth]{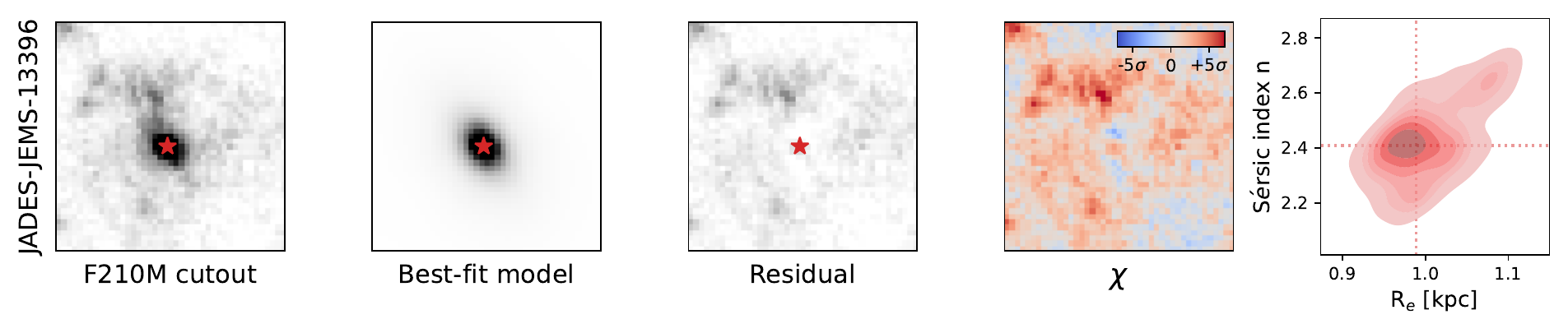}
    \includegraphics[width=0.87\textwidth]{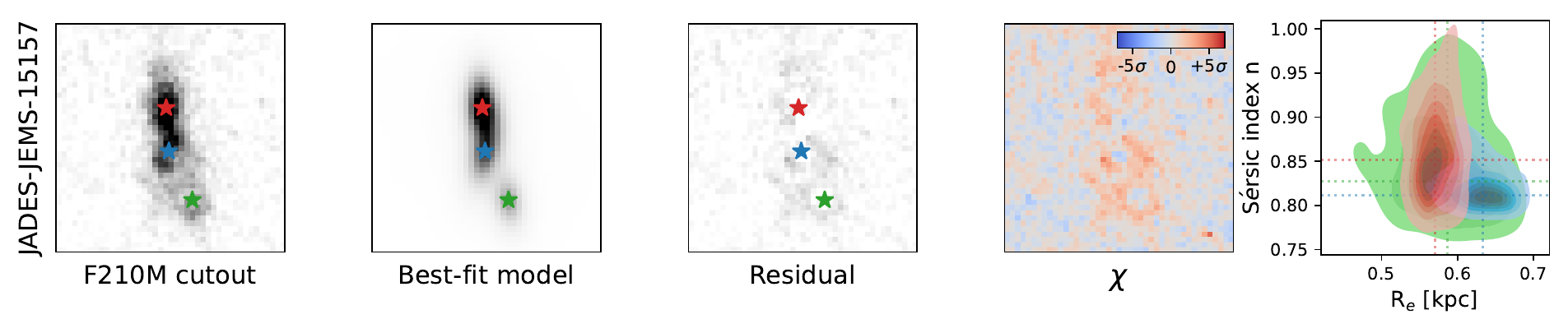}
    \includegraphics[width=0.87\textwidth]{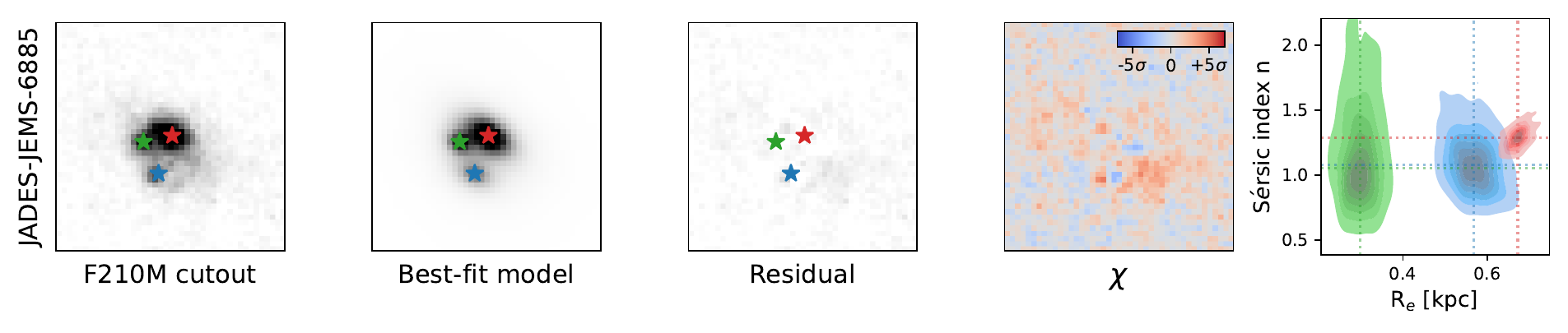}
    \includegraphics[width=0.87\textwidth]{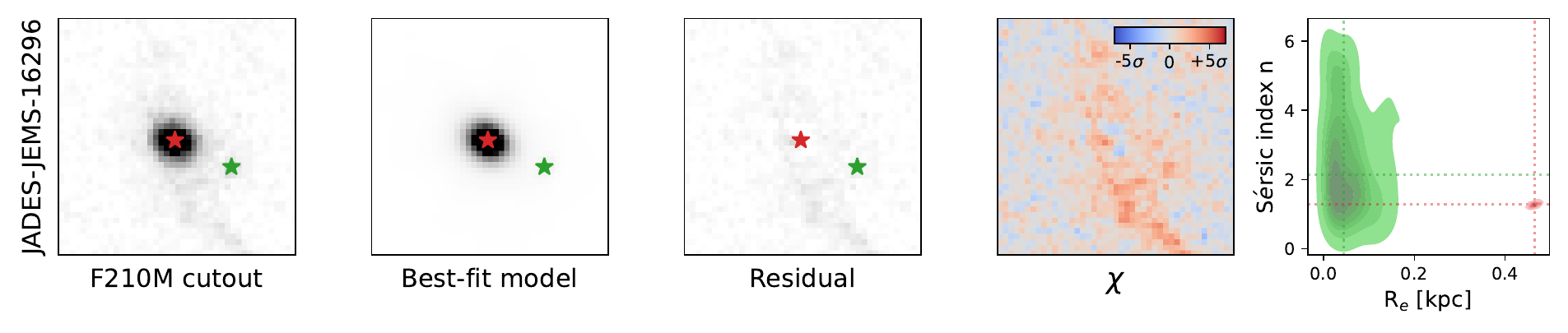}
    \includegraphics[width=0.87\textwidth]{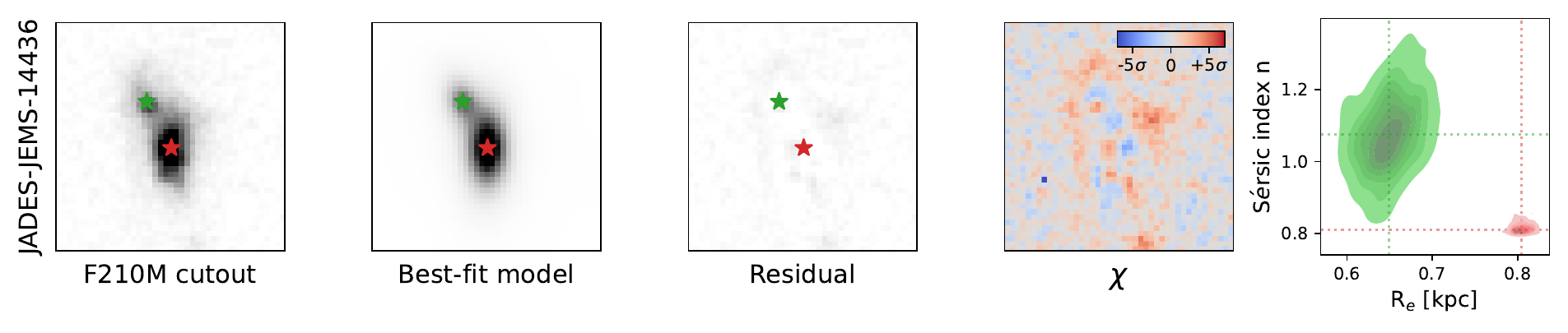}
    \includegraphics[width=0.87\textwidth]{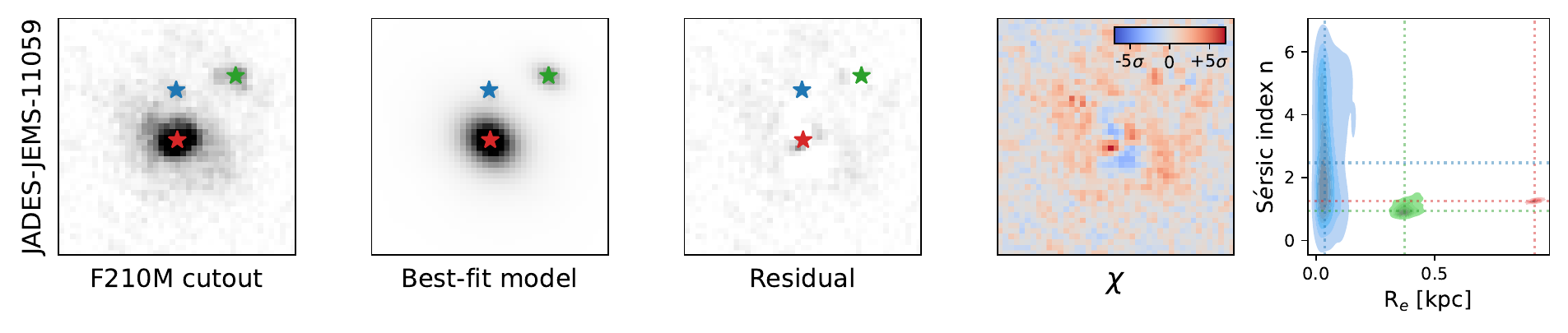}
    \caption{Morphological analysis for the individual substructures. Each substructure is modelled as a S\'{e}rsic profile using the code {\sc forcepho} that adopts an MCMC framework to perform PSF-convolved analysis in individual exposures to better estimate the parameter uncertainties (Section \ref{sec:masses}). The best-fit centroids of individual substructures are marked with the star symbols. The left three columns show the F210M cutout, best-fit model and residual, respectively. The fourth column shows the residual $\chi$ map, which is defined as $\chi=$ (data $-$ model)/error, from which we clearly see the Smooth component surrounding the clumpy substructures. The last column shows the posterior corner plot of S\'{e}rsic index $n$ vs circularized effective (i.e., half-light) radius R$_e$.}
    \label{fig:fpho}
\end{figure*}

By default, we use the best-fit S\'{e}rsic profile from the {\sc forcepho} fitting (Section \ref{sec:morph_mea}) to derive the aperture correction factor $\mu$. As the posterior plots of Figure \ref{fig:fpho} show, the S\'{e}rsic profiles of C1 are tightly constrained, with typical uncertainties of S\'{e}rsic index $n$ and effective radius R$_e$ of $<10\%$. C1 are resolved/marginally resolved in NIRCam/F210M images (PSF FWHM $\approx0''.07$), spanning a range in R$_e$ of $0''.07-0''.14$, corresponding to $0.5-1$ kpc at the median redshift $\bar{z}=3.7$. For minor clumps C2 and C3, their S\'{e}rsic profiles are much less constrained. 
We also check our results by adopting a rather conservative aperture correction, namely, assuming the clumps are point sources, which we already knew does not accurately describe C1's morphologies. Using the point source aperture correction can lead to a decrease in stellar mass by a factor of $\sim 1.5-2.1$, which, still, cannot substantially change our conclusions because all C1 have $\rm{M_{C}^{int}/M_{Toomre}\gtrsim2}$ (see Figure \ref{fig:mt} and Section \ref{sec:ori} below).

\subsection{Origins of the Substructures} \label{sec:ori}

A prominent feature of high-redshift galaxies is the presence of giant stellar clumps which are far more massive than individual star clusters observed in nearby galaxies \citep[e.g.,][]{Elmegreen2007, Genzel2008, Genzel2011, Guo2012, Guo2018, Wuyts2012}. One scenario to form such giant clumpy substructures is through in-situ fragmentation due to violent disk instabilities \citep[e.g.,][]{Noguchi1999,Immeli2004,Bournaud2007,Elmegreen2008,Dekel2009,Ceverino2010,Inoue2016}. In a rotating gaseous disk, gravity needs to overcome both gas pressure and shear forces produced by the relative motion between fluid parcels. \citet{Toomre1964} showed that in-situ fragmentation can happen if the parameter $Q$ is less than unity (Section \ref{sec:intro}). Consequently, clumps have a characteristic mass which represents the largest stellar mass that a clump can form via Toomre instability, a.k.a. the Toomre mass. Following \citet{Genzel2011}, the Toomre mass can be calculated as
\begin{equation}
    M_{\rm{Toomre}}\approx 5\times10^9 \left(\frac{f_{\rm{young}}}{0.4}\right)^2 \left(\frac{M_{\rm{disk}}}{10^{11}M_\sun}\right) M_\sun 
    \label{eqn:mt}
\end{equation}
where $f_{\rm{young}}$ is the mass fraction of components forming stars and $M_{\rm{disk}}$ is the total baryonic mass of a disk. 

To estimate $M_{\rm{Toomre}}$ for our sample, we assume $f_{\rm{young}}\approx f_{\rm{gas}}$, where $f_{\rm{gas}}$ is the gas-to-baryonic mass fraction, i.e., $M_{\rm{gas}}/(M_{\rm{gas}}+M_*)$. Strictly speaking, $M_{\rm{gas}}$ here should be the summation of molecular and atomic gas masses in galaxy disks, but we assume $M_{\rm{gas}}\approx M_{\rm{molgas}}$ for the following reason. While directly measuring atomic gas in galaxies beyond the local universe is technically challenging, studies using the damped Lyman-alpha absorbers toward high-redshift quasars infer a generally very weak, if any, evolution of the cosmic atomic Hydrogen density ($\rho_{\rm{H I}}$) up to $z\sim5$ \citep[e.g.,][]{Rao2006,Prochaska2009,Peroux2020,Walter2020}. Meanwhile, observations with different probes of molecular gas, including CO, FIR SED and sub-mm dust continuum, have reached a consensus, at least up to  $z=4$, that the molecular-gas mass of galaxies rapidly evolves at a rate of $(1+z)^{3\pm1}$ \citep[e.g.,][]{Genzel2015,Scoville2017,Tacconi2018,Liu2019}. Putting together the weak $\rho_{\rm{H I}}$ and strong molecular-gas evolutions, because the mass ratio of atomic gas divided by molecular gas is $\approx 2-3$ at $z\sim0$ \citep[e.g.,][]{Saintonge2011,Catinella2013}, it is reasonable to assume $M_{\rm{gas}}\approx M_{\rm{molgas}}$ (hence $f_{\rm{gas}}\approx f_{\rm{molgas}}$) in galaxies at $z\gtrsim1$ (see a similar argument from \citealt{Tacconi2018} for this assumption).

We estimate $f_{\rm{molgas}}$ for the 6 galaxies using the best-fit relationship from \citet{Tacconi2020} (their Table 3), who combined a number of surveys of galaxies' molecular gas content across cosmic time, and also included the dependence of the relationship on stellar mass and the distance to the star-forming main sequence. We note, however, that almost all those high-redshift surveys were designed for galaxies with M$_*>10^{10}M_\sun$, either on or above the star-forming main sequence. Because 5 out of the 6 galaxies presented in this work have M$_*\approx10^{9.3-10}M_\sun$ (Table \ref{tab:basic_info}), we use the $f_{\rm{molgas}}$ derived for M$_*=10^{10}M_\sun$ galaxies.  As we will show in Section \ref{dis:quench}, all 6 galaxies are below the star-forming main sequence. Using their distances to the star-forming main sequence, we find $f_{\rm{gas}}\approx 0.5$ for our sample, similar to those observed in star-forming galaxies on the main sequence at $z\sim2$ \citep[e.g.,][]{Tacconi2008,Tacconi2010,Daddi2010}. We note that $f_{\rm{molgas}}$ increases with decreasing stellar mass. If the scaling relationship of $f_{\rm{molgas}}$ from \citet{Tacconi2020} can be safely used for galaxies with M$_*<10^{10}M_\sun$, we would get $f_{\rm{molgas}}\sim0.6$ assuming M$_*=10^{9.3}M_\sun$, the lowest stellar mass of the 6 galaxies. This would increase $M_{\rm{Toomre}}$ by a factor of $(0.6/0.5)^2=1.44$, which however is small compared to the stellar mass of C1 and hence does not change our conclusion about the formation of C1 detailed below.

\begin{figure*}
    \centering
    \includegraphics[width=1\textwidth]{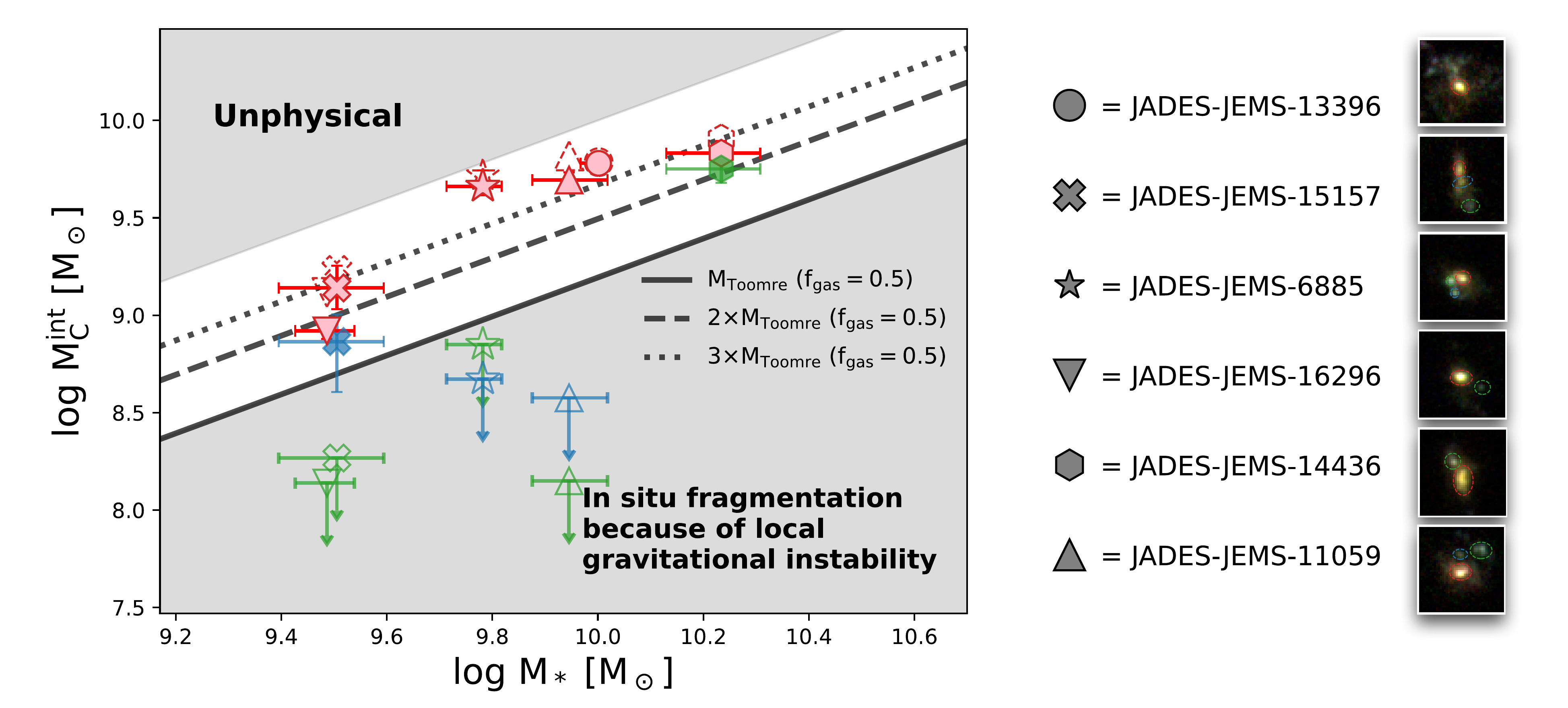}
    \caption{The scatter plot of the intrinsic stellar masses of clumpy substructures ($y$-axis) vs. the total stellar masses of their host galaxies ($x$-axis). The black solid line marks the prediction of Toomre mass $M_{\rm{Toomre}}$, which represents the largest stellar mass that a clump can form through Toomre instability in an unstable, gaseous disk (see Section \ref{sec:ori}). The dashed and dotted lines mark $2\times$ and $3\times M_{\rm{Toomre}}$, respectively. The upper-left grey shaded region marks the unphysical part where $\rm{M_{C}^{int} > M_*}$. The symbols used for individual galaxies, as well as the corresponding color scheme used for individual clumpy substructures are labelled on the right. For low-mass clumps whose stellar masses are already less than $M_{\rm{Toomre}}$ before subtracting the contribution from the surrounding smooth component (Section \ref{sec:masses}), we plot the upper limits of their $\rm{M_{C}^{int}}$. Similar to previous plots, the stellar cores C1 are color-coded in red. The filled symbols with error bars show the $\rm{M_{C}^{int}}$ estimated via the  method of Section \ref{sec:mcint_submass}, while the hollow symbols with dashed edges show the $\rm{M_{C}^{int}}$ estimated via a different method described in of Section \ref{sec:mcint_subflux}. We find that the intrinsic stellar masses of C1 are significantly larger than $M_{\rm{Toomre}}$ in all the six galaxies, suggesting that C1 may not form through in-situ fragmentation due to Toomre instability.}
    \label{fig:mt}
\end{figure*}

With $f_{\rm{gas}}$ in hand, we then get $M_{\rm{disk}} = M_*+M_{\rm{gas}} = M_*/f_{\rm{gas}}$, and finally use Equation \ref{eqn:mt} to calculate $M_{\rm{Toomre}}$. In Figure \ref{fig:mt}, we plot the relationship between $M_{\rm{Toomre}}$ and $M_*$ as the black solid line. In what follows, we discuss possible physical origins of the clumpy substructures.

To begin, Figure \ref{fig:mt} shows that the most minor clumps (C2 and C3) are less massive than $M_{\rm{Toomre}}$, hence they are consistent with in-situ formation through gravitational fragmentation in gaseous, turbulent disks. These clumps are bluer, and have enhanced specific star formation rates compared to C1 and the Smooth component (Table \ref{tab:sed_info}). These properties are similar to the star-forming clumps found in galaxies at lower redshifts $1<z<3$ from earlier HST studies \citep[e.g.,][]{Elmegreen2005, Elmegreen2007,Wuyts2012}. Quantitatively, these individual minor clumps contribute $5-50\%$ of the total star formation rate, and $<10\%$ of the total stellar mass of their host galaxies, in broad agreement with earlier studies of UV-bright clumps at $1<z<3$ \citep[e.g.,][]{Guo2012}. In addition, they have overall rising SFHs started $\approx0.3$ Gyr ago, which is also consistent with the young stellar ages found in $2<z<3$ star-forming clumps \citep{Elmegreen2009,ForsterSchreiber2011,Genzel2011,Guo2018}. 

Apart from C1, two other clumpy substructures are more massive than $M_{\rm{Toomre}}$, i.e., the C3 of JADES-JEMS-15157 (blue X in Figure \ref{fig:mt}) and the C2 of JADES-JEMS-14436 (green hexagon in Figure \ref{fig:mt}). Regarding the former, its intrinsic stellar mass is $\sim1.5\times$ $M_{\rm{Toomre}}$. We notice, however, that it might contain two distinct clumps, which is particularly clear in the F210M image but becomes less obvious in other filters (Figure \ref{fig:poststamps}). After taking this into account, its formation is then still consistent with in-situ fragmentation. Regarding the latter, i.e. the C2 of JADES-JEMS-14436, we see a clear tidal feature associated with this substructure in the NIRCam SW's filter images (Figure \ref{fig:poststamps_cont} and Appendix \ref{app:ind}). Therefore, the physical origin of this substructure is very likely related to a minor merger with a mass ratio of 1:10 (Table \ref{tab:sed_info}). The minor mergers can have two broad effects on the clump formation. First, minor mergers can disturb the gas distribution in galaxy disks, trigger local starbursts and hence induce clump formation \citep{Bournaud2014,Inoue2016}. Second, the clumps can be minor mergers themselves, i.e., the clumps have an ex-situ origin. Using cosmological hydrodynamical simulations, \citet{Mandelker2014} found that the ex-situ clumps have a typical stellar age of $0.5-3$ Gyr and specific star formation rate of $0.1-2$ Gyr$^{-1}$, which are in quantitative agreement with the C2 of JADES-JEMS-14436 (Figure \ref{fig:seds_photz} and Appendix \ref{app:ind}).  

Finally, at the core of this study is the physical origin of C1 -- the stellar core. In each one of the 6 galaxies, C1 is redder than the other parts of the galaxy, and has significantly larger stellar mass than $M_{\rm{Toomre}}$, with a typical ratio of $\rm{M_{C}^{int}/M_{Toomre}\gtrsim2}$. This excessive stellar mass is strong evidence that C1 may not formed via a single in-situ fragmentation in gaseous disks. Relative to those  minor clumps discussed above, C1 has a lower specific star formation rate and is closer to the light centroid of NIRCam images at $\sim 4\micron$. All these are consistent with C1 being a forming stellar core/(pseudo-)bulge, which will be discussed in detail in the remainder of this paper.

\section{Discussion} \label{sec:diss}

\subsection{The Formation of C1: Early (Pseudo-)Bulge Formation?} \label{dis:bulge}

In the local universe, bulges and pseudo-bulges are believed to have different origins, with the former primarily formed through major mergers \citep[e.g.][]{Hernquist1989,Barnes1996,Hopkins2009} and the latter formed through secular evolution in galaxy disks \citep[e.g.][]{Debattista2004,Okamoto2013,Athanassoula2015}.  According to the classification for $z\sim0$ galaxies, the light distribution of a classic bulge is similar to a spheroid ($n=4$), while it becomes flattened (smaller $n$) for a pseudo-bulge \citep{Kormendy2004}. Therefore, from a purely morphological perspective (Figure \ref{fig:fpho}), the C1 of JADES-JEMS-13396 ($n>2$) is more like a classic bulge, while in the other 5 galaxies C1 are consistent with pseudo-bulges, although we stress that without dynamical measures it remains difficult to definitively distinguish the two.

At high redshifts, early numerical simulations showed that stellar cores formed through the coalescence of in-spiral stellar clumps should have $n\sim2-4$, being consistent with classic bulges \citep{Elmegreen2008}. Later simulations however, after taking into account the gas kinematics of individual clumps, showed that the remnant of the clump coalescence should still have significant rotation, arguing that the high-redshift cores are more consistent with being pseudo-bulges \citep{Ceverino2012,Inoue2012}. Regardless, those simulations consistently showed that the coalescence of clumps is an efficient way to grow dense central cores in gas-rich galaxies at high redshifts. Indeed, the coexistence of C1 with other minor, off-center star-forming clumps (C2 and C3) in the 6 galaxies is consistent with this picture.  

The reconstructed SFHs provide another key constraint on the physical origin of C1. If they form through the coalescence of in-spiral clumps, the timescale for the clumps' inward migration will be $\approx0.5$ Gyr \citep[e.g.,][]{Noguchi1999,Immeli2004,Bournaud2007,Genzel2008,Genzel2011}. This is consistent with the SED fitting that the mass-weighted stellar ages of C1 span a range of $0.4-0.7$ Gyr (Section \ref{sec:sfh_ind}). The clump migration scenario also predicts the radial age gradient of clumps. Indeed, those off-center minor clumps (C2 and C3, if present) have significantly younger SFHs (Figure \ref{fig:seds_specz} and \ref{fig:seds_photz}), spanning a range in mass-weighted stellar age of $0.1-0.5$ Gyr (Table \ref{tab:sed_info}). These are in quantitative agreement with the scenario of growing the stellar core of massive galaxies at high redshift through VDI, i.e. one of the processes associated with gas-rich compaction initially proposed by \citet[][also see the Section \ref{sec:intro} for a detailed description]{Dekel2009} which predicts that clumps should have stellar ages $<0.5$ Gyr at the outskirt while they are older at the center. 

One important question about the efficacy of the inward migration scenario is the uncertain lifetime of clumps. Can clumps formed in disks survive stellar feedback until they coalesce at the center? Some simulations with intense stellar feedback models showed that clumps are short-lived, and should be self-destroyed within $<50-100$ Myr \citep{Genel2012,Hopkins2012}. In contrast, other simulations with different stellar feedback models showed that they can produce relatively longer-lived stellar clumps with a typical age of a few hundred Myr \citep{Perez2013,Bournaud2014,Ceverino2014}. The stellar ages of $0.1-0.5$ Gyr measured for the minor clumps C2 and C3 favor a relatively longer lifetime of clumps than the predictions from intense stellar feedback models, similar conclusions were also reached by earlier studies of $z\sim2$ star-forming clumps \citep{Wuyts2012,Guo2012}. Arguably, however, the measured stellar ages of C2 and C3 still seem to be shorter than the $\approx0.5$ Gyr required for the inward migration. As pointed out by \citet{Bournaud2016}, however, the observed stellar ages of clumps are only lower limits to their true ages, because during their evolution clumps continuously exchange mass with the surrounding ISM through outflows and inflows \citep{Bournaud2007,Dekel2013,Bournaud2014,Perret2014}. According to the zoom-in simulations from \citet{Bournaud2014}, a clump with an observed stellar age of $0.1-0.2$ Gyr corresponds to a true age of $0.3-0.5$ Gyr. Taking this into account, the inferred stellar populations for the minor clumps thus are still in line with the scenario where the central regions of high-redshift galaxies are built through the in-spiral of clumps. 

We also note that gas-rich major mergers (another process associated with gas-rich compaction), if present, can help efficiently grow central cores/bulges at high redshifts as well. However we do not see clear evidence of major mergers in the 6 galaxies presented here. The effect induced by recent minor mergers, e.g., the formation of bars, can also contribute to the buildup of (pseudo-)bulges \citep[e.g.,][]{Bournaud2005,ElicheMoral2011,Guedes2013}. Given the complex morphologies of the galaxies, however, it is hard to distinguish an in-situ stellar clump from a merging low-mass galaxy. Minor mergers thus might also contribute to the development of central regions in the six galaxies.

Finally, we notice that C1 has a comparable/younger mass-weighted stellar age than the Smooth component in each one of the galaxies (Table \ref{tab:sed_info}). As the reconstructed SFHs show (see Figures \ref{fig:seds_specz} and \ref{fig:seds_photz}), this is because the Smooth component started its major mass assembly either at about the same time as C1, or earlier, i.e., $>1$ Gyr prior to the time of observation. Interestingly, recent studies of galactic archaeology found that the Milky Way's bulge is predominately composed of $\gtrsim9-10$ Gyr-old stars \citep[corresponding to a formation redshift $z\gtrsim2-3$, e.g.,][]{Barbuy2018,Hasselquist2020}, while the formation of thick-disk stars likely started as early as at $z>6$, i.e. 13 Gyr ago \citep[][]{Xiang2022}. These are seemingly similar to the reconstructed SFHs of C1 and the Smooth component. It is thus possible that we are witnessing the early formation of structural components that will eventually evolve into the bulge and thick-disk components in the Milky Way-like galaxies at $z\sim0$, although we caution that the 6 galaxies are likely hosted by dark matter halos $\approx0.5-1.5$ dex more massive\footnote{We base this estimate on the stellar-to-halo mass ratio, and the halo mass growth history from \citet{Behroozi2013a,Behroozi2013b}. For a $10^{9.7}M_\sun$ galaxy at $z=3$, we find it is hosted by a $\sim10^{12.7}M_\sun$ halo at $z\sim0.1$, which is $\approx0.7$ dex more massive than the Milky Way's dark matter halo \citep{Posti2019}.} than the Milky Way's.

\subsection{Implications for Galaxy Quenching} \label{dis:quench}

We now advance the discussion by connecting what we found about the 6 galaxies to the general picture of galaxy quenching. 

\begin{figure*}
    \centering
    \includegraphics[width=0.77\textwidth]{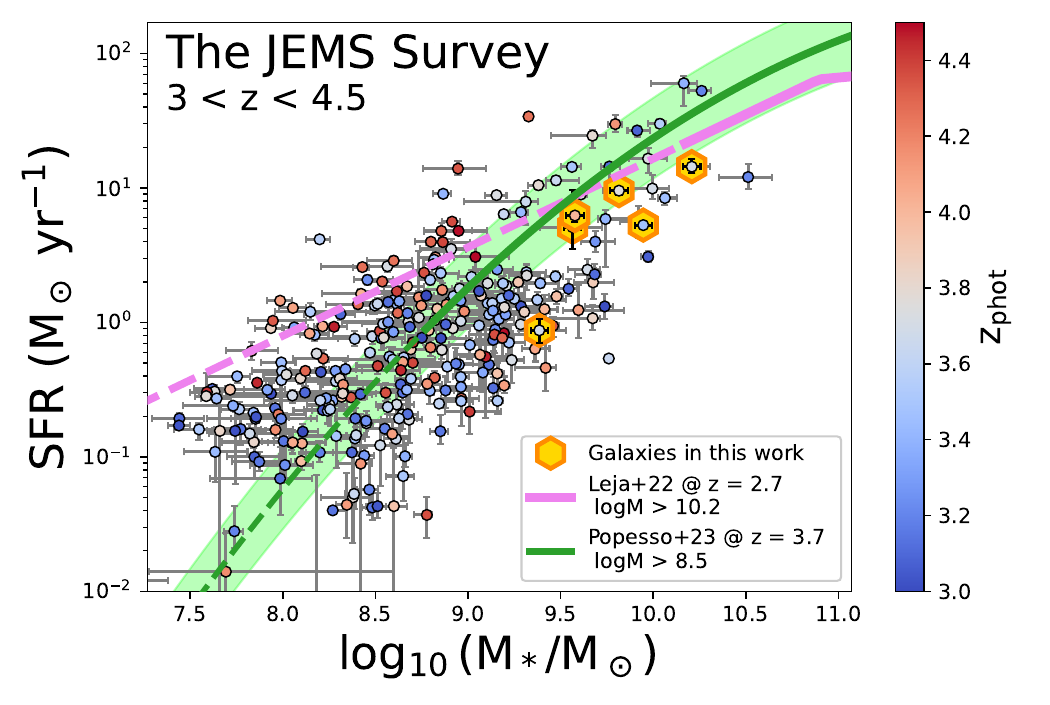}
    \caption{The star-forming main sequence of $3<z<4.5$ galaxies from JEMS. The stellar mass and star formation rate are derived from SED fitting with integrated flux. Each galaxy is color-coded using its best-fit photometric redshift. The green solid line shows the star-forming main sequence at $z=3.7$ for $M_*>10^{8.5}M_\sun$ galaxies from \citet{Popesso2023}. We extrapolate it to $M_*<10^{8.5}M_\sun$, and plot it as the green dashed line. The green shaded region marks the  $\pm0.3$ dex range. Also plotted as the magenta solid line is the star-forming main sequence at $z=2.7$ for $M_*>10^{10.2}M_\sun$ galaxies from \citet{Leja2022}, who also used \prospector with a similar SED model of this work. Similarly, we also extrapolate it to $M_*<10^{10.2}M_\sun$, and plot it as the magenta dashed line. The six galaxies presented in this work are marked with golden hexagons. They are below the star-forming main sequence by $\approx0.2-0.7$ dex. }
    \label{fig:sfms}
\end{figure*}

To begin, in Figure \ref{fig:sfms} we present the star-forming main sequence, i.e. star formation rate vs stellar mass, for all 301 galaxies at $3<z<4.5$ in our parent sample from JEMS (Section \ref{sec:sample}). The star formation rate and stellar-mass measures shown in the plot are from our \prospector SED fitting with the same model as that assumed for the 6 galaxies (Section \ref{sec:sed}), measured with integrated flux. We compare the star-forming main sequence with that from \citet{Popesso2023}\footnote{As noted by \citet{Popesso2023}, they made a $-0.3$ dex correction for the stellar-mass measures from the \prospector fitting with nonparametric SFHs. This systematic offset in stellar-mass measures between nonparametric and parametric SFHs has also been reported by others \citep{Leja2019,Leja2022,Ji2022a,Ji2023}. For a proper comparison, we therefore enlarge the stellar mass in the best-fit star-forming main sequence of \citet{Popesso2023} (their Equation 10) by 0.3 dex.} who conducted a comprehensive literature search, and used a consistent method to derive  the star-forming main sequence over $1<z<6$ for galaxies with $M_*>10^{8.5}M_\sun$. As Figure \ref{fig:sfms} shows, our measurements are in excellent agreement with the distribution measured by \citet{Popesso2023}. We also plot the star-forming main sequence from \citet{Leja2022}, who also used \prospector with a very similar SED model to ours, but only measured the relation for galaxies up to $z=2.7$ with $M_*>10^{10.2}M_\sun$ owing to the stellar-mass incompleteness. As Figure \ref{fig:sfms} shows, the 6 galaxies presented in this work are below the star-forming main sequence by $0.2-0.7$ dex, suggesting a possible link between the presence of a massive stellar core (e.g., C1) and galaxy quenching.

\begin{figure}
    \centering
    \includegraphics[width=0.47\textwidth]{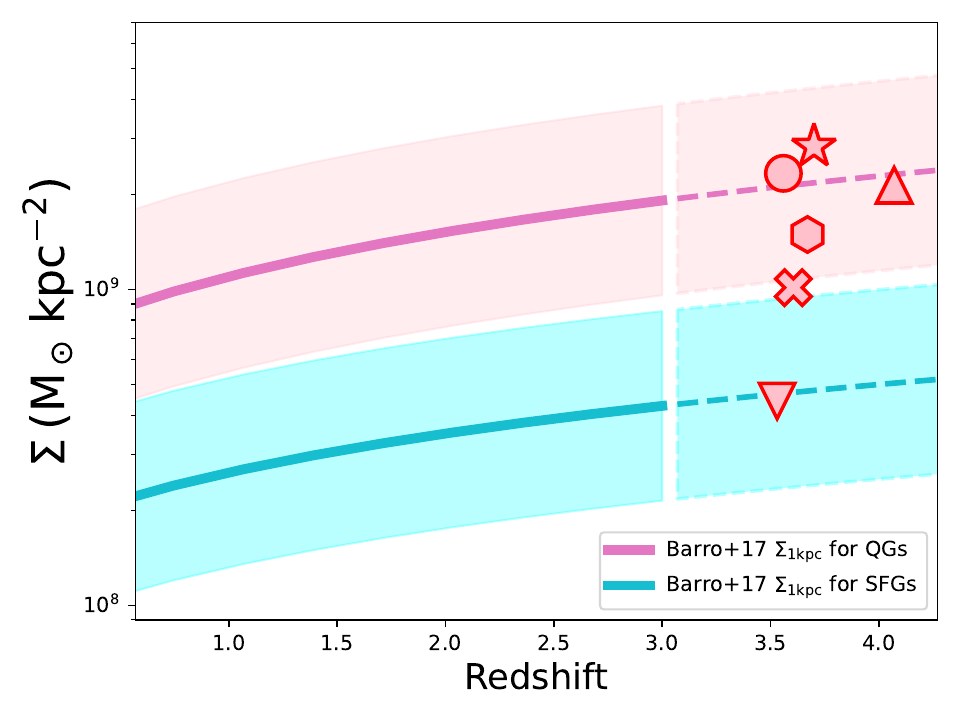}
    \caption{The stellar-mass surface density at the location of C1. The symbol used for each galaxy is the same as Figure \ref{fig:mt}. The solid line shows the $\Sigma_{\rm{1kpc}}$ evolution of galaxies at $0.5<z<3$ measured by \citet{Barro2017}. We extrapolate this evolution to $z>3$ and plot it as the dashed line. The evolution of star-forming galaxies and quiescent galaxies is plotted in cyan and magenta, respectively. The shaded region marks the $\pm0.3$ dex range. Although the six galaxies have not quenched yet, their stellar-mass densities within the region of C1 are already high, approaching the central stellar-mass surface density of high-redshift quiescent galaxies (see Section \ref{dis:quench} for detailed discussions and caveats of this comparison).}
    \label{fig:sigma}
\end{figure}

In Figure \ref{fig:sigma}, we continue the discussion by comparing the stellar-mass surface densities at the location of C1 ($\rm{\Sigma_{C1}}$)  with the prediction based on the evolution of $\Sigma_{\rm{1kpc}}$, i.e., the stellar-mass surface density within the central radius of 1 kpc, of $0.5<z<3$ galaxies  measured by \citet{Barro2017} using HST data from CANDELS. Specifically, we compare with the $\Sigma_{\rm{1kpc}}$ evolution of star-forming (the cyan line in Figure \ref{fig:sigma}) and quiescent (the magenta line in Figure \ref{fig:sigma}) galaxies with $M_*=10^{9.7}M_\sun$, i.e., the median stellar mass of the galaxies presented in this work. 

Before proceeding to discuss the results, we clarify how we estimate $\rm{\Sigma_{C1}}$ for a fair comparison with the existing $\Sigma_{\rm{1kpc}}$ measures, and caution about the caveats in this comparison. First, because C1 are close to the center of stellar-light distributions (NIRCam images at $\sim4\micron$ in Figure \ref{fig:poststamps} and Figure \ref{fig:poststamps_cont}), and their sizes are $\approx0.5-1$ kpc (effective radius, Figure \ref{fig:fpho}), $\rm{\Sigma_{C1}}$ is a good proxy for the central stellar-mass surface densities of the galaxies presented in this work. Second, we note that the existing $\Sigma_{\rm{1kpc}}$ measures simply assumed a single S\'{e}rsic profile for the entire galaxy, meaning that those $\Sigma_{\rm{1kpc}}$ measures included all stellar masses in the central regions, regardless of their origins (e.g., cores vs stars in disks). For a fair comparison, instead of using $\rm{M_{C1}^{int}}$ (Equation \ref{eqn:m_int}), we thus include all masses at the location of C1, i.e., $\rm{\Sigma_{C1} = \eta M_{C1}/A_{C1}}$, where $\rm{A_{C1}}$ is the aperture size used for C1. Third, the earlier studies were only able to study the evolution of $\Sigma_{\rm{1kpc}}$  up to $z=3$.  In this work, we therefore need to extrapolate the $\Sigma_{\rm{1kpc}}$ evolution from \citet{Barro2017}  to a higher redshift range, i.e., $3<z<4.5$. Finally, at $z\sim3$, while the stellar-mass completeness of CANDELS is still $\sim 10^{9.5}M_\sun$ for star-forming galaxies \citep[e.g.,][]{Ji2018,Santini2022}, it rises to $\sim 10^{10.3}M_\sun$ for quiescent galaxies \citep[][]{Barro2017}. We therefore also need to extrapolate the $\Sigma_{\rm{1kpc}}$ evolution for quiescent galaxies from \citet{Barro2017} to a lower stellar mass $10^{9.7}M_\sun$. 

As Figure \ref{fig:sigma} shows, except JADES-JEMS-16296, all the other 5 galaxies have large central stellar-mass surface densities, namely, that their $\rm{\Sigma_{C1}}$ are already similar to the $\Sigma_{\rm{1kpc}}$ of quiescent galaxies of similar redshifts and masses. Interestingly, JADES-JEMS-16296 in fact is the galaxy that shows evidence of recent rejuvenation of central star formation  which will be further discussed in Section \ref{dis:wetcompaction}. Although the galaxies presented in this work are not quenched yet, they are below the star-forming main sequence. On the one hand, although we cannot be sure that these 6 galaxies will not rejuvenate their star-formation activities in the future, the findings here suggest that galaxies have already formed significant amount of their mass in a spheroidal central component while they are still in the star-forming phase, which is consistent with what has been found in the IllustrisTNG simulations \citep{Pillepich2019,Tacchella2019}. On the other hand, it is also possible that the galaxies are actually heading toward quiescence. The presence of a massive stellar core with a surface density comparable to  quenched galaxies then suggests that the suppression of star formation may occur along with the development of dense central regions. The implication is that there is either a causal link between quenching and central structural transformations, or an as-yet unidentified physical process that controls both events. 

The conclusion above is in broad agreement with the studies at lower redshifts \citep[e.g.][]{Fang2013,Barro2017,Whitaker2017,Tacchella2018,Suess2021,Ji2023}. What is different here, though, is that we are now able to study the buildup of central regions in the act at $3<z<5$ in the progenitor of lower-redshift massive quiescent galaxies, thanks to the deep NIRCam imaging obtained by JADES and JEMS. This, in our view, is an important step forward because those lower-redshift quiescent galaxies are already old -- the typical stellar age of $z\sim1-2$ massive quiescent galaxies is about 1 Gyr \citep[e.g.,][]{Carnall2019b,Tacchella2022,Ji2022a}. This fact complicates the interpretation for the correlation between the star-formation properties and central stellar-mass surface density that is well established at $z=1-2$ \citep[e.g.][]{Belli2015, Williams2017, EstradaCarpenter2020}, because post-quenching processes can alter galaxies' structures. Thus, it is difficult to assess if the correlation is really due to a causal link between the development of dense cores and quenching, or is simply a byproduct of another yet-to-be identified physical process. Pushing observations closer to higher redshifts $z>3$, i.e. closer to the time of quenching, enables a more direct view of the relative timing between structural transformations and quenching, hence guarantees a more accurate constraint on the relationship between the two phenomena.

\subsection{Consistency between our Observations and the Gas-rich Compaction Scenario} \label{dis:wetcompaction}

Finally, we discuss our constraints on the physical mechanisms responsible for the correlation between the star-formation properties and central stellar-mass surface density observed in galaxies at cosmic noon \citep[e.g.,][]{Franx2008,Cheung2012,Whitaker2017,Barro2017}. Our observations are fully in line with the so-called ``gas-rich compaction'' scenario, which, as we already mentioned in Section \ref{sec:intro}, is a class of processes whose main feature is highly dissipative gas accretion toward the center of galaxies. This process promotes enhanced central star-formation activity, followed by the central gas depletion and finally the quenching of central star formation \citep{Dekel2009,Ceverino2010,Dekel2014,Wellons2015,Zolotov2015,Tacchella2016}. 

One important prediction of gas-rich compaction is that, right after a compaction event, we should observe a younger stellar population at the center as a result of recent central starburst triggered by the highly dissipative gas accretion. This is observed in the 6 galaxies where C1 are comparable to/younger than the Smooth component (Section \ref{sec:sfh_ind}). Additionally, we find that C1 feature a common SFH whose star formation rate declines over the past $0.1-0.3$ Gyr, following a major star-forming episode  $\approx0.5-1$ Gyr prior, and peaked $\approx0.2-0.5$ Gyr ago. This SFH is consistent with the 6 galaxies experiencing a gas compaction event, followed by a suppression of star formation likely as the result of central gas depletion. If there is future gas replenishment, this chain of events, i.e. gas compaction followed by gas depletion, will repeat several times before the final quiescence of the entire galaxy at $z<3$ \citep{Tacchella2016}. Indeed, we see evidence of  star-formation rejuvenation in the C1 of JADES-JEMS-16296 by the time of observation. This rejuvenation event happens  at $\approx0.3$ Gyr after the peak of the galaxy's most recent major star-formation episode. Interestingly, this timescale is very similar to the model predictions for the oscillation timescale  ($0.2-0.5\,{\rm{t_H}}$, corresponding to $0.3-0.8$ Gyr at $z=3.7$) of star formation triggered by gas replenishment \citep{Tacchella2016}. However,  we caution that this result is sensitive to the assumed prior of nonparametric SFHs (see Appendix \ref{app:prior}).

\section{Summary} \label{sec:sum}

This is the first paper of the series that uses deep 14-filter JWST/NIRCam imaging data from the JEMS and JADES surveys to study in detail the spatially resolved stellar populations of galaxies at redshifts $3<z<4.5$, a key epoch of the universe that can provide unique insight into the role that substructures among massive galaxies may play in the transition from star-forming disks to bulge-dominated, quiescent galaxies.

We visually inspected color maps at rest-frame $3600-4100$ \AA\ produced by JWST/NIRCam imaging for all $3<z<4.5$ galaxies in JEMS, from which we identified a sample of 37 galaxies having substructures with distinct rest-frame colors. These 37 galaxies show diverse spatially resolved stellar populations. In this first paper, we present the spatially resolved stellar populations in 6 galaxies with massive stellar cores, representing a specific stage of galaxy structural transformation.

We divided each one of the 6 galaxies into different structural components, including (1) a bright central stellar core C1, the focus of this study, (2) off-center, minor clumps C2 and C3 (if present), and (3) the Smooth component representing the remaining, extended light distribution of the galaxy. Together with 9-filter ancillary imaging from HST, we fit a total of 23-filter photometry covering truly panchromatic swathes of observed wavelength range $\lambda=0.4-5\micron$, corresponding to the rest-frame $0.1-1\micron$, using the SED fitting code \prospector with nonparametric SFHs. We then studied the stellar-population properties of those different structural components.

In each one of the galaxies, C1 has a redder color at rest-frame 4000 \AA\ and a significantly lower specific star formation rate than off-center clumps C2 and C3. We showed that C1 is $\gtrsim2$ times more massive than $M_{\rm{Toomre}}$, i.e., the maximum stellar mass that a clump can form via Toomre instability, suggesting that it may not have formed via single in-situ fragmentation in an unstable, gaseous disk. Instead, because C1 has a typical stellar age of $\sim 0.5$ Gyr which is similar to the timescale of clump inward migration as the result of dynamical friction, we argued that C1 likely formed through the coalescence of giant stellar clumps, a model prediction from the violent disk instability which is one of the process associated with the gas-rich compaction scenario at high redshift.

We found that all 6 galaxies are below the star-forming main sequence by $0.2-0.7$ dex. We showed that the stellar-mass surface densities of C1 are already comparable to the central stellar-mass surface density of quenched galaxies of similar masses and redshifts. In addition, we found that the stellar populations of C1 are either comparable to or younger than the Smooth component. Putting these together, we thus concluded that we are likely witnessing the coeval development of dense central cores, along with the beginning of galaxy-wide quenching at $z > 3$, likely triggered by gas-rich compaction.

As a closing remark, we stress that the sample presented here only contains 6 galaxies and hence has no statistical power. In addition, the visual selection means that this study is biased toward bright substructures in relatively bright galaxies. To obtain a robust statistical characterization of massive stellar cores in high-redshift galaxies, and a comprehensive picture of its relationship with quenching, a much larger sample with a more uniform selection method is required. We, however, view this work as a pilot study to demonstrate the power of JWST, especially its imaging capability with NIRCam, in constraining the spatially resolved mass assembly history of galaxies at $z>3$ in unprecedented detail. We reiterate that the galaxies here were selected from JEMS, whose sky coverage is relatively small ($\sim10$ arcmin$^2$). The ongoing JADES survey will eventually produce NIRCam imaging and NIRSpec spectroscopy of similar depth over a sky area of $\gtrsim200$ arcmin$^2$ in the GOODS fields, promising a statistically significant galaxy sample at $z>3$ that will dramatically advance our understanding of the physics of galaxy structural transformation and quenching in the early universe.

\section*{acknowledgments}
ZJ, BDJ, BR, FS, DJE, MR, GR, KH, CNAW, EE, ZC, JMH and JL acknowledge funding from JWST/NIRCam contract to the University of Arizona NAS5-02015.
The research of CCW is supported by NOIRLab, which is managed by the Association of Universities for Research in Astronomy (AURA) under a cooperative agreement with the National Science Foundation.
WMB, TJL, RM and LS acknowledge support by the Science and Technology Facilities Council (STFC), ERC Advanced Grant 695671 ``QUENCH''.
AJB and JC acknowledge funding from the ``FirstGalaxies'' Advanced Grant from the European Research Council (ERC) under the European Union’s Horizon 2020 research and innovation programme (Grant agreement No. 789056).
DJE is supported as a Simons Investigator.
RH acknowledges funding for this research was provided by the Johns Hopkins University, Institute for Data Intensive Engineering and Science (IDIES).
SC acknowledges support by European Union’s HE ERC Starting Grant No. 101040227 - WINGS.
ECL acknowledges support of an STFC Webb Fellowship (ST/W001438/1).
RM acknowledges funding from a research professorship from the Royal Society.
This research of KB is supported in part by the Australian Research Council Centre of Excellence for All Sky Astrophysics in 3 Dimensions (ASTRO 3D), through project number CE170100013.

This work is based on observations made with the NASA/ESA/CSA James Webb Space Telescope. The data were obtained from the Mikulski Archive for Space Telescopes at the Space Telescope Science Institute, which is operated by the Association of Universities for Research in Astronomy, Inc., under NASA contract NAS 5-03127 for JWST. This research is based (in part) on observations made with the NASA/ESA Hubble Space Telescope obtained from the Space Telescope Science Institute, which is operated by the Association of Universities for Research in Astronomy, Inc., under NASA contract NAS 5–26555. 

This material is based upon High Performance Computing (HPC) resources supported by the University of Arizona TRIF, UITS, and Research, Innovation, and Impact (RII) and maintained by the UArizona Research Technologies department. This work made use of the {\it lux} supercomputer at UC Santa Cruz which is funded by NSF MRI grant AST 1828315, as well as the High Performance Computing (HPC) resources at the University of Arizona which is funded by the Office of Research Discovery and Innovation (ORDI), Chief Information Officer (CIO), and University Information Technology Services (UITS).

\vspace{5mm}
\facilities{HST, JWST}

\software{Prospector \citep{Johnson2021}, FSPS \citep{Conroy2009,Conroy2010}, MIST \citep{Choi2016,Dotter2016}, MILES \citep{Falcon-Barroso2011}, Webbpsf \citep{Perrin2012,Perrin2014}, Photutils \citep{larry_bradley_2022_6825092}}

\appendix

\section{Comparing effective PSF with webbpsf-predicted model PSF} \label{app:psf}

\begin{figure*}
    \centering
    \includegraphics[width=0.97\textwidth]{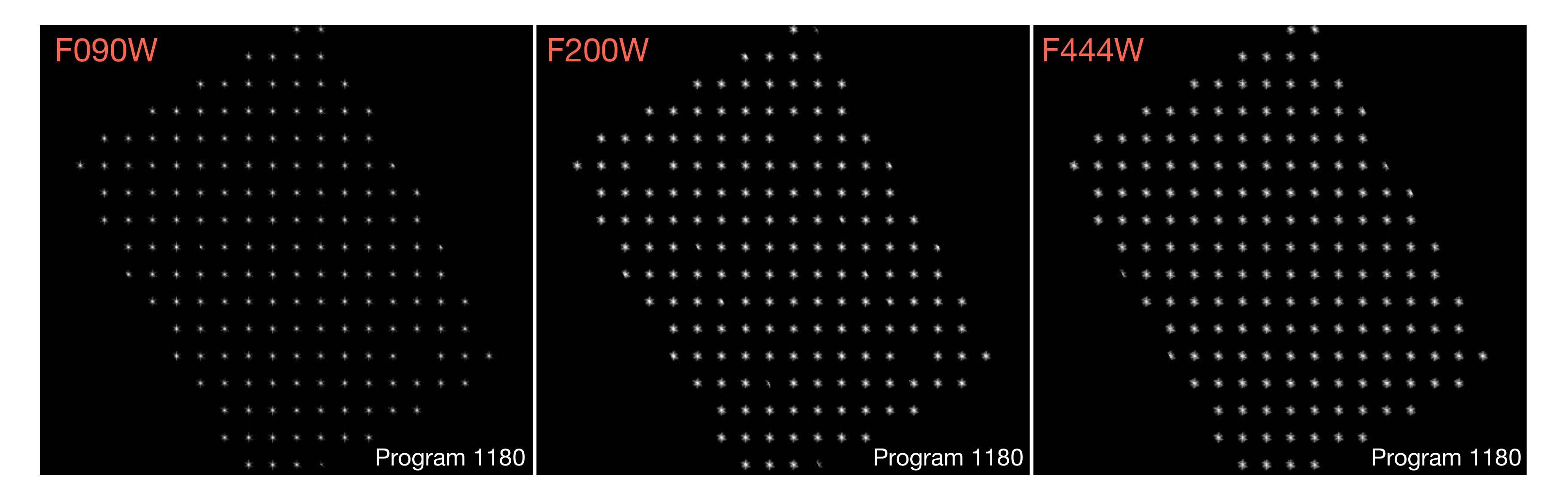}
        \caption{JADES PSF mosaics of NIRCam/F090W (left), F200W (middle) and F444W (right) filters produced following the procedure described in Appendix \ref{app:psf}. }
    \label{fig:psf_mosaic}
\end{figure*}

Knowledge of the Point Spread Function (PSF) plays an essential role in getting accurate PSF-matched aperture photometry, and hence the inferences of physical properties of stellar populations. Per JWST User Documentation\footnote{\url{https://jwst-docs.stsci.edu}}, NIRCam's PSFs are under-sampled in both $<2\micron$ SW filters and $<4\micron$ LW filters. To overcome this, we have constructed empirical and theoretical PSF models for JADES and JEMS with the following three different approaches:
\begin{itemize}
    \item Effective PSF (ePSF) -- We construct ePSFs using the \texttt{EPSFBuilder} in {\sc Photutils}, which adopts the empirical method from \citet{Anderson2000} to iteratively solve the centroids and fluxes of a list of input point sources, and then stack them together.  For the list of point sources, we visually selected 13 isolated point sources from JADES and JEMS. These point sources are bright, but unsaturated across all NIRCam filters used in this study. 
    \item Model PSF (mPSF) -- Two effects are taken into account during the construction of mPSFs. The first effect comes from instrumental responses, which are taken care by the software {\sc Webbpsf}. The other effect comes from data reduction, especially from the mosaicking process. To include both effects to the mPSF construction, we assume a list of synthetic point sources with known WCS coordinates. For each one of the point sources, we first project its WCS coordinate to individual exposures. In each exposure, we then convert the WCS coordinate to the corresponding detector coordinate, and use the {\sc Webbpsf} to predict the PSF at that location, during which we adopt the JWST on-orbit measured Optical Path Difference (OPD) map closest to the date of that exposure. We repeat these for all synthetic point sources. Finally, we use the mosaicking pipeline of JADES to combine individual exposures together and get the mosaicked PSF images. We have made the PSF mosaics for all NIRCam filters used in this work (three examples are shown in Figure \ref{fig:psf_mosaic}). These mosaics allow us to check astrometric accuracy, image quality and to study the spatial variation of PSFs, which will be presented in detail in Ji et al 2023 in preparation. 
    \item Simple {\sc Webbpsf} -- Unlike mPSFs, simple {\sc Webbpsf} models do not include any observational and data reduction effects.
\end{itemize}

\begin{figure*}
    \centering
    \includegraphics[width=0.307\textwidth]{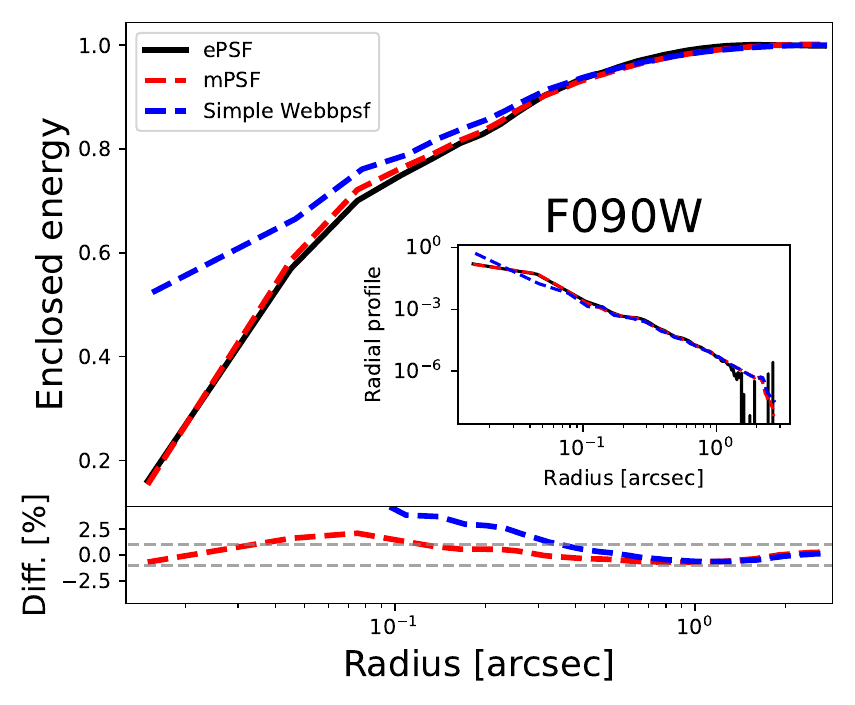}
    \includegraphics[width=0.307\textwidth]{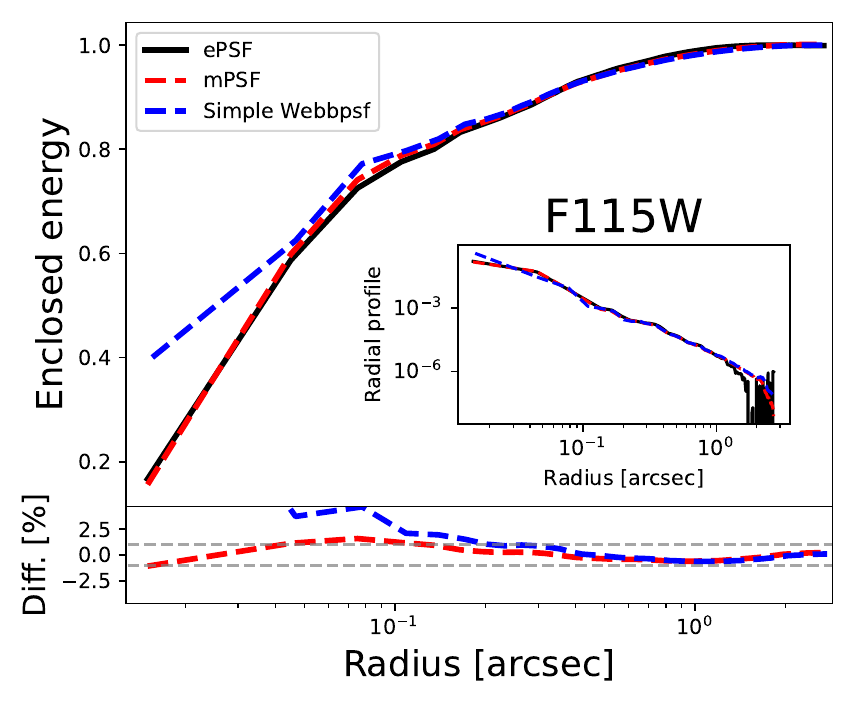}
    \includegraphics[width=0.307\textwidth]{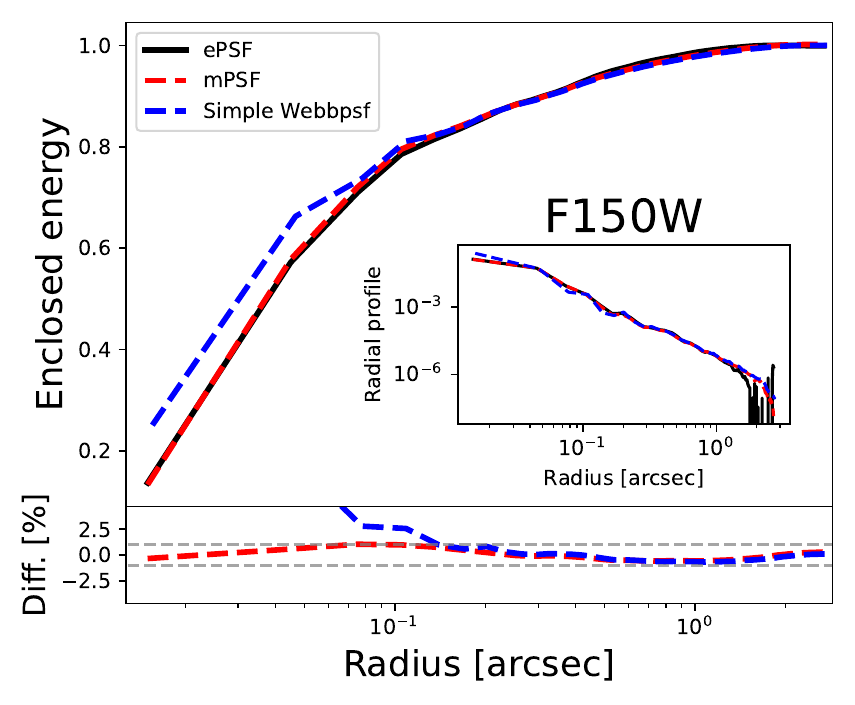}
    \includegraphics[width=0.307\textwidth]{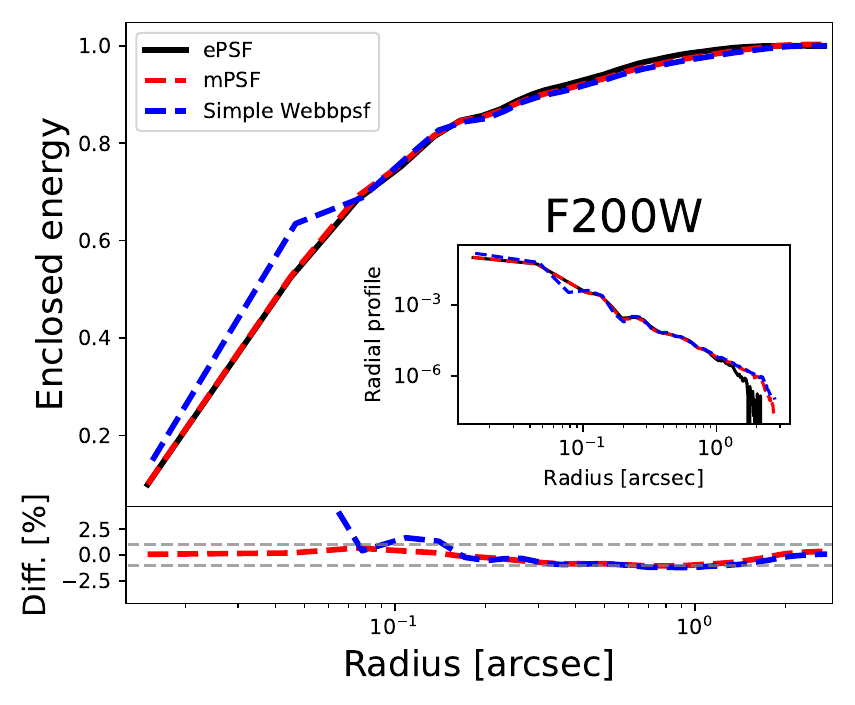}
    \includegraphics[width=0.307\textwidth]{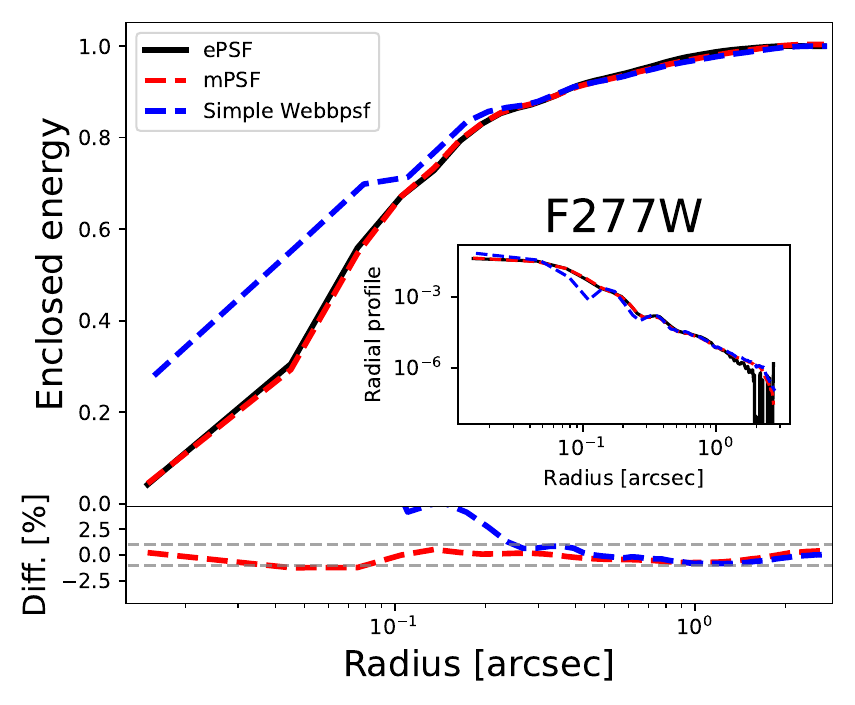}
    \includegraphics[width=0.307\textwidth]{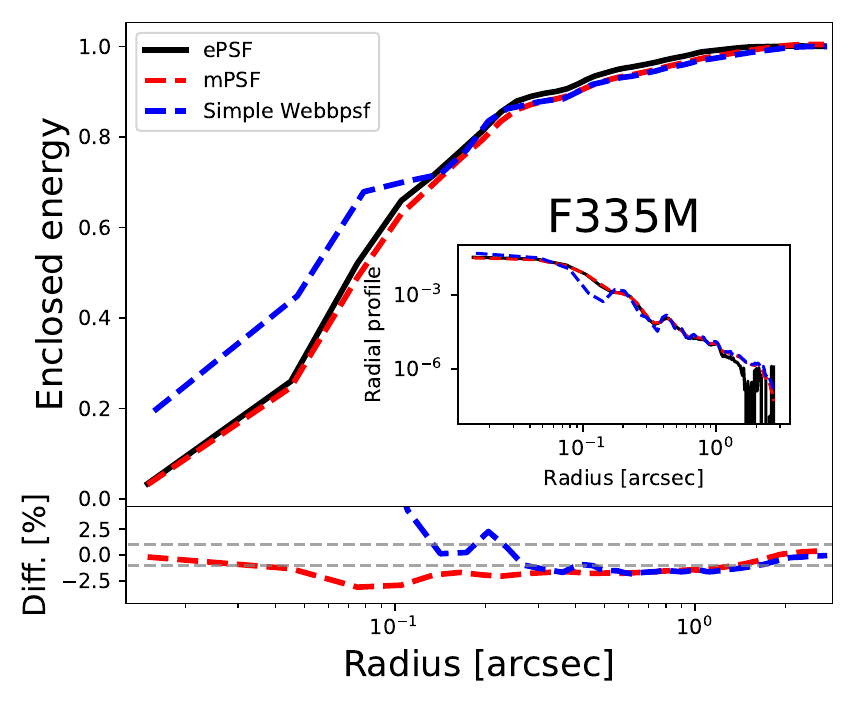}
    \includegraphics[width=0.307\textwidth]{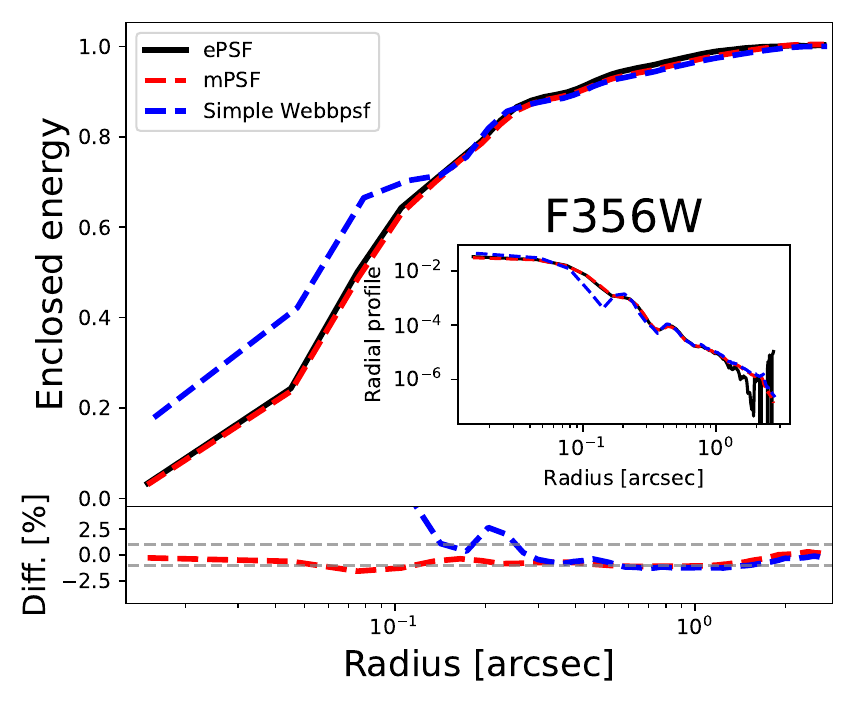}
    \includegraphics[width=0.307\textwidth]{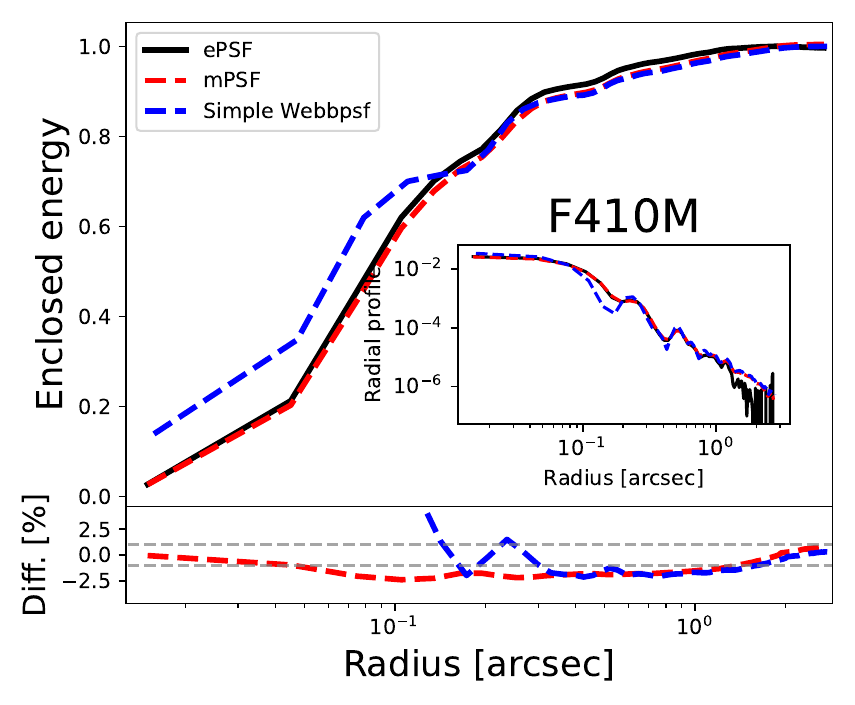}
    \includegraphics[width=0.307\textwidth]{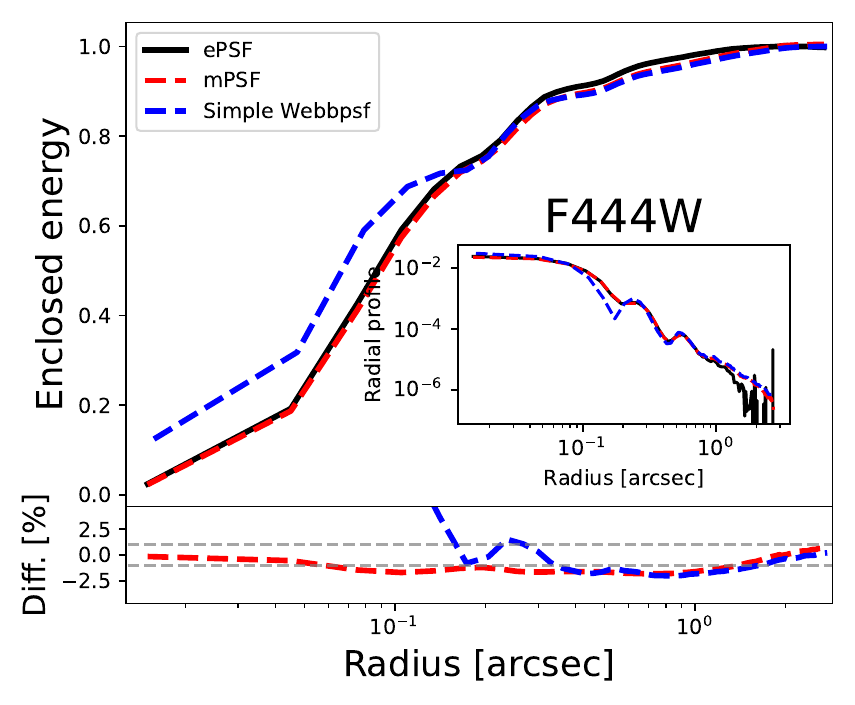}
    \caption{Comparisons among three different PSF models (Appendix \ref{app:psf}) for the NIRCam filters used by JADES. Plotted in the bottom of each panel are the differences between mPSFs and ePSFs (red dashed line), and between  simple {\sc Webbpsf} models and ePSFs (blue dashed line), where the horizontal, grey dashed lines mark $\pm 1\%$ range. Excellent agreement is observed between ePSFs and mPSFs, while simple {\sc Webbpsf} models significantly deviate from ePSFs at small angular scales.}
    \label{fig:PSFs}
\end{figure*}

\begin{figure*}
    \centering
    \includegraphics[width=0.307\textwidth]{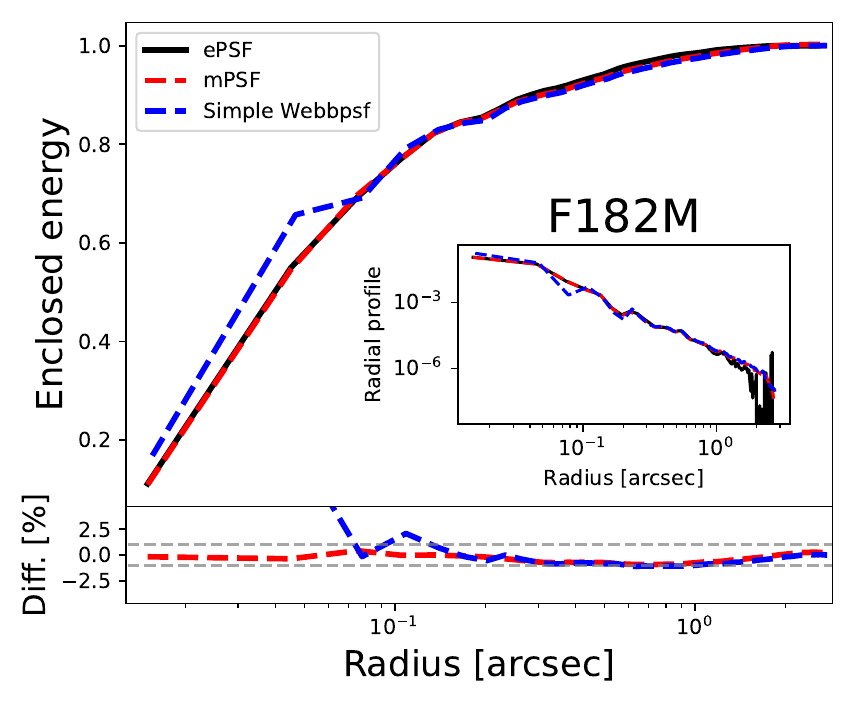}
    \includegraphics[width=0.307\textwidth]{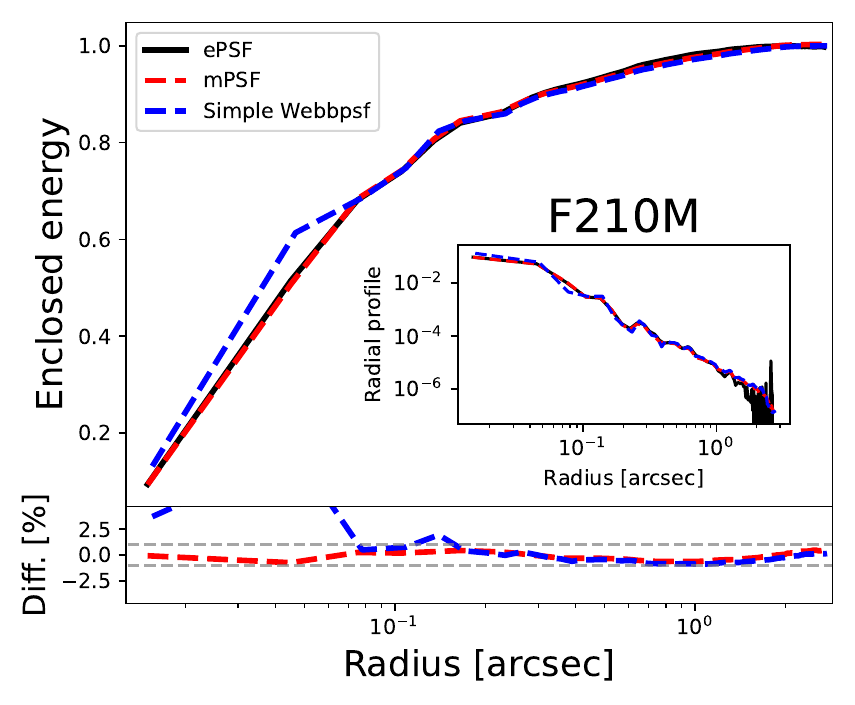}
    \includegraphics[width=0.307\textwidth]{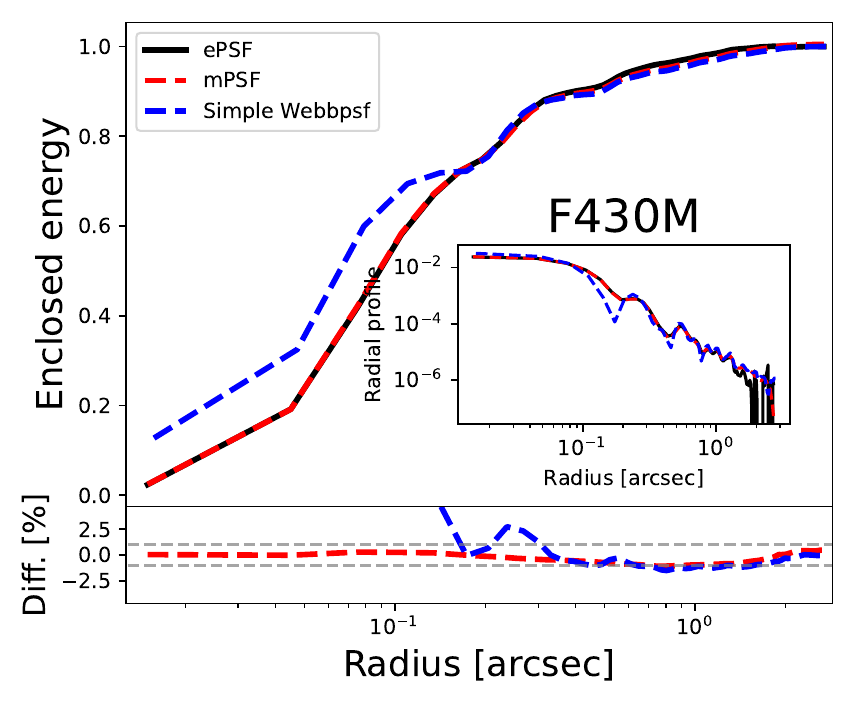}
    \includegraphics[width=0.307\textwidth]{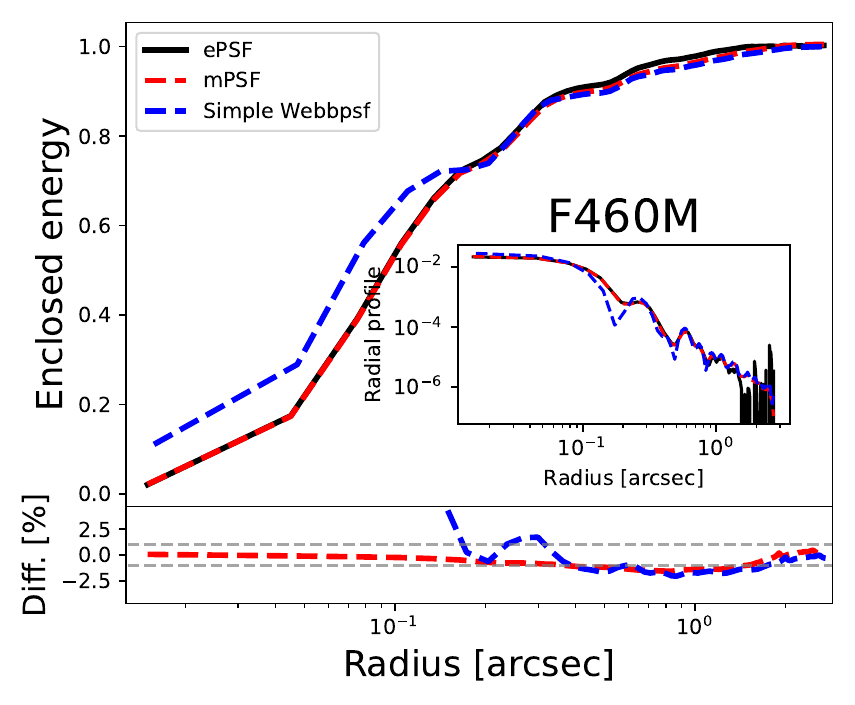}
    \includegraphics[width=0.307\textwidth]{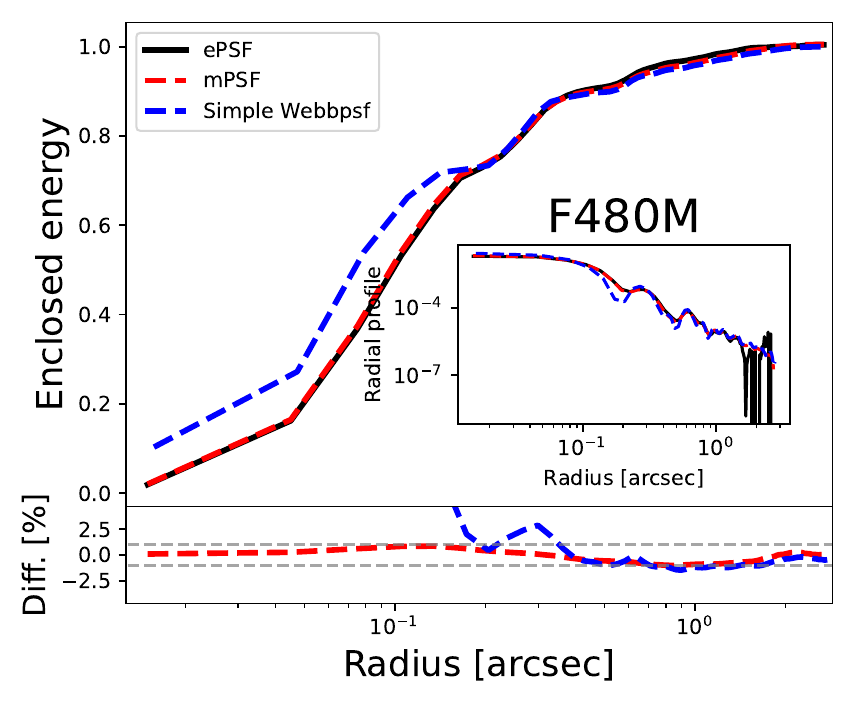}
    
    \caption{Similar to Figure \ref{fig:PSFs}, but for the 5 medium-band NIRCam filters used by JEMS.} \label{fig:PSFs_MB}
\end{figure*}

In Figure \ref{fig:PSFs} and \ref{fig:PSFs_MB}, we compare the above three PSF models. Excellent agreement is seen between ePSFs and mPSFs, with a typical difference of $\lesssim 1\%$ in radial profiles of enclosed energy. We have tested our PSF-matched photometry using the mPSFs, and found no substantial changes at all for the results of this paper. 

We note, however, that the analysis here demonstrates that simple {\sc Webbpsf} models can significantly deviate from ePSFs, especially at small angular scales of $<0.1$ arcsec. Therefore, using simple {\sc Webbpsf} models can lead to significant systematics in the PSF-related analysis, if observational and data reduction effects are not taken into account. Once these effects are properly modelled, the prediction of {\sc Webbpsf} is very accurate. This last point is very important for extragalatic deep surveys like JADES, whose observational configurations (e.g. dithering patterns) can be complex and where not enough high S/N, unsaturated stars are in the field for constructing high-quality ePSF models to large angular scales.

\section{Comparing co-added SFHs with SFHs from SED fitting of integrated flux} \label{app:sfh_comparison}
\begin{figure*}
    \centering
    \includegraphics[width=0.97\textwidth]{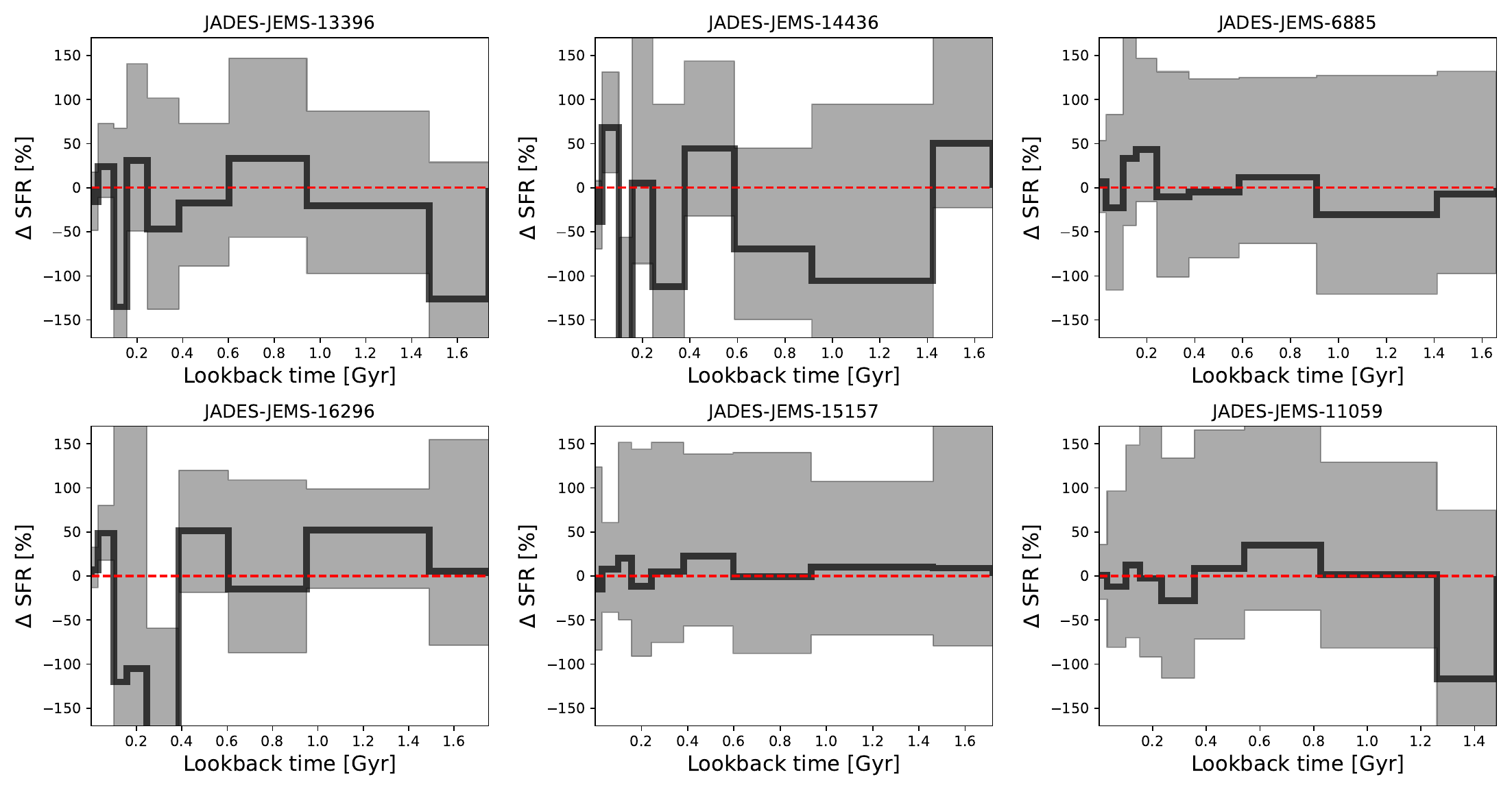}
    \caption{Comparison of the coadded SFH of individual structural components with the SFH derived from SED fitting of integrated flux. The y-axis shows the relative difference, i.e. (coadded - integrated) / integrated, of the two SFHs. Grey shaded region shows $1\sigma$ uncertainty.} \label{fig:sfh_compare}
\end{figure*}

Apart from comparing the coadded stellar mass and SFR with those measured from SED fitting of integrated flux (Figure \ref{fig:coadd_total}), here we compare the coadded SFH of individual structural components, with the SFH from SED fitting of integrated flux. The results are shown in Figure \ref{fig:sfh_compare}. The two SFHs are consistent with each other within uncertainty, further demonstrating the consistency and robustness of our SED fitting results. 

\section{SED fitting with different priors of star formation history} \label{app:prior}

Apart from the fiducial SED fitting with nonparametric SFHs assuming the continuity prior (Section \ref{sec:sed}), we have also tested our \prospector fitting results using the other two SFHs, namely, nonparametric SFHs with the Dirichlet prior \citep{leja2017} and parametric SFHs with a delayed-tau functional form. For the difference between the continuity and Dirichlet priors, we refer readers to \citet{leja2017} for  details. In short, these two priors for nonparametric SFHs are very similar except that the continuity prior is strongly against sudden changes in star formation rate in adjacent lookback time bins. For the delayed-tau model, i.e. $\rm{SFR}(t)\propto (t-t_{age})e^{-(t-t_{age})/\tau}$, we fit $\tau$ with a logarithmically flat prior of $\log (\tau/\rm{Gyr})\in (-2,1)$, and $t_{\rm{age}}$ with a flat prior of $t_{\rm{age}}/\rm{Gyr}\in(10^{-3},t_{\rm{H}})$.

\begin{figure*}
    \centering
    \includegraphics[width=0.77\textwidth]{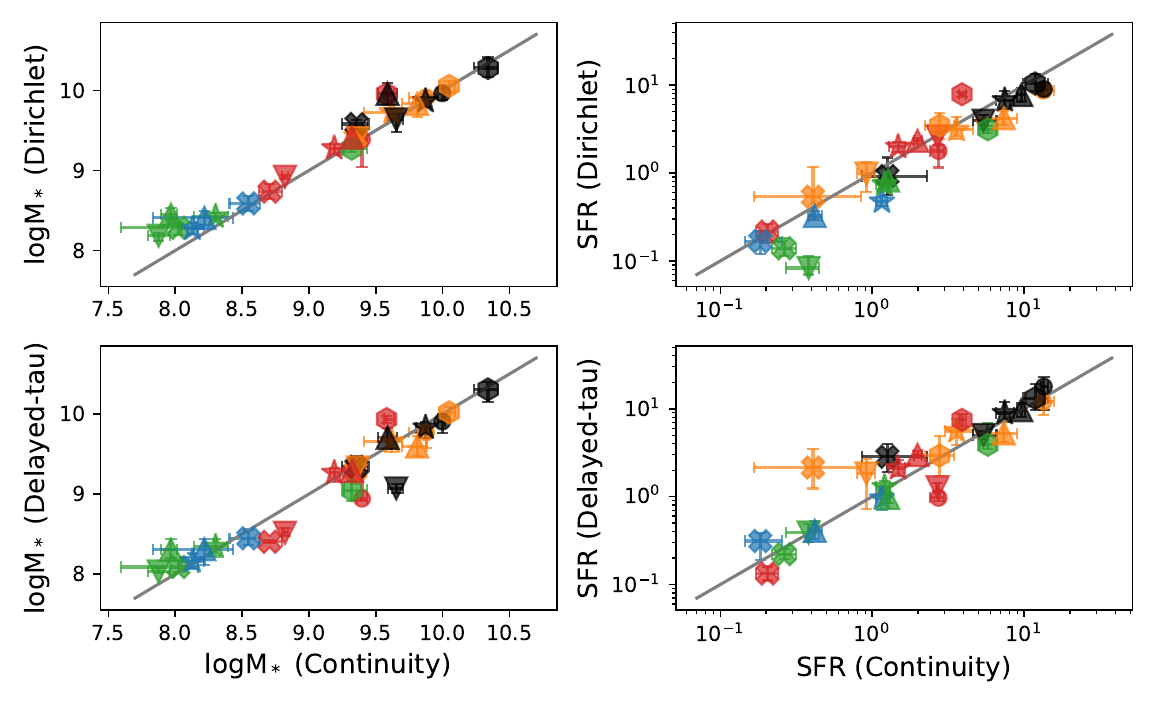}
    \caption{Comparisons of stellar-mass and star formation rate measures from different SFHs for the individual structural components of the 6 galaxies presented in this work. Color-coding is the same as Figure \ref{fig:seds_specz} in the main text (C1 are in red). Symbols used for individual galaxies are the same as Figure \ref{fig:mt} in the main text. The black solid line in each panel marks the one-to-one relationship. }\label{fig:comp_params}
\end{figure*}

\begin{figure*}
    \centering
    \includegraphics[width=\textwidth]{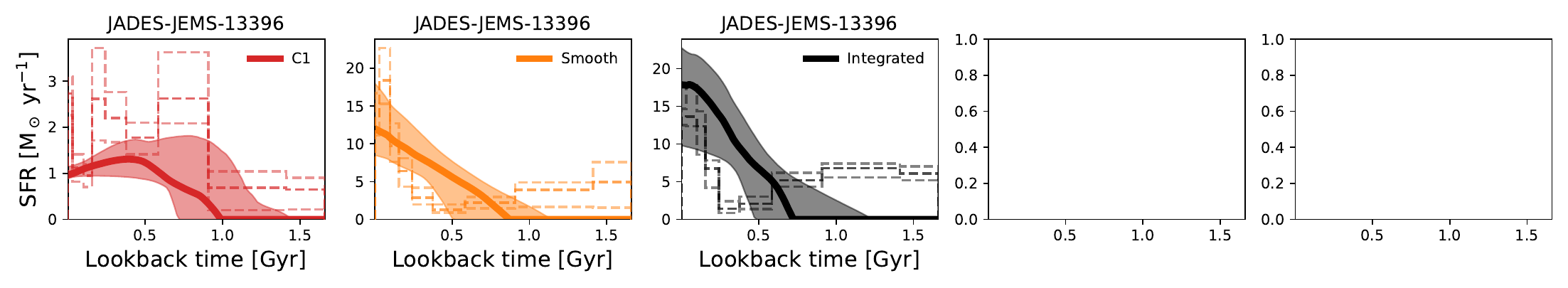}
    \includegraphics[width=\textwidth]{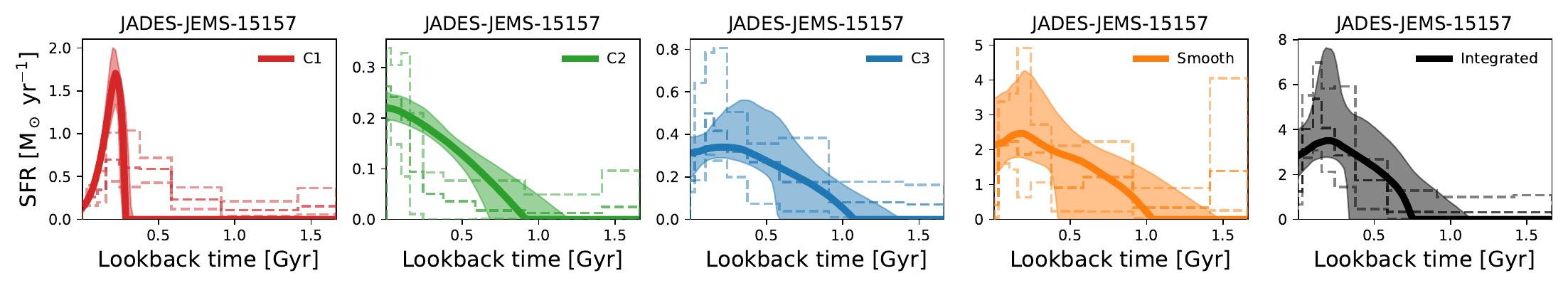}
    \includegraphics[width=\textwidth]{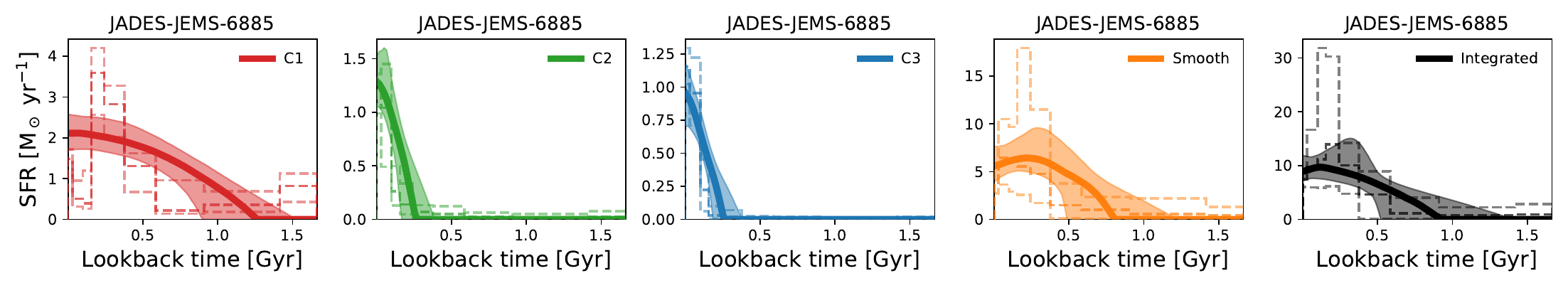}
    \includegraphics[width=\textwidth]{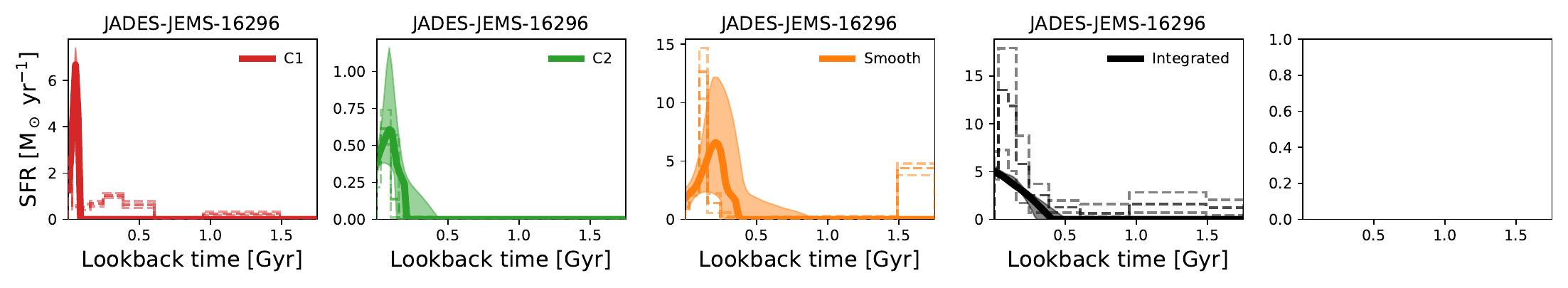}
    \includegraphics[width=\textwidth]{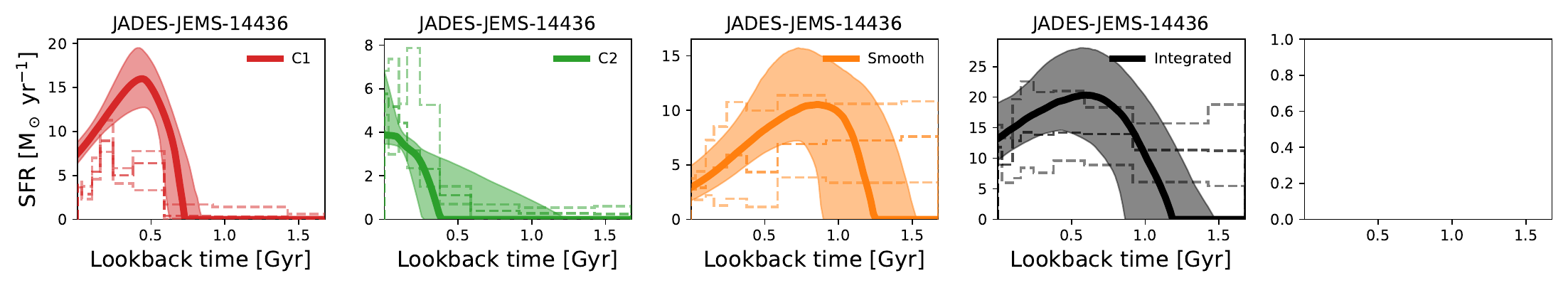}
    \includegraphics[width=\textwidth]{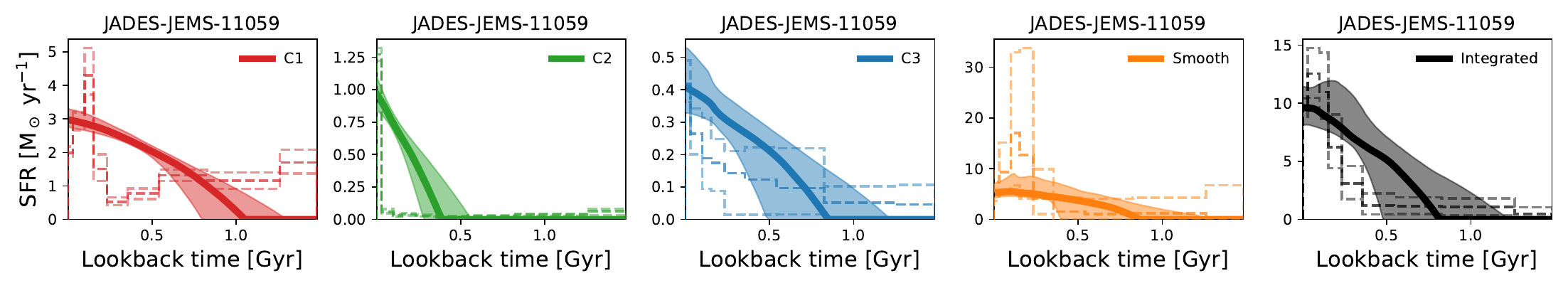}
    \caption{The comparison of the nonparametric SFH reconstructed by our fiducial model (Section \ref{sec:sed}), plotted as dashed lines, with the SFH derived from a parametric delayed-tau model, plotted as solid lines (the best-fit) and shaded regions (1$\sigma$ uncertainty).}
    \label{fig:compare_sfh_param}
\end{figure*}

\begin{figure*}
    \centering
    \includegraphics[width=\textwidth]{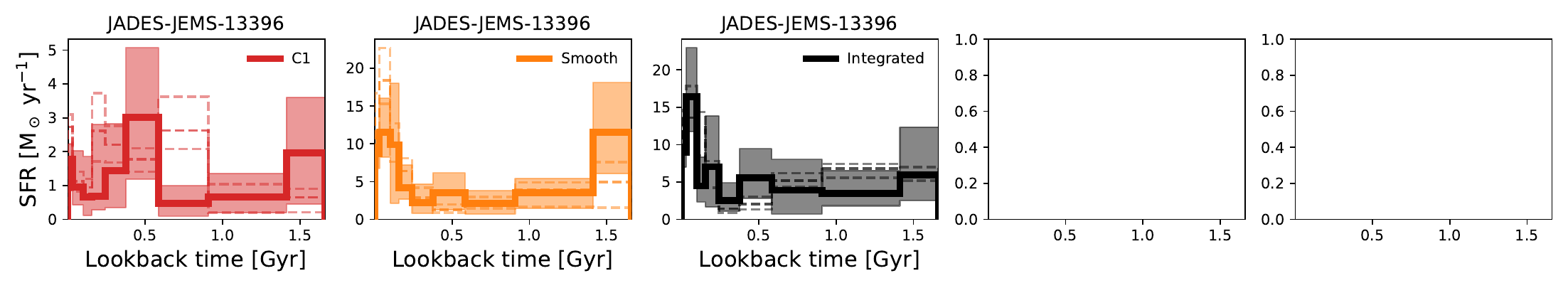}
    \includegraphics[width=\textwidth]{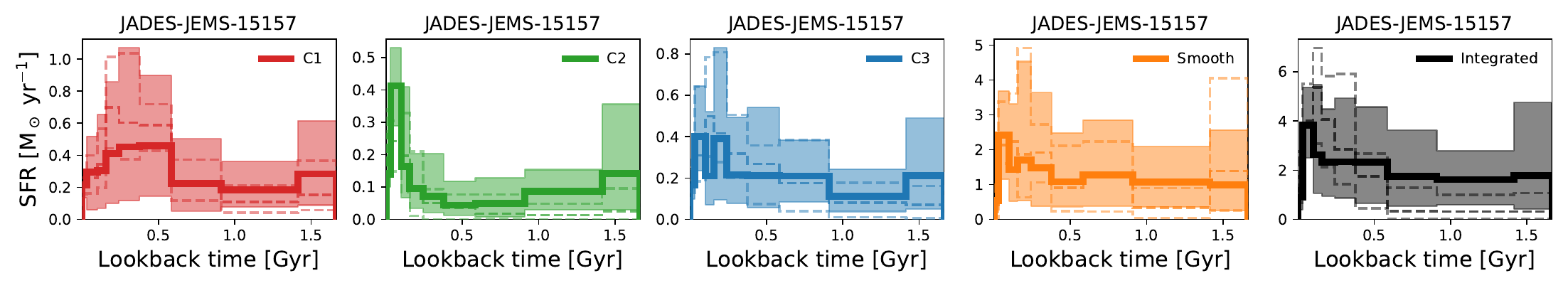}
    \includegraphics[width=\textwidth]{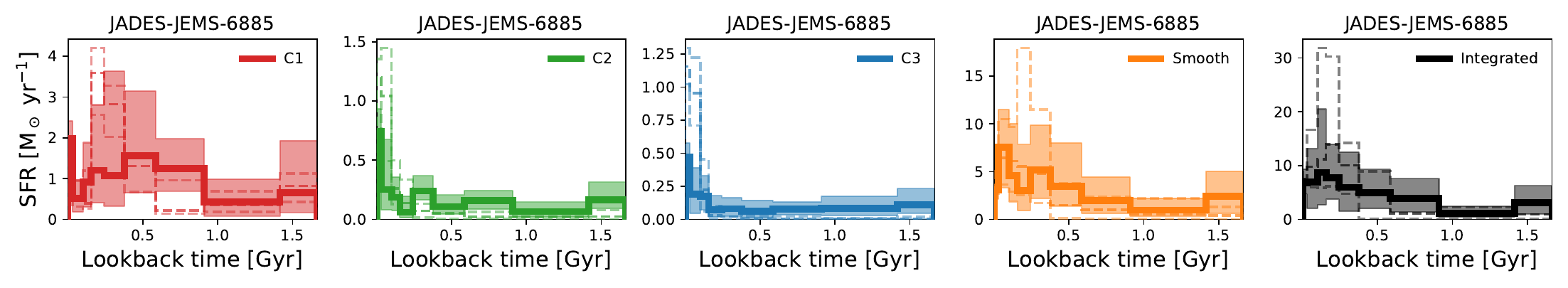}
    \includegraphics[width=\textwidth]{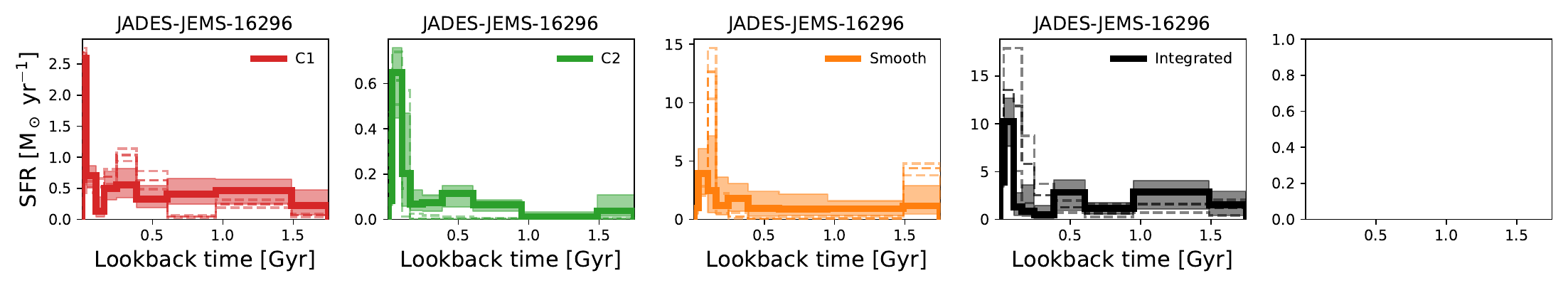}
    \includegraphics[width=\textwidth]{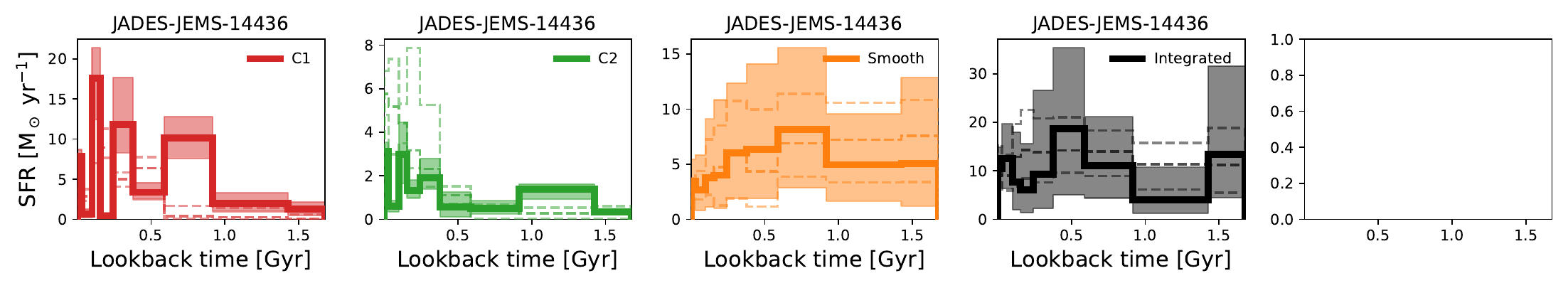}
    \includegraphics[width=\textwidth]{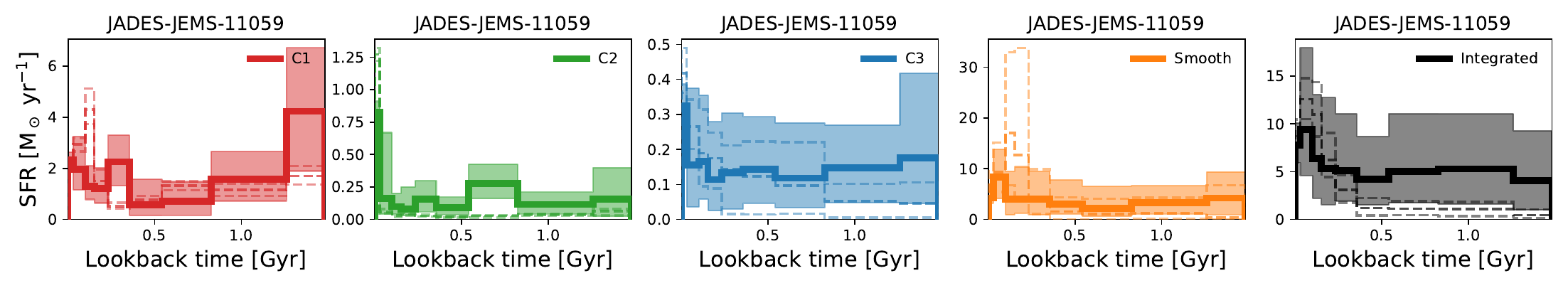}
    \caption{Similar to Figure \ref{fig:compare_sfh_param}, but now comparing with the nonparametric SFH reconstructed using the Dirichlet prior. }
    \label{fig:compare_sfh_prior}
\end{figure*}

In Figure \ref{fig:comp_params}, we compare the measurements of stellar mass and star formation rate from different SFH assumptions. Very good agreement is seen across the three SFH models. In Figure \ref{fig:compare_sfh_param} and \ref{fig:compare_sfh_prior}, we compare the fiducial reconstructed SFHs with the delayed-tau and Dirichlet nonparametric SFHs, respectively. As expected, the nonparamtric SFHs are able to capture more complex shapes than the delayed-tau ones, but the overall shapes of these two SFHs are very similar. Meanwhile, relative to the fiducial SFHs from the continuity prior, the Dirichlet nonparametric SFHs show larger fluctuations in star formation rate in adjacent lookback time bins, but these two nonparametric SFHs are in general mirror each other in terms of their overall shapes. The tests here show that the results of this work are not sensitive to the  assumptions of SED models.

\section{Detailed descriptions of the spatially resolved stellar populations of the 6 galaxies} \label{app:ind}
{\bf JADES-JEMS-13396.} The top row of Figure \ref{fig:seds_specz} shows the SED fitting result of this galaxy. Despite that the mass-weighted stellar age of C1 is consistent with that of the Smooth component, the shapes of their SFHs are different. For the Smooth component, its mass assembly was at a relatively low, but non-negligible rate in early times of $t_{\rm{lookback}}>0.4$ Gyr, followed by a rapid increase in SFR. For C1, however, its most mass formed through a recent star-formation episode during $t_{\rm{lookback}}=0.2\sim0.8$ Gyr. By the time of observation, the Smooth component is forming stars at a higher rate than C1, with the sSFR of the former being $\approx$ 1.6 times larger (Table \ref{tab:sed_info}). These differences in SFH shapes qualitatively remains if we instead use a parametric delayed-tau or nonparametric Dirichlet SFH priors (Figure \ref{fig:compare_sfh_param} and \ref{fig:compare_sfh_prior}). We also notice that a color gradient is clearly present in the Smooth component, with the upper-left side (likely contains several faint clumps in it) being bluer than its rest parts.

{\bf JADES-JEMS-15157.} The middle row of Figure \ref{fig:seds_specz} shows the SED fitting result of this galaxy. Unlike C2 having a monotonically rising SFH, the SFH of C1 gradually rises toward a relatively recent peak at \tlookback $\approx0.3$ Gyr, followed by a rapid decline over the past 0.2 Gyr. By the time of observation, the sSFR of C1 is only $\approx0.4$ Gyr$^{-1}$, $\approx$ 7.5 times smaller than that of C2 (Table \ref{tab:sed_info}). We note that C3 has a similar stellar population to that of C1, but this is because the spatial separation between C1 and C3 is so small -- about 0.1" that is smaller than the angular resolution of NIRCam/F444W imaging -- that the aperture photometry (Section \ref{sec:phot}) for one substructure is significantly affected by the other. Finally, while C1 and the Smooth component share a similar SFH in early times, they differentiate from each other at late times: unlike C1 whose SFR quickly declined after the \tlookback $\approx0.3$ Gyr peak, the Smooth component retained that peak SFR toward a later time of \tlookback $<0.1$ Gyr. All these remain qualitatively unchanged if we instead use a parametric delayed-tau or nonparametric Dirichlet SFH priors (Figure \ref{fig:compare_sfh_param} and \ref{fig:compare_sfh_prior}).

{\bf JADES-JEMS-6885.} The bottom row of Figure \ref{fig:seds_specz} shows the SED fitting result of this galaxy. There is no significant mass assembly before \tlookback $>1$ Gyr for all structural components, but their SFHs are different over the recent 1 Gyr. Compared to C2 and C3, C1 is older by $0.3\sim0.5$ Gyr  (Table \ref{tab:sed_info}). Moreover, C1 has a poststarburst-like SFH, namely, it experienced a burst of star formation during \tlookback $=0.2\sim0.6$ Gyr, followed by a rapid decline of SFR. Differing from C1, C2 and C3 are experiencing on-going bursts of star formation started $\approx$ 0.2 Gyr ago. By the time of observation, the sSFR of C1 is 6.3 and 9.3 times smaller than that of C2 and C3, respectively. Finally, while the Smooth component generally mirrored the shape of C1's SFH before the peak (\tlookback $\approx0.2$ Gyr), they differed from each other afterward: the SFR of C1 quickly declined to a very low level, while the Smooth component retained the peak SFR to a later time of \tlookback $<0.1$ Gyr. All these differences in the SFHs of different structural components remain qualitatively unchanged if we instead use the other two SFH priors (Figure \ref{fig:compare_sfh_param} and \ref{fig:compare_sfh_prior}).

{\bf JADES-JEMS-16296.} The top row of Figure \ref{fig:seds_photz} shows the SED fitting result of this galaxy. Both C1 and C2 are experiencing bursts of star formation started $\lesssim0.1$ Gyr ago, and their instantaneous sSFRs are similar, $\approx 5$ Gyr$^{-1}$. While it is not seen from the delayed-tau SFH (Figure \ref{fig:compare_sfh_param}), the reconstructed nonparametric SFH (Figure \ref{fig:seds_photz} and \ref{fig:compare_sfh_prior}) suggests that older (\tlookback $\gtrsim0.5$ Gyr) stellar populations are also present in C1, leading to its larger mass-weighted stellar age than C2 (Table \ref{tab:sed_info}).  

{\bf JADES-JEMS-14436.} The middle row of Figure \ref{fig:seds_photz} shows the SED fitting result of this galaxy. The major mass assembly event of C1 started at \tlookback $\approx0.6$ Gyr, peaked at \tlookback $\approx0.2$ Gyr and gradually declined afterward. For C2, its SFR monotonically increased since \tlookback $\approx0.8$ Gyr, before which its SFR was consistent with zero. There also seems to be a tidal tail associated with C2 (see e.g. NIRCam F150W cutouts in Figure \ref{fig:poststamps_cont}), so this is a possible merging system. Also noticed for this galaxy is that the Smooth component has a very extended SFH, namely, its mass assembly began at \tlookback $>1$ Gyr, corresponding to redshift $z>7$, and the SFR started to decline since \tlookback $\approx0.6$ Gyr. All these features in SFHs remain qualitatively unchanged if we instead use the other two SFH priors (Figure \ref{fig:compare_sfh_param} and \ref{fig:compare_sfh_prior}).

{\bf JADES-JEMS-11059.} The bottom row of Figure \ref{fig:seds_photz} shows the SED fitting result of this galaxy. C1 contains old stellar populations with age $>1$ Gyr, and it also has a recent burst of star formation that started $\approx0.2$ Gyr ago, peaked at \tlookback $\approx0.1$ Gyr and declined afterward. Meanwhile, C3 has an overall rising SFH. For C2, although it is not in the segmentation map of the galaxy (Figure \ref{fig:poststamps_cont}), it is still included to the analysis because its photometric redshift solution is consistent with that of the galaxy JADES-JEMS-11059. C2 is experiencing a starburst started $\lesssim0.3$ Gyr ago.

\bibliography{ji_jades_mb_2022}{}
\bibliographystyle{aasjournal}

\end{document}